\documentclass[fleqn,usenatbib]{mnras}
\usepackage{import}

\usepackage{afterpage}
\usepackage{placeins}

\import{./front_matter/}{preamble.sty}

\title[LEGOII: 3 mm molecular lines across 100 pc of W49]{LEGO II: A 3\,mm molecular line study covering 100\,pc of one of the most actively star-forming portions within the Milky Way Disc}

\author[A.T.~Barnes et al.]{A.T.~Barnes,$^{1}$\thanks{E-mail: ashleybarnes.astro@gmail.com}
J.~Kauffmann,$^{2}$ 
F.~Bigiel,$^{1}$
N.~Brinkmann,$^{3}$
D.~Colombo,$^{3}$ \and
A.E~Guzm\'{a}n,$^{4}$ 
W.J.~Kim,$^{5}$
L.~Sz\H{u}cs,$^{6}$
V.~Wakelam,$^{7}$
S.~Aalto,$^{8}$ 
T.~Albertsson,$^{3}$ \and 
N.J.~Evans~II,$^{9}$ 
S.C.O.~Glover,$^{10}$  
P.F.~Goldsmith,$^{11}$
C.~Kramer,$^{12}$  \and
K.~Menten,$^{3}$  
Y.~Nishimura$^{13, 14}$,
S.~Viti,$^{15}$ 
Y.~Watanabe,$^{16}$
A.~Weiss,$^{3}$  \and 
M. Wienen,$^{3,17}$ 
H.~Wiesemeyer,$^{3}$ 
and F.~Wyrowski$^{3}$ 
\\
Affiliations are listed at the end of the paper. 
}

\date{Accepted 2020 June 12; Revised 2020 June 09; Received 2020 April 25}

\pubyear{2020}

\def \araa {{\rm {ARA\&A}}}
\def \apj {{\rm {ApJ}}}

\def \apjs {{\rm {ApJS}}}
\def \apjl {{\rm {ApJL}}}

\def \aap {{\rm {A\&A}}}

\def \mnras {{\rm {MNRAS}}}

\def \pasp {{\rm {PASP}}}

\def \jaa {{\rm JAA}}

\begin{document}


\label{firstpage}
\pagerange{\pageref{firstpage}--\pageref{lastpage}}
\maketitle

\begin{abstract}
The current generation of (sub)mm-telescopes has allowed molecular line emission to become a major tool for studying the physical, kinematic, and chemical properties of extragalactic systems, yet exploiting these observations requires a detailed understanding of where emission lines originate within the Milky Way. In this paper, we present 60\arcsec\ ($\sim$\,3\,pc) resolution observations of many 3mm-band molecular lines across a large map of the W49 massive star-forming region ($\sim$\,100\,$\times$\,100\,pc at 11\,kpc), which were taken as part of the ``LEGO'' IRAM-30m large project. We find that the spatial extent or brightness of the molecular line transitions are not well correlated with their critical densities, highlighting abundance and optical depth must be considered when estimating line emission characteristics. We explore how the total emission and emission efficiency (i.e. line brightness per H$_{2}$ column density) of the line emission vary as a function of molecular hydrogen column density and dust temperature. We find that there is not a single region of this parameter space responsible for the brightest and most efficiently emitting gas for all species. For example, we find that the HCN transition shows high emission efficiency at high column density ($10^{22}$\,\cmsq) and moderate temperatures (35\,K), whilst e.g. \NtHp\ emits most efficiently towards lower temperatures ($10^{22}$\,\cmsq; <20\,K). We determine $X_{\mathrm{CO} (1-0)} \sim 0.3 \times 10^{20} \,\mathrm{cm^{-2}\,(K\,km\,s^{-1})^{-1}}$, and $\upalpha_{\mathrm{HCN} (1-0)} \sim 30\,\mathrm{M_\odot\,(K\,km\,s^{-1}\,pc^2)^{-1}}$, which both differ significantly from the commonly adopted values. In all, these results suggest caution should be taken when interpreting molecular line emission.
\end{abstract}





\begin{figure*}
	\includegraphics[width=\textwidth]{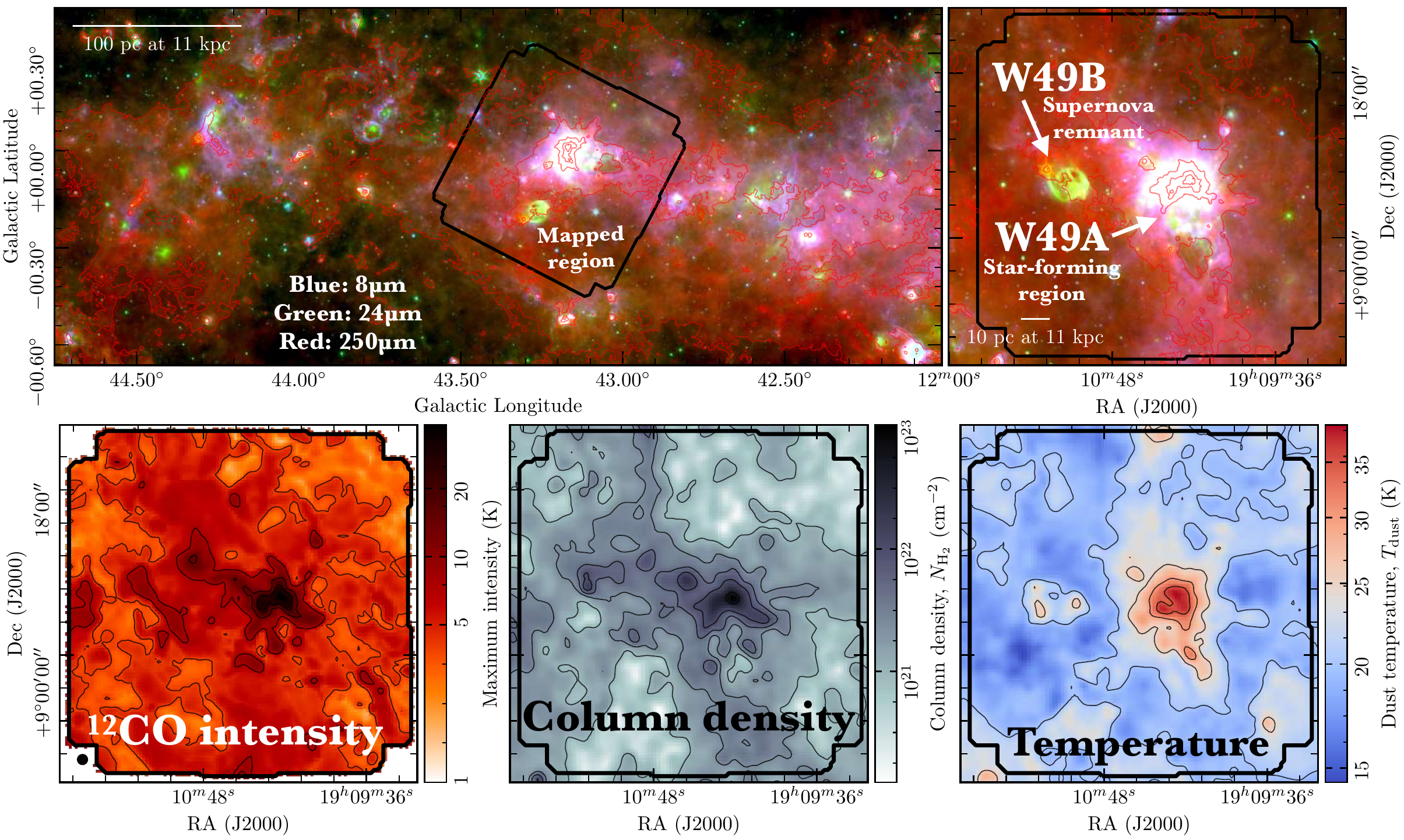}
	\caption{The region around W49 within the plane of the Milky Way. The upper panel shows a three-colour image across the Galactic plane, and a zoom-in on the mapped region of W49 indicated by the heavy black line (see Section\,\ref{sec:observations}). Shown are the {\it Spitzer} 8~\micron\ in blue, {\it Spitzer} 24~\micron\ in green, and {\it Herschel} 250~\micron\ in red \citep{carey_2009, churchwell_2009, molinari_2010}. Overlaid as red contours is the 250~\micron\ map in levels of 100, 200, 400, 1000, 2000, 10000~MJy\,sr$^{-1}$, which were chosen to best highlight the dense dust/gas distribution \citep{molinari_2010}. The labels on the zoom-in panel show the positions of the W49A star-forming region and the W49B supernova remnant. Shown in the main and zoom-in panels are scale bars of 100\,pc and 10\,pc at a distance of 11\,kpc, respectively. The lower row, left panel shows a map of the peak brightness temperature from the $^{12}$CO molecular line, overlaid with black contours of 4, 8, 15, 22, 25~K. This $^{12}$CO map highlights the extended nature of the molecular emission, which completely fills the mapped region. The angular beam size of 60\arcsec, or $\sim$3\,pc at the distance of W49A ($\sim$11\,kpc; \citealp{gwinn_1992, zhang_2013}), is shown as a black circle in the lower left of the lower left panel. The lower row, centre and right panels show maps of the molecular hydrogen column density \NHtwo\ (centre panel) and dust temperature \Tdust\ (right panel) derived from {\it Herschel} observations. These maps have been smoothed to an angular resolution of 60\arcsec\ to match the resolution of the IRAM-30m observations. Black contours of \NHtwo\ have been overlaid on the \NHtwo\ map in levels of 1, 2.5, 5, 10, 50, 100\,$\times 10^{21}$~\cmsq, and black contours of \Tdust\ have been overlaid on the \Tdust\ map in levels of 10, 15, 20, 25 and 30\,K.}
 	\label{fig:w49_rgb}
\end{figure*}

\section{Introduction}

A full understanding of the process of star formation from dense molecular clouds is still one of the major outstanding problems in astronomy. Many fundamental open questions are still hotly debated, such as: understanding the relative balance of turbulence, magnetic field, and gravitational energy (e.g. \citealp{krumholz_2005a, padoan_2011, hennebelle_2013}), how these properties can be observationally determined (e.g. \citealp{federrath_2010, orkisz_2017}), how these properties may vary with Galactic environment (e.g. \citealp{federrath_2016, barnes_2017}), and if they are important for the fraction of ``dense'' gas that is eventually converted into stars (e.g. \citealp{lada_2010, lada_2012}). 

The luminosities, $L_\mathrm{Q}$, from the now extensive collection of known molecular line transitions, $Q$,\footnote{As of November 2019, around 210 molecules have been detected in the interstellar medium \citep[][\url{https://cdms.astro.uni-koeln.de/classic/molecules}]{endres_2016}.} originating from molecular clouds are useful tools for studying the above-mentioned properties. In particular, transitions with high critical densities are regularly used in the literature to selectively study the relatively denser gas within molecular clouds. For example, it is typically assumed that the HCN\,($J=1-0$) molecular line transition is a particularly good tracer of ``dense gas'' (number density of H$_2$ above 10$^{4}$\cmcb), owing to its high critical density for emission, and a chemical formation pathway that produces abundant HCN molecules within cold, dense gas (e.g. \citealp{boger_2005}). Under specific physical assumptions, the luminosity of HCN\,($J=1-0$), $L_\mathrm{HCN(1-0)}$ can be used to estimate the dense molecular gas mass, $M_\mathrm{dg}$, within star-forming regions \citep{gao_2004a, gao_2004b, wu_2010}. A measure of the dense gas mass, along with the star formation rate (SFR), $\dot{M_{*}}$, has been used to constrain the models for star formation through, for example, the star formation efficiency (SFE) of dense gas ($\dot{M_{*}} / M_\mathrm{dg}$). However, with the advent of more sensitive radio telescopes that are capable of efficiently mapping large regions of the sky, it has been pointed out that the $L_\mathrm{HCN(1-0)} \propto M_\mathrm{dg}$ relation may not hold across environments (e.g. \citealp{kauffmann_2017, pety_2017, shimajiri_2017, evans_2020, nguyenluong_2020}). The results from these authors suggest that the underlying assumption that molecular line emission originates solely from gas around and above the critical density is fundamentally flawed, despite this being made by many molecular line studies. Indeed, \citet{evans_1989} highlighted three decades ago that such an assumption breaks down out of the Rayleigh-Jeans limit in the sub-mm regime, where sub-thermal emission is the typical mode for emission before quickly becoming optically thick just above the critical density. If anything then, the opposite is true, and emission from molecular lines within the sub-mm regime traces gas with densities up to the critical density. Moreover, there could be a contribution to $L_\mathrm{Q}$ from various additional excitation mechanisms (e.g. electron excitation; \citealp{goldsmith_2018}).

Investigating the emission properties of these dense gas tracers -- and many other commonly observed molecular line transitions -- across a broad range of environments is, therefore, of current critical importance (e.g. \citealp{ nishimura_2017, watanabe_2017}). This is of particular relevance now more than ever, as these dense molecular gas tracers are becoming routinely observable in other galaxies (e.g. \citealp{bigiel_2016}), where, even at the highest achievable spatial resolution for the closest disc galaxies, each line-of-sight contains a very broad range of environments (e.g. gas densities and temperatures; e.g. \citealp{jimenezdonaire_2019}).

One of the primary aims of the Molecular Line Emission as a Tool for Galaxy Observations (LEGO) project is to further investigate the above issue by obtaining large maps of many commonly observed 3\,mm-band molecular lines, towards a selection of Milky Way star-forming regions that span a range of environmental properties (e.g. cloud density, star formation feedback, galactic environment, and metallicity). Building on the initial results from the Orion nebula presented by \citet{kauffmann_2017}, we present in this work observations of the W49 massive star-forming region, which represents the most distant and, in terms of star formation rate, extreme source within the LEGO sample.


A three-colour image of the W49 region and its position within the galactic plane of the Milky Way is presented in the upper row of Figure\,\ref{fig:w49_rgb}. The bright blues ({\it Spitzer} 8~\micron\ emission) and greens ({\it Spitzer} 24~\micron\ emission) within this image highlight the regions of ongoing star formation, whilst the reds ({\it Herschel} 250~\micron\ emission) show the cold dust (\citealp{carey_2009, churchwell_2009, molinari_2010}). The mapped region (indicated by the black box) is of particular interest for a study of how the gas emission properties vary with environment, thanks to the large range of physical regimes present within such a relatively small projected area on the sky. Table\,\ref{table:source_info} summarises several properties of interest for the mapped region.
 
The W49A region highlighted in Figure\,\ref{fig:w49_rgb} is a very well studied actively star-forming region that contains high-mass stars and associated \ion{H}{ii} regions (total stellar mass of 10$^{4-5}$\Msol; e.g. \citealp{homeier_2005}). Thanks to these \ion{H}{ii} regions, it is one of the brightest known radio sources outside of the Galactic Centre (e.g. \citealp{westerhout_1958}). Therefore, W49A has been the subject of many radio continuum and recombination line investigations that have tried to characterise the embedded stellar objects and ionised gas properties \citep{anantharamaiah_1985, depree_1997, depree_2000, rugel_2018b}. Moreover, and of particular interest to the study presented here, the central part of this region has been the subject of several mm-wavelength spectroscopic observations, which find a wealth of molecular line emission \citep{roberts_2011, nagy_2012, galvan-madrid_2013, nagy_2015}. Parallax distance measurements using H$_2$O maser emission towards W49A place it at a distance of $11.11^{+0.79}_{-0.69}$\,kpc, locating it in a distant section of the Perseus arm near the solar circle in the first Galactic quadrant \citep{gwinn_1992, zhang_2013}. The W49B supernova remnant is also highlighted in Figure\,\ref{fig:w49_rgb}, and stands out as a bright feature in 24\,\micron\ emission (i.e. green in the upper right panel).
 
The paper is organised as follows. Section\,\ref{sec:observations} describes the IRAM-30m observations, the procedure for creating the integrated intensity maps, and the {\it Herschel} derived molecular hydrogen column density and dust temperature maps. Section\,\ref{sec:results} presents the results of the comparison of integrated intensities of the various molecular lines to the column density. Section\,\ref{sec:analysis} presents an analysis of where the majority of the emission is emitted within our mapped region, created by applying regional and dust temperature masks. Section\,\ref{sec:discussion} places these in the context of our understanding of molecular line emission. In Section\,\ref{sec:conclusions}, we summarise the main results of this work. 

\begin{table*}
	\caption{Source properties of the W49 region.}
	\begin{tabular}{lcc}
\hline
Parameter & Value & Notes \\
\hline
Distance & $11.11^{+0.79}_{-0.69}$\,kpc & 60\arcsec $\approx$ 3pc \citep{gwinn_1992, zhang_2013} \\
Velocity range of all emission & [-20.0 to 90.0]\,\kms & LSR, radio convention \\
Systemic velocity of W49A & [-10.0 to 20.0]\,\kms & LSR, radio convention (e.g. \citealp{galvan-madrid_2013}) \\
Projection centre & 19$^\mathrm{h}$10$^\mathrm{m}$32.487$^\mathrm{s}$, 9\degree05\arcmin36.532\arcsec & RA, Dec (J2000), or $l,b$\,=\,43.19, -0.06 (see Figure\,\ref{fig:w49_rgb})\\
Field of view & 0.54\degree\, $\times$ \,0.54\degree & $\sim$\,100\,$\times$\,100 \,pc at 11\,kpc (see Figure\,\ref{fig:w49_rgb})\\
Column density, \NHtwo  & [0.1, 2.1, 210] $\times10^{21}$\cmsq & $N_\mathrm{H_2}^\mathrm{min}$, $N_\mathrm{H_2}^\mathrm{mean}$, $N_\mathrm{H_2}^\mathrm{max}$ (Section\,\ref{subsec:NhTdust} and Appendix\,\ref{appendix_colerr}) \\
Dust temperature, \Tdust & [14.4, 20.5, 38.9]\,K & $T_\mathrm{dust}^\mathrm{min}$, $T_\mathrm{dust}^\mathrm{mean}$, $T_\mathrm{dust}^\mathrm{max}$ (Section\,\ref{subsec:NhTdust} and Appendix\,\ref{appendix_colerr}) \\
\hline
	\end{tabular}
	\label{table:source_info}
\end{table*}

\section{Observations}\label{sec:observations}

\begin{table*}
	\caption{Information on the selected observed molecular lines, ordered by increasing rest frequency. Section~\ref{sec:line-selection} describes how the lines recorded in this table were selected, and how the line characteristics recorded here were obtained. Columns 1 to 10 show the name of each molecule, the transition information (used to refer to each transition throughout), the frequency of the transition, the upper energy level of the transition, the Einstein spontaneous decay coefficient, the collisional deexcitation rate coefficients at a kinetic temperature of 20\,K, and the critical and effective densities for emission.$^{1}$ Transitions that are not available within the LAMDA database have the corresponding information blanked. Additional information on the molecular line database used within this work can be found in Table\,\ref{table:line_database}. The full, machine-readable version of this Table can be obtained from the supplementary online material.}
	\begin{tabular}{lccccccccc}
\hline
Species & Transition & Frequencies & $E_{\rm u} / k_{\rm B}$ & $A_{ul}$ & $C_{ul}^\mathrm{2lvl}$ & $C_{ul}$ & $n_{\rm crit}^\mathrm{2lvl}$ & $n_{\rm crit}$ & $n_{\rm eff}$ \smallskip\\
 &  & ($\mathrm{GHz}$) & ($\mathrm{K}$) & ($\mathrm{s^{-1}}$) & ($\mathrm{cm^{3}\,s^{-1}}$) & ($\mathrm{cm^{3}\,s^{-1}}$) & ($\mathrm{cm^{-3}}$) & ($\mathrm{cm^{-3}}$) & ($\mathrm{cm^{-3}}$) \\
\hline

H$^{13}$CN & $1-0$ & 86.3401764 & \dots &  \dots &  \dots &  \dots &  \dots &  \dots & $1.6 \times 10^{5}$ \\
H$^{13}$CO$^{+}$ & $1-0$ & 86.7542880 & 4.16 & $3.9 \times 10^{-5}$ & $2.3 \times 10^{-10}$ & $9.3 \times 10^{-10}$ & $1.7 \times 10^{5}$ & $4.1 \times 10^{4}$ & $2.2 \times 10^{4}$ \\
SiO & $2-1$ & 86.8469950 & 6.25 & $2.9 \times 10^{-5}$ & $1.1 \times 10^{-10}$ & $2.8 \times 10^{-10}$ & $2.7 \times 10^{5}$ & $1.0 \times 10^{5}$ & \dots \\
HN$^{13}$C & $1-0$ & 87.0908590 & \dots &  \dots &  \dots &  \dots &  \dots &  \dots  & \dots  \\
CCH & $N = 1-0, J = 3/2-1/2$ & 87.3169250 & 4.193 & $1.5 \times 10^{-6}$ & $1.4 \times 10^{-11}$ & $5.8 \times 10^{-11}$ & $1.1 \times 10^{5}$ & $2.7 \times 10^{4}$ & \dots \\
CCH & $N = 1-0, J = 1/2-1/2$ & 87.4020040 & 4.197 & $1.3 \times 10^{-6}$ & $8.5 \times 10^{-12}$ & $5.9 \times 10^{-11}$ & $1.5 \times 10^{5}$ & $2.2 \times 10^{4}$ & \dots \\
HNCO & $J_{K_{a},K_{c}} = 4_{0,4} - 3_{0,3}$ & 87.9252380 & 10.55 & $9.0 \times 10^{-6}$ & $6.8 \times 10^{-11}$ & $4.3 \times 10^{-10}$ & $1.3 \times 10^{5}$ & $2.1 \times 10^{4}$ & \dots \\
HCN & $1-0$ & 88.6318473 & 4.25 & $2.4 \times 10^{-5}$ & $1.9 \times 10^{-11}$ & $8.1 \times 10^{-11}$ & $1.3 \times 10^{6}$ & $3.0 \times 10^{5}$ & $4.5 \times 10^{3}$ \\
HCO$^{+}$ & $1-0$ & 89.1885260 & 4.28 & $4.3 \times 10^{-5}$ & $2.3 \times 10^{-10}$ & $9.2 \times 10^{-10}$ & $1.8 \times 10^{5}$ & $4.6 \times 10^{4}$ & $5.3 \times 10^{2}$ \\
HNC & $1-0$ & 90.6635640 & 4.35 & $2.7 \times 10^{-5}$ & $9.0 \times 10^{-11}$ & $2.6 \times 10^{-10}$ & $3.0 \times 10^{5}$ & $1.1 \times 10^{5}$ & $2.3 \times 10^{3}$ \\
H & 41$\alpha$$(42-41)$ & 92.0344340 & \dots &  \dots &  \dots &  \dots &  \dots &  \dots  & \dots \\
N$_{2}$H$^{+}$ & $1-0$ & 93.1737770 & 4.47 & $3.6 \times 10^{-5}$ & $2.3 \times 10^{-10}$ & $8.9 \times 10^{-10}$ & $1.6 \times 10^{5}$ & $4.1 \times 10^{4}$ & $5.5 \times 10^{3}$ \\
C$^{34}$S & $2-1$ & 96.4129500 & \dots &  \dots &  \dots &  \dots &  \dots &  \dots  & \dots  \\
CH$_3$OH-E & $J_K = 2_{-1} - 1_{-1}$ & 96.7393630 & 12.5 & $2.6 \times 10^{-6}$ & $9.0 \times 10^{-11}$ & $2.9 \times 10^{-10}$ & $2.8 \times 10^{4}$ & $8.7 \times 10^{3}$ & \dots \\
CH$_3$OH-A & $J_K = 2_{0} - 1_{0}$ & 96.7413770 & 7.0 & $3.4 \times 10^{-6}$ & $1.0 \times 10^{-10}$ & $3.5 \times 10^{-10}$ & $3.4 \times 10^{4}$ & $9.8 \times 10^{3}$ & \dots \\
CS & $2-1$ & 97.9809530 & 7.1 & $1.7 \times 10^{-5}$ & $4.6 \times 10^{-11}$ & $1.6 \times 10^{-10}$ & $3.6 \times 10^{5}$ & $1.0 \times 10^{5}$ & $1.2 \times 10^{4}$ \\
SO & $J_{K} = 3_{2} - 2_{1}$ & 99.2999050 & 9.2 & $1.1 \times 10^{-5}$ & $3.9 \times 10^{-11}$ & $3.4 \times 10^{-10}$ & $2.9 \times 10^{5}$ & $3.3 \times 10^{4}$ & \dots \\
HC$_{3}$N & $12-11$ & 109.1736380 & 34.058 & $1.0 \times 10^{-4}$ & $5.2 \times 10^{-11}$ & $6.4 \times 10^{-10}$ & $2.0 \times 10^{6}$ & $1.6 \times 10^{5}$ & $1.1 \times 10^{5}$ \\
C$^{18}$O & $1-0$ & 109.7821760 & 5.27 & $6.3 \times 10^{-8}$ & $3.2 \times 10^{-11}$ & $1.3 \times 10^{-10}$ & $1.9 \times 10^{3}$ & $4.8 \times 10^{2}$ & \dots \\
HNCO & $J_{K_{a},K_{c}} = 5_{0,5} - 4_{0,4}$ & 109.9057530 & 15.82 & $1.8 \times 10^{-5}$ & $7.4 \times 10^{-11}$ & $4.4 \times 10^{-10}$ & $2.4 \times 10^{5}$ & $4.1 \times 10^{4}$ & \dots \\
$^{13}$CO & $1-0$ & 110.2013540 & 5.29 & $6.3 \times 10^{-8}$ & $3.2 \times 10^{-11}$ & $1.3 \times 10^{-10}$ & $1.9 \times 10^{3}$ & $4.8 \times 10^{2}$ & \dots \\
C$^{17}$O & $1-0$ & 112.3589880 & 5.39 & $6.7 \times 10^{-8}$ & $3.2 \times 10^{-11}$ & $1.3 \times 10^{-10}$ & $2.1 \times 10^{3}$ & $5.2 \times 10^{2}$ & \dots \\
CN & $N = 1-0, J = 1/2-1/2$ & 113.1913170 & 5.43 & $1.2 \times 10^{-5}$ & $7.2 \times 10^{-12}$ & $6.2 \times 10^{-11}$ & $1.6 \times 10^{6}$ & $1.9 \times 10^{5}$ & \dots \\
CN & $N = 1-0, J = 3/2-1/2$ & 113.4909820 & 5.45 & $1.2 \times 10^{-5}$ & $7.2 \times 10^{-12}$ & $5.0 \times 10^{-11}$ & $1.7 \times 10^{6}$ & $2.4 \times 10^{5}$ & $1.7 \times 10^{4}$ \\
CO & $1-0$ & 115.2712020 & 5.53 & $7.2 \times 10^{-8}$ & $3.2 \times 10^{-11}$ & $1.3 \times 10^{-10}$ & $2.2 \times 10^{3}$ & $5.7 \times 10^{2}$ & \dots \\

\hline
	\end{tabular}
	
	\begin{minipage}{0.98\textwidth}
    \vspace{1mm}
    $^{1}$ The two-level approximation of the critical density has been calculated by accounting for only the single downward collisional rate coefficient from the initial to final energy level ($C_{ul}^\mathrm{2lvl}$ and $n_{\rm crit}^\mathrm{2lvl}$). The full critical density calculation accounts for all the possible downward transition collisional rate coefficients to and from the final energy level ($C_{ul}$ and $n_{\rm crit}$). The effective excitation densities ($n_{\rm eff}$) are taken from \citet{shirley_2015}, and have been defined by radiative transfer modelling as the density that results in a molecular line with an integrated intensity of 1\,\Kkms\ (also see \citealp{evans_1999}). \\ 
    \end{minipage}
    
	\label{table:line_info}
\end{table*}

\begin{table}
\centering
	\caption{Observational properties across the mapped region (i.e. that covered with both vertical and horizontal on-the-fly scans). The columns show the molecule name, the average cube rms (in a 1\kms\ channel), mean uncertainty of the integrated intensity, and mean value of the integrated intensity. The final column shows the area percentage, $A_\mathrm{cov}$, within the mapped region that has an integrated intensity above five times the uncertainty; i.e. $W_Q$\,$>$\,$5\sigma_{W_Q}$. The table has been ordered by decreasing $A_\mathrm{cov}$ value (c.f. Table\,8 of \citealp{pety_2017}). The information given is for maps that have been smoothed to an angular resolution of 60\arcsec, and have a spectral resolution of 0.6\,\kms. See Table\,\ref{table:obs_stats} for additional statistical properties of the molecular line integrated intensity maps. The full, machine-readable version of this Table can be obtained from the supplementary online material.}
	\begin{tabular}{lcccccc}
\hline
Line & $\sigma_\mathrm{rms}$ (1\,\kms) & $\sigma_{W}$ & $W^\mathrm{mean}_Q$ & $A_\mathrm{cov}$ \smallskip  \\
 & (K) & \multicolumn{2}{c}{ (\Kkms) } & (\%) \\
\hline


 CO\,(1-0) & 0.15 & 0.62 & 84.56 & 100.0 \\
 $^{13}$CO\,(1-0) & 0.06 & 0.26 & 13.75 & 99.9 \\
 C$^{18}$O\,(1-0) & 0.06 & 0.24 & 1.05 & 34.1 \\
 HCN\,(1-0) & 0.05 & 0.22 & 1.38 & 31.9 \\
 HCO$^{+}$\,(1-0) & 0.06 & 0.23 & 0.94 & 21.3 \\
 CS\,(2-1) & 0.05 & 0.21 & 0.93 & 21.1 \\
 HNC\,(1-0) & 0.06 & 0.23 & 0.61 & 14.0 \\
 CCH\,(1-0,3/2-1/2) & 0.05 & 0.22 & 0.63 & 12.8 \\
 CN\,(1-0,3/2-1/2) & 0.09 & 0.36 & 0.93 & 8.9 \\
 SO\,(3-2) & 0.05 & 0.21 & 0.36 & 8.4 \\
 CCH\,(1-0,1/2-1/2) & 0.05 & 0.22 & 0.39 & 7.1 \\
 N$_{2}$H$^{+}$\,(1-0) & 0.04 & 0.15 & 0.17 & 7.0 \\
 H41$\alpha$\,(42-41) & 0.05 & 0.22 & 0.15 & 4.1 \\
 CH$_3$OH\,(2-1) & 0.05 & 0.21 & 0.06 & 3.2 \\
 CN\,(1-0,1/2-1/2) & 0.08 & 0.34 & 0.21 & 3.1 \\
 SiO\,(2-1) & 0.06 & 0.23 & 0.06 & 2.9 \\
 C$^{34}$S\,(2-1) & 0.05 & 0.22 & 0.09 & 2.7 \\
 HC$_{3}$N\,(12-11) & 0.04 & 0.18 & 0.02 & 2.4 \\
 H$^{13}$CO$^{+}$\,(1-0) & 0.06 & 0.23 & -0.01 & 2.2 \\
 H$^{13}$CN\,(1-0) & 0.06 & 0.26 & -0.01 & 2.2 \\
 HN$^{13}$C\,(1-0) & 0.05 & 0.22 & 0.05 & 1.6 \\
 HNCO\,(4-3) & 0.05 & 0.22 & -0.02 & 0.5 \\
 C$^{17}$O\,(1-0) & 0.16 & 0.65 & 0.11 & 0.5 \\
 HNCO\,(5-4) & 0.06 & 0.25 & -0.01 & 0.1 \\

\hline
	\end{tabular}
	\label{table:obs_info}
\end{table}

\subsection{Observations with the IRAM--30m telescope}

\begin{figure}
\centering
	\includegraphics[trim = 0mm 0mm 0mm 0mm, clip, width=0.95\columnwidth]{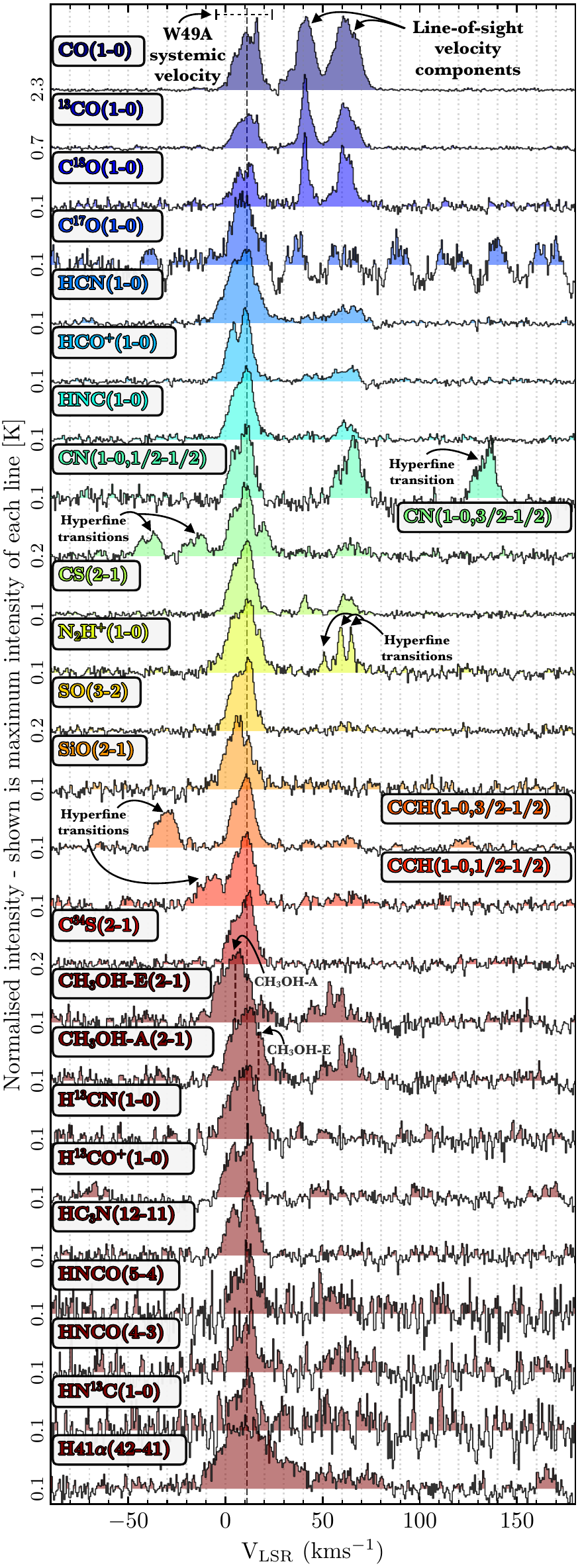}\vspace{-2mm}
    	\caption{Normalised spectra from each molecule averaged over the mapped region, for pixels with a integrated intensity $>3\,\sigma_{W_Q}$ (see Figure\,\ref{fig:momentmaps_smoothmask}). The peak brightness temperature of each spectrum is given on the left axis. The coloured shaded region of each spectrum shows $>0$\,K intensities. \note{Highlighted are several features of interest within the spectra, such as the systemic velocity of W49A, hyperfine transitions, and line blending.}}
    	\label{fig:w49_spec}
\end{figure}

\subsubsection{Data reduction}
The observations presented here were taken as part of the Molecular Line Emission as a Tool for Galaxy Observations (LEGO) survey (IRAM project code: 183--17). The full observational information of this survey, which covers $\sim{}25$ regions throughout the Galactic Disk, will be summarised in a future publication (Kauffmann et al. in preparation). This survey was conducted using the Eight Mixer Receiver (EMIR; \citealp{carter_2012}) on the 30m telescope of the Instituto de Radioastronom\'ia Milim\'etrica (IRAM) on Pico Veleta, Spain. Data in the horizontal and vertical polarizations were acquired with the Fast Fourier Transform Spectrometer (FTS) operating in a wide--band mode that provides a spectral resolution of 200~kHz (i.e., $0.6~{\rm{}km\,s^{-1}}\cdot{}[\nu/100~{\rm{}GHz}]^{-1}$ in velocity). The EMIR receiver was operated in two setups, centred at local oscillator frequencies of 97.830~GHz and 103.840~GHz. As a result, the observations cover in combination continuous frequency ranges of 86.10~to~99.89~GHz and 101.78~to~115.57~GHz. Relatively bright emission lines from a variety of astrophysically relevant molecules can be found at these frequencies. The transitions investigated by LEGO are listed in Table~\ref{table:line_info}, which is created as described in Sec.~\ref{sec:line-selection}. The telescope's intrinsic beam size\footnote{\url{http://www.iram.es/IRAMES/mainWiki/Iram30mEfficiencies}\label{footnote:IRAM-calibration}} is $24\farcs{}6\cdot{}(\nu/100~{\rm{}GHz})^{-1}$. The angular resolution of our data is lower, though, due to the processing described below.

The observations cover a total area of $\sim{}30\arcmin\times{}30\arcmin$ size around a central reference coordinate of 19$^\mathrm{h}$10$^\mathrm{m}$32.487$^\mathrm{s}$, 9\degree05\arcmin36.532\arcsec. The On-The-Fly (OTF) mapping technique was used to create the mosaic image, using a dump time of 0.25~s and a scanning speed of $90\arcsec\,{\rm{}s}^{-1}$. The rows of the OTF map were spaced by $10\arcsec$, corresponding to less than half the telescope's intrinsic resolution at the frequencies $\lesssim{}115~\rm{}GHz$ covered by our project. The OTF maps were taken by scanning along either the RA or Dec. axis of the equatorial coordinate system. Every point in the map was covered by at least two orthogonal scans, in order to reduce artefacts due to scan patterns.

The LEGO data reduction pipeline uses functionality from the Python programming language and from IRAM's GILDAS software suite (i.e., the CLASS package).\footnote{\url{http://www.iram.fr/IRAMFR/GILDAS}} This approach permits use of Python's superior bookkeeping functionality to control the data flow, while it also takes advantage of the rich and well-tested data processing functionality within CLASS for low-level tasks. The pipeline, as well as the derived data products, will be released as part of future publications. In the very first processing step, observing logs are generated on the basis of header information of observed scans and data from IRAM's Telescope Access for Public Archive System (TAPAS).\footnote{\url{https://www.iram-institute.org/EN/content-page-247-7-55-177-247-0.html}} These logs constitute the basis for all further data processing steps.

The pipeline processes the data separately for every emission line recorded in Table~\ref{table:line_info}. Python scripts are used to identify, for a given emission line, the observations covering a particular source. CLASS is then employed to extract a velocity range of $\gtrsim{}350~\rm{}km\,s^{-1}$ total width around a given spectral line (a larger range is extracted around transitions with substructure in order to contain all lines), to grid all data to a common set of velocity channels of $0.6~\rm{}km\,s^{-1}$ width, and to subtract a spectral baseline of third order from the data. Velocities in the range $-50~{\rm{}to}~+80~{\rm{}km\,s^{-1}}$ are ignored when fitting the baseline. This range is further expanded in case of emission lines with hyperfine structure, provided this line substructure is more compact than $20~\rm{}km\,s^{-1}$. The original velocity range of $-50~{\rm{}to}~+80~{\rm{}km\,s^{-1}}$ is not modified for baseline fitting in cases where line substructure extends over a range $>20~\rm{}km\,s^{-1}$: we hypothesise that line substructure spread out over a large velocity range is less likely to bias baseline fits, and the absence of substantial artefacts in manually inspected processed data validates this approach. Future versions of the pipeline may include more sophisticated algorithms for automatic flagging of significant line emission during baseline fitting, and future publications will include a detailed assessment of the baseline quality. The spectra delivered by the telescope’s data processing system are calibrated in forward-beam brightness temperatures (the $T_{\rm{}A}^{\ast}$-scale). We convert these to the main beam brightness scale using the relation $T_{\rm{}mb}=F_{\rm{}eff}/B_{\rm{}eff}\cdot{}T_{\rm{}A}^{\ast}$. We adopt frequency-independent forward and main beam efficiencies of $F_{\rm{}eff}=0.95$ and $B_{\rm{}eff}=0.8$, respectively, consistent with the available calibration data$^{\ref{footnote:IRAM-calibration}}$. The baselined data were eventually gridded into a data cube with pixels of $10\arcsec\times{}10\arcsec$ size. This effectively also averages data from the two independent polarisations in every part of the map. This step uses a frequency-dependent Gaussian gridding kernel that increases the intrinsic resolution of the data (i.e., impact of frequency-dependent telescope beam and gridding) to $30\arcsec$.


The angular resolution of our data is, however, further reduced due to the high scan speed and slow sampling pattern employed by our project. We pursue this fast mapping strategy because it allows us to cover a large area on the sky at the expense of somewhat reduced angular resolution. To be specific, along the scan direction, we obtain one data sample every $90\arcsec\,{\rm{}s}^{-1}\cdot{}0.25~{\rm{}s}=22\farcs5$. The impact of this sampling pattern on the angular resolution can, to first order, be gauged by modelling it by a Gaussian smoothing kernel of $22\farcs5$ width at half peak value that is extended along the scan direction. The resulting angular resolution in this model -- along the sampling direction of a single scan -- would be $([30\arcsec]^2+[22\farcs{}5]^2)^{1/2}=37\farcs{}5$. The angular size of the beam would be smaller when averaging perpendicular scans. A future data-oriented LEGO paper will provide a more detailed discussion of this issue.

To allow direct comparison to the molecular hydrogen column density and dust temperature maps (see Section\,\ref{subsec:NhTdust}), we further smooth data cubes with a Gaussian kernel of $52\arcsec$ width at half peak value to achieve an angular resolution of 60\,\arcsec, and re-sample the dataset onto the same 20\arcsec\ pixel size spatial grid. Further smoothing the dataset also has the benefit of increasing the signal-to-noise of the lines across the mapped region, whilst somewhat mitigating the beam smearing effect caused by the fast mapping speed.

\begin{figure*}
	\includegraphics[width=1\textwidth]{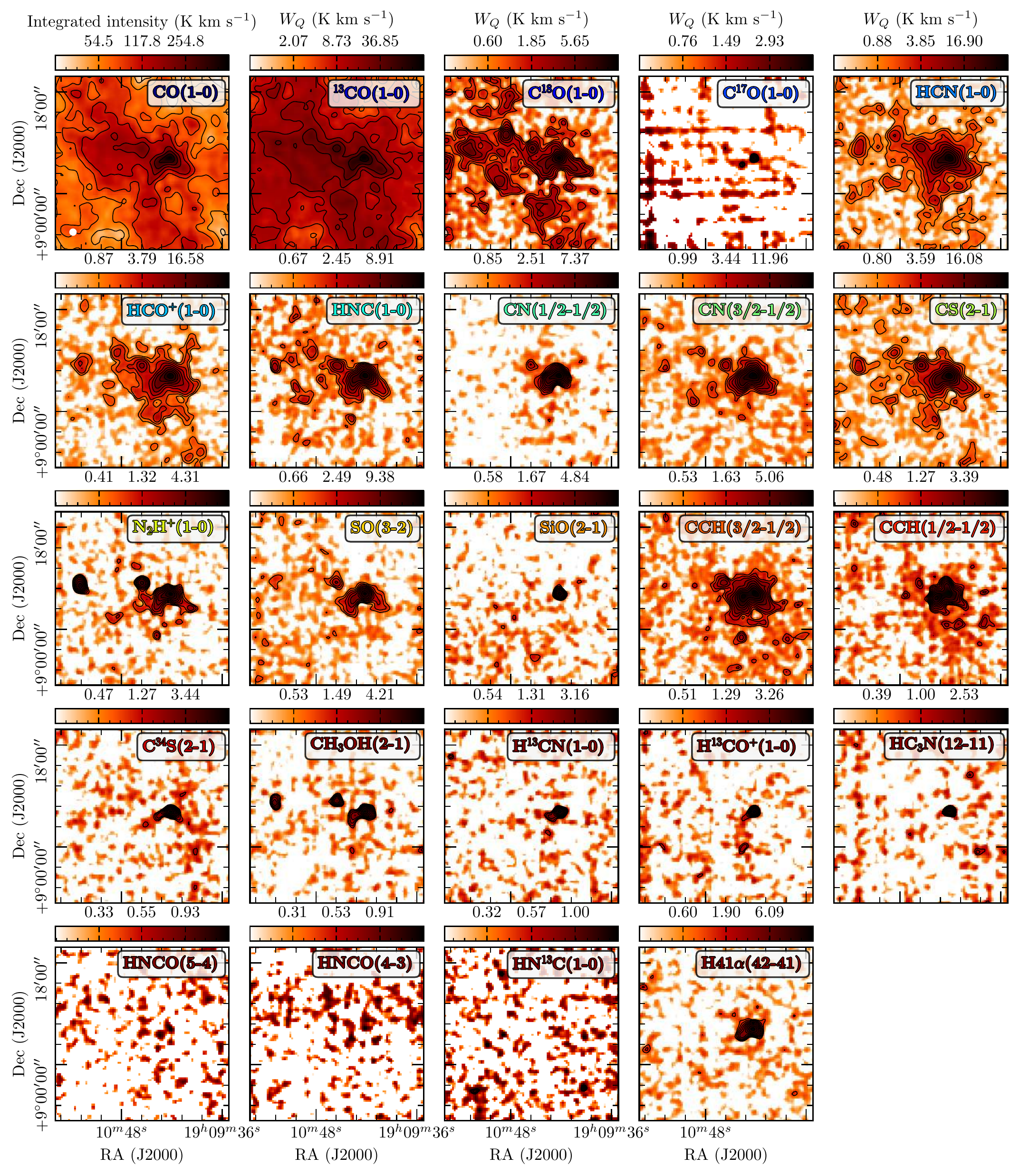}
        \caption{Maps of the integrated intensity, \WQ, for the selected molecular transitions, $Q$, across the W49 region. Overlaid on each panel are signal-to-noise contours in levels of 5\,$\sigma_{W_Q}$, increasing up to the maximum $\sigma_{W_Q}$ value within the mapped region (see Table\,\ref{table:obs_info}). The angular beam size of 60\arcsec, or $\sim$3\,pc at the distance of W49A ($\sim$11\,kpc; \citealp{gwinn_1992, zhang_2013}), is shown as a white circle in the lower left of the first panel.}
    \label{fig:momentmaps_smoothmask}
\end{figure*}


\subsubsection{Selection of emission lines\label{sec:line-selection}}

The LEGO pipeline generates data cubes spanning a few hundred channels for a selected set of emission lines. This approach is necessary because processing the full data set would be computationally too demanding. Table~\ref{table:line_info} lists the transitions for which data cubes are produced. Additional details on quantum numbers and frequencies are presented in Table~\ref{table:line_database}. All of the lines are emitted by molecules, with the exception of the H41$\alpha$ hydrogen recombination line. 

The catalogue is systematically constructed using the following steps. At a minimum, LEGO should cover all emission lines that can be studied in nearby galaxies. We thus start our line catalogue by including all lines that \citet{watanabe_2014} detect in M51 at a signal-to-noise ratio $>2$. To this we add all transitions studied by \citet{pety_2017} in Orion B, to be consistent with their work. We then include rare isotopologues of species where the abundant line is potentially optically thick. (This is not relevant in practice, since rare forms of HCN, HNC, and $\rm{}HCO^+$ are already included due to previous selection steps.) We further include bright lines from selected species that intuitively seem interesting, such as long and simple carbon chains. In practice, these lines are taken from \citet{watanabe_2015}. Finally, the lowest-lying hydrogen $\alpha$ recombination line above a frequency of 86~GHz is included in the table.

We are dealing with a large number of transitions, and the nature of our experiments do not require a particularly high level of accuracy in transition properties. We, therefore, obtain transition parameters from catalogues that cover a large number of lines, but we do not make any effort to refine the accuracy of parameters on the basis of the latest literature. Rest frequencies are taken from the Lovas/NIST database \citep{lovas_2004}, as accessed via the Splatalogue\footnote{\url{https://www.cv.nrao.edu/php/splat/index.php}} database for astronomical spectroscopy. This catalogue is also used to characterize the substructure of a transition. Provided splitting is present in a given transition, the emission lines were split into groups of $<100~\rm{}MHz$ width (i.e., $<300~{\rm{}km\,s^{-1}}\cdot[\nu/100~{\rm{}GHz}]^{-1}$ in velocity). The group's rest frequency, $\nu_\mathrm{rest}$, is taken to be the frequency of the line with the highest ``intensity'' in the \citet{lovas_2004} catalogue. The full quantum numbers of this transition are recorded in Table~\ref{table:line_database}. The lowest and highest frequencies of lines in a given group (i.e. $\nu_\mathrm{min}$ and $\nu_\mathrm{max}$) are also collected in Table~\ref{table:line_database}, for example, to guide baseline subtraction and to build data cubes that are wide enough to contain all transitions within a given line group. Table~\ref{table:line_info} lists the quantum numbers common within a given line group, as well as $\nu_\mathrm{rest}$.

The Leiden Atomic and Molecular Database (\citealp{Schoier_2005}; LAMDA, accessed in November 2019) is used to obtain the upper energy levels, Einstein coefficients and downward collisional rates coefficients for the lines covered by LEGO. When calculating the collisional rates coefficients, we ignore line splitting, where present, and use LAMBDA data files collapsing transition substructure in those cases. The critical density for each of the transitions has been calculated using two approximations. The first is the simple two-level approximation, which only accounts for the downward collisional rate of the initial upper ($u$) to final lower ($l$) energy level such that $n_{\rm{}crit}^\mathrm{2lvl}=A_{ul}/C_{ul}$. The second approximation accounts for all the possible upward and downward transition collisional rate coefficients from the initial energy level, which when neglecting background contribution and assuming optically thin conditions can be defined as $n_{\rm{}crit}=A_{ul}/\sum_{u\neq{}k}C_{uk}$ (subscript $uk$ represents a transition from the ``upper'' energy level, $u$, to any allowed energy level, $k$, even where $\mathrm{E}_k>\mathrm{E}_u$). Collisional rates of upward transitions have been calculated using the formalism outlined by \citet[][equation 4]{shirley_2015}. In both the calculations, we use the collision rates determined at a kinetic temperature 20\,K, assuming that H$_2$ is the dominant collisional partner, and assume that the emission is optically thin. These results are presented in Table~\ref{table:line_info}.

Also given in Table~\ref{table:line_info} are the effective excitation densities ($n_{\rm eff}$) taken from \citet{shirley_2015}. These have been determined using the RADEX radiative transfer modelling code \citep{vandertak_2007}, and are defined as the density that results in a molecular line with an integrated intensity of 1\,\Kkms\ (also see \citealp{evans_1999}). These estimates then account for radiative trapping effects that lower the critical density within the optically thick regime. In Table~\ref{table:line_info}, we show $n_{\rm eff}$ for a kinetic temperature 20\,K and the reference column densities for each molecule from \citet[][table\,1]{shirley_2015}.


\subsubsection{Spectra, noise properties and integrated intensity maps}\label{sec:IImap}

A brief inspection of the cubes shows that they contain a varying degree of complexity in their position-position-velocity space structure. Figure\,\ref{fig:w49_spec} shows the spectrum from each molecular transition averaged across the mapped region. Here, we can see that several of the lines contain multiple velocity components, which vary in spatial structure and peak intensity (also see channel maps presented in Figure\,\ref{channel_maps}). 


\note{Highlighted at the top of Figure\,\ref{fig:w49_spec} is the range of velocities ($\sim$-10\,--\,20\,\kms) observed towards W49A, and represented as a vertical dashed black line is the mean systemic velocity of W49A ($\sim$11\,\kms; e.g. \citealp{nagy_2012, nagy_2015}). Within this velocity range, we find asymmetric line profiles, and, in some cases, multiple narrowly separated ($<10$\,\kms) velocity components. These line profiles are likely due to the unresolved internal structure and kinematics (e.g. outflows) within W49A, which is currently being heavily influenced by stellar feedback (e.g. see \citealp{galvan-madrid_2013}). Also labelled in Figure\,\ref{fig:w49_spec} are examples of several more broadly spaced ($>10$\,\kms) velocity components from sources unrelated to W49A, which, given their velocities, are thought to arise from emission in intervening spiral arms (e.g. W49B at $\sim$60\,\kms).} 

\note{There are a couple of additional potential causes for the multiple velocity components seen in averaged spectra. The first is hyperfine splitting of molecular transitions (also see Table\,\ref{table:line_database}). Labelled in Figure\,\ref{fig:w49_spec} are several examples for lines that have narrow (e.g. N$_2$H$^+$) and broadly (e.g. CCH) spaced hyperfine transitions. The second is self-absorption within regions of high optical-depth, which can result in the narrow ($<10$\,\kms) spaced velocity components. This would be particularly relevant towards the centre of W49A, and for the more abundant molecules (e.g. CO; \citealp{nagy_2012, galvan-madrid_2013}).}

\note{The final feature to note within Figure\,\ref{fig:w49_spec}, is the blending of the CH$_3$OH spectra. The CH$_3$OH-E and CH$_3$OH-A transitions shown here are very close in frequency (corresponding to a velocity difference of $\sim$5\,\kms), and cannot be fully resolved towards W49A (also see Figure\,\ref{fig:momentmaps_smoothmask}).\footnote{It is worth considering that there are several (typically weaker) CH$_3$OH transitions within the frequency range that we have not considered within this work. These include the 2(0)-1(0) and 2(1)-1(1) transitions of CH$_3$OH-E at rest frequencies of 96.744545\,GHz and 96.755501\,GHz, respectively (e.g. \citealp{menten_1988, leurini_2004}). This is noted within Table\,\ref{table:line_database} within the Appendix of this work.} Further analysis of the CH$_3$OH lines in this work is, therefore, limited to the significantly brighter A-type line, and all references to CH$_3$OH\,(2-1) henceforth correspond to the CH$_3$OH-A\,(2-1) line.} 

\note{Despite identifying several velocity components within the spatially averaged spectra, when examining the cubes we find that, in general, there is a single velocity component that dominates the emission along a given line-of-sight.} Such that the total emission at any position within a molecular line cube can typically be attributed to a single velocity range (see Figure\,\ref{channel_maps}). It then appears the brightest sources within the mapped region are generally distinct in both velocity space and spatially. Throughout this work, we then make the simple assumption that at each position there is a single source that is responsible for the emission seen in molecular line emission. Additionally, the analysis presented in this work relies on a comparison to the total hydrogen column density and dust temperature maps, which also integrate emission from all molecular material along the line-of-sight (see Section\,\ref{subsec:ancillary_observations}). Hence, not limiting our analysis to a given velocity range allows us to make a consistent comparison across all the available datasets.

The aim of this work is to produce intensity maps integrated along all velocities that a) trace the compact and bright emission, b) recover the diffuse and extended emission, c) have a constant noise profile where emission is not present. To do so, we follow a two-step masking procedure for the data cubes to reduce the noise within positions with significant line emission. Briefly, for this procedure, we firstly create a mask including only the voxels (velocity pixels) with significant CO emission, which we then apply to a given line cube. As the CO line shows significant emission within the majority of velocity channels where emission is found from the other lines, this masking procedure provides a convenient way of producing a generous integration velocity range at each line-of-sight such that both the noise is reduced in pixels containing emission and pixels without emission show a flat noise profile. In the case of lines with hyperfine structure that extends outside of the CO mask (e.g. CCH), we choose to expand the mask to also include voxels with significant emission from the line cube (i.e. producing a final mask where each position has either significant emission from the line hyperfine structure or CO emission). 

We follow the method outlined by \citet{dame_2011} to produce the initial CO data cube mask. For this method, we firstly smooth the cubes by a factor of two spatially (i.e. to 120\arcsec), and a factor of 10 spectrally (i.e. to 6\kms). We then determine the rms value, $\sigma_{\rm rms}$, along each line of sight using two spectral windows, covering velocities of -150\,--\,-20\,\kms\ and 90\,--\,150\,\kms. These velocity ranges were chosen for their lack of significant emission (see Figure\,\ref{fig:w49_spec}). An initial smoothed CO cube mask is then produced to include all channels above a 5\,$\sigma_{\rm rms}$ threshold. If hyperfine structure is present within a given cube, but outside of this CO mask, then this smoothing procedure is applied to the cube. Significant emission from this smoothed cube is then included in the final mask for that cube only. These final smoothed masks are then applied to the full resolution cubes, which are used to create the integrated intensity maps ($W_Q = \int T_{Q,{\rm mb}}\,{\rm d}v$). The uncertainty on the integrated intensity at each pixel was calculated as,
\begin{equation}
\sigma_{W_Q} = \sigma_\mathrm{rms} \Delta v_{\mathrm{res}} (N_\mathrm{ch})^{1/2},
\label{sigma}
\end{equation}
where $\Delta_{v_\mathrm{res}}$ is the velocity resolution, and $N_\mathrm{ch}$ is the number of unmasked channels used to create the moment maps. Mean values for the $\sigma_\mathrm{rms}$ and $\sigma_{W_Q}$ for each molecular line are shown in Table\,\ref{table:obs_info}. The final moment maps that are used throughout this work are displayed in Figure\,\ref{fig:momentmaps_smoothmask}.

\subsection{Dust column density and temperature maps}\label{subsec:ancillary_observations}

 \begin{figure}
    	\includegraphics[width=1\columnwidth]{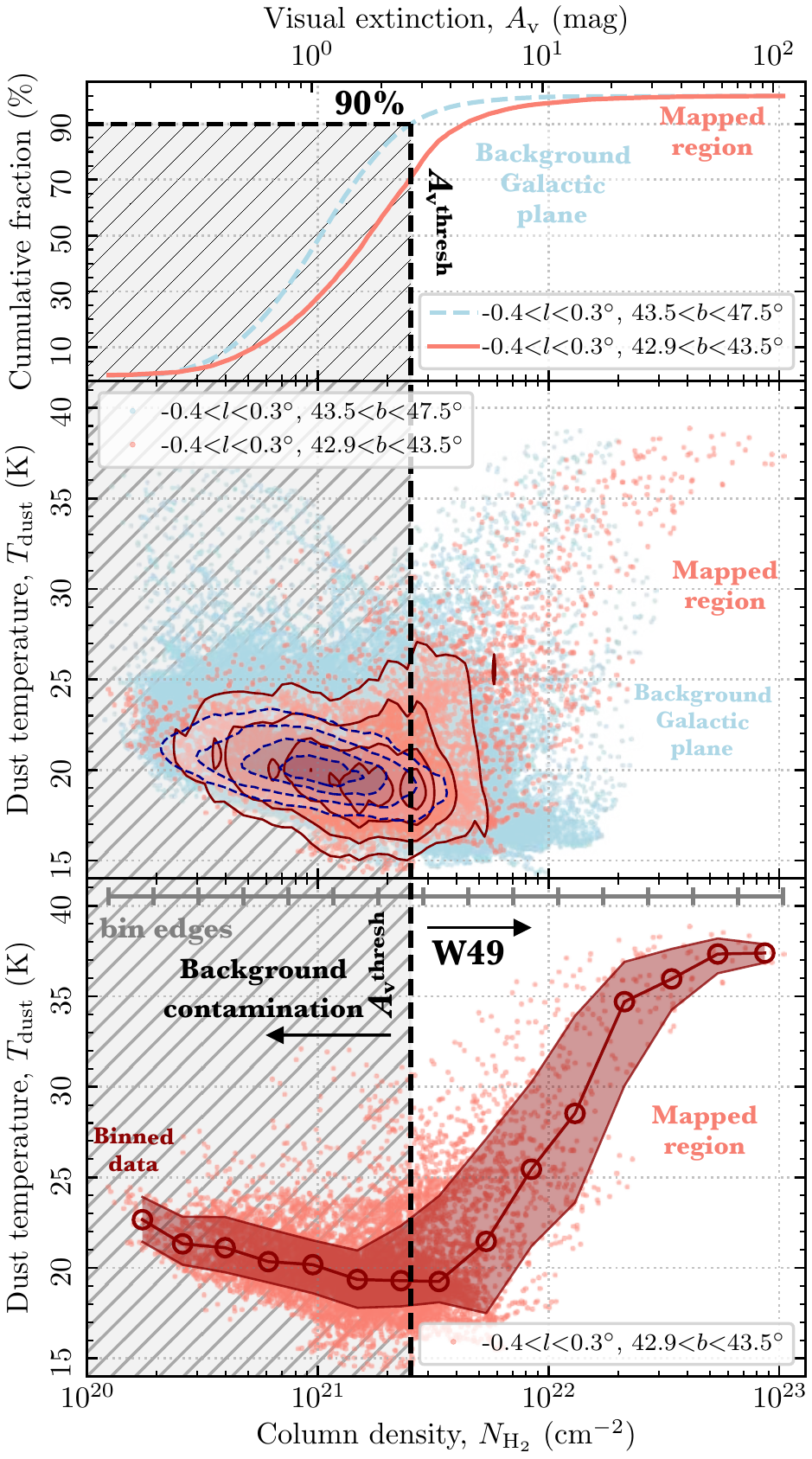}
    	\vspace{-5mm}
    	\caption{Line-of-sight contamination in the molecular hydrogen column density map. The upper panel shows the cumulative (pixel) distribution of molecular hydrogen column density (extinction on upper axis) across the mapped region (red) and Galactic plane (blue) region (see Figure\,\ref{fig:w49_rgb}). Highlighted as a horizontal dashed line and shaded region is the 90\% percentile for the Galactic plane, which corresponds to a column density $N_\mathrm{H_2}^\mathrm{thresh}$\,=\,$2.5\times10^{21}$\,cm$^{-2}$ ($A_\mathrm{v}^\mathrm{thresh}$\,=\,2.7\,mag). The central panel shows the dust temperature distribution as a function of column density across both the mapped region and Galactic plane. The overlaid contours show the pixel density of all the points in levels of 10, 25, 50, and 75\,per cent. The lower panel shows the dust temperature distribution as a function of column density for only the mapped region. The joined open circles show the median values within equally logarithmically spaced column density (extinction) bins, as shown at the top of the panel. The red shaded region shows the standard deviation within the bins.} 
\label{fig:w49_column-temp}
\end{figure}

The molecular hydrogen column density and dust temperature maps, together with their uncertainties, were obtained by fitting modified blackbody models on a pixel-by-pixel basis to the {\it Herschel} Space Observatory large program Hi-Gal data \citep{molinari_2011}. For that, we follow a similar procedure as described in \citet{guzman_2015}. We also subtract dust emission arising from diffuse gas in the Galactic plane attributable to clouds located in the foreground or background of W49 using the same method as in \citet{guzman_2015}. \note{Here, however, the background is smoothed to a much larger angular scale of 660\arcsec (34\,pc at a distance of 11\,kpc), to approximately match the size of the mapped region. The subtracted maps balance preserving small-scale structures with an angular resolution equal to the longest {\it Herschel} wavelength map (32\arcsec), whilst removing large-scale variations due to the background ($>660$\arcsec). A smoother background would leave too much diffuse emission not associated with W49 but to the Galactic plane, and a background with more structure on smaller scales would filter out emission associated with W49 itself.} The column density and temperature maps produced using this method, however, suffer from minor small-scale artefacts, such as saturated values (e.g. towards the centre of W49A). To correct for these issues, we spatially smooth the maps using the {\sc astropy.convolution} package, which accurately interpolates over bad or missing values within an image. We find that a Gaussian kernal of 52\arcsec\ provided the optimal trade-off between final map resolution and a complete interpolation over missing values. The final molecular hydrogen column density and dust temperature have an angular resolution of 60\arcsec, and have been resampled onto a 20\arcsec\ pixel grid. These maps are presented within the lower row of Figure\,\ref{fig:w49_rgb}.
 
 In Appendix\,\ref{appendix_colerr}, we present an analysis of the column density and dust temperature uncertainties produced by random noise in the {\it Herschel} measurements. We find that mean fractional uncertainties across the mapped region studied in this work are $\Delta N_\mathrm{H_2}\,/\,N_\mathrm{H_2}\,=\,\pm\,30\%$ and $\Delta T_\mathrm{dust}/T_\mathrm{dust} = \pm 10\%$ (see Figure\,\ref{realnoise_pspace2}).
 
 It is useful to convert the H$_2$ column density map into units of visual extinction, $A_\mathrm{v}$, for comparison with other works (e.g. \citealp{kauffmann_2017}). To do so, we adopt the commonly used extinction relation $A_\mathrm{v} = 1.1 \times 10^{-21} N_\mathrm{H_2}$, where $A_\mathrm{v}$ is in units of mag, and $N_\mathrm{H_2}$ is in units of cm$^{-2}$ \citep{bohlin_1978, fitzpatrick_1999, lacy_2017}.
%

\section{Results}\label{sec:results}

In this section, we present the results from the molecular hydrogen column density, dust temperature, and molecular line integrated intensity maps. For ease of comparison, all the data presented in this work have been smoothed to the same angular resolution of 60\arcsec, which corresponds to a spatial resolution of $\sim$\,3\,pc at the distance of the W49A star-forming region (11\,kpc; \citealp{zhang_2013}). It is worth noting that, given the low spatial resolution of these observations, we are most likely not fully resolving any structures within our source; e.g the central star-forming core of the W49A is contained within a single pixel (e.g. \citealp{rugel_2018b}). Each pixel within our maps is expected, therefore, to contain a range of physical properties (e.g. densities and temperatures), and should be considered as parsec scale averages across a representative star-forming region of the Galactic Plane. The maps presented throughout this work contain all emission integrated along the line-of-sight. Similar to the effect of the low spatial resolution, this may cause averaging of physical properties within any given pixel.

\subsection{Distribution of the molecular hydrogen and dust temperature}\label{subsec:NhTdust}

In Section\,\ref{sec:observations}, we discussed the H$_2$ column density uncertainty produced by the random noise in the {\it Herschel} maps. There is, however, a second systematic uncertainty produced by the unrelated fore- and back-ground clouds along the line of sight. This has already been somewhat accounted for by subtracting the diffuse emission from the individual {\it Herschel} wavelength maps, yet unrelated Galactic plane emission may be still present within the final column density map. 

To account for the remaining fore- and back-ground contamination we implement a location-dependent minimum $N_\mathrm{H_2}$ ($A_\mathrm{v}$) threshold for Galactic plane contamination. This threshold represents the Galactic background level on top of which the column density enhancements can be reliably attributed to the emission within the molecular line cubes. This threshold is defined, $N_\mathrm{H_2}^\mathrm{thresh}$ ($A_\mathrm{v}^\mathrm{thresh}$), such that 90\% of the pixels within a area of the Galactic plane have column densities less than this threshold (i.e. $N_\mathrm{H_2}(90\%) < N_\mathrm{H_2}^\mathrm{thresh}$). This analysis has been limited to the -0.4$<$$l$$<$0.3$^{\circ}$, 43.5$<$$b$$<$47.5$^{\circ}$ region for the Galactic plane (see Figure\,\ref{fig:w49_rgb}). This is a large section of the Galactic plane directly adjacent to the LEGO mapped region, and hence should provide a representative background Galactic plane column density distribution for the same latitude range of the mapped region.

The upper panel of Figure\,\ref{fig:w49_column-temp} shows the cumulative (pixel) distribution of molecular hydrogen column density (extinction on upper axis) across the mapped region and the Galactic plane. Highlighted as a horizontal dashed line is the 90 percentile for the Galactic plane, which corresponds to a column density $N_\mathrm{H_2}^\mathrm{thresh}$\,=\,$2.5\times10^{21}$\,cm$^{-2}$ ($A_\mathrm{v}^\mathrm{thresh}$\,=\,2.7\,mag). We find minimum -- mean -- maximum values of the column density within the mapped region of 0.1 -- 2.5 -- 104.2\,$\times10^{21}$\cmsq, and dust temperatures of 14.4 -- 20.6 -- 38.8\,K. Across the Galactic plane region we find minimum -- mean -- maximum values of 0.1 -- 1.4 -- 29.0\,$\times10^{21}$\cmsq, and 14.3 -- 20.5 -- 38.7\,K. Whilst above the column density threshold within the mapped region we find 2.5 -- 5.5 -- 104.2\,$\times10^{21}$\cmsq, and 14.7 -- 21.8 -- 38.8\,K. 

The central panel of Figure\,\ref{fig:w49_column-temp} shows the dust temperature distribution as a function of column density for the mapped region and the Galactic plane. The overlaid contours highlight where the majority of data points lie within both distributions. Comparing these contours we see that both regions appear to be qualitatively similar, with the mapped region shifted higher by a factor of a few in column density. Moreover, below the column density threshold, there appears to be a shallow negative correlation between the column density and dust temperature for both regions. Such a negative correlation is typically expected for quiescent molecular gas, where the highest column densities have the lowest temperatures due to the lack of a heating source and efficient cooling mechanisms (i.e. via dust radiation). Given the aforementioned line-of-sight contamination at low column densities, it is, however, difficult to assess the significance of this relation.

Finally, the lower panel of Figure\,\ref{fig:w49_column-temp} shows the dust temperature distribution as a function of column density for only the mapped region. This clearly shows that above the column density threshold there is an increase in the dust temperature with increasing column density, which originates from positions associated with the W49A massive star-forming region (c.f. Figure\,\ref{fig:w49_rgb}). This is then indicative of an internal heating source associated with star formation (an increased interstellar radiation field; e.g. \citealp{pety_2017}).

We now compare to an independent measure of the molecular hydrogen column density. In doing so, we aim to test the fore- and back-ground contamination threshold of $N_\mathrm{H_2}^\mathrm{thresh}$\,=\,$2.5\times10^{21}$\,cm$^{-2}$ ($A_\mathrm{v}^\mathrm{thresh}$\,=\,2.7\,mag) in the {\it Herschel} molecular hydrogen column density map. To do so, we follow the method presented in \citet[][their equations 11 to 13]{heiderman_2010} for converting the $^{13}$CO($1-0$) integrated intensity into a $^{13}$CO column density. This method solves for the excitation temperature and optical depth of $^{13}$CO($1-0$) using both $^{12}$CO($1-0$) and $^{13}$CO($1-0$) peak intensities, by assuming LTE conditions, that both lines have equal excitation temperatures, and that the $^{12}$CO($1-0$) and $^{13}$CO($1-0$) are optically thick and thin, respectively. We then convert to a molecular hydrogen column density assuming a $^{13}$CO abundance of $5\,\times\,10^{-6}$.

\begin{figure}
    	\includegraphics[width=1\columnwidth]{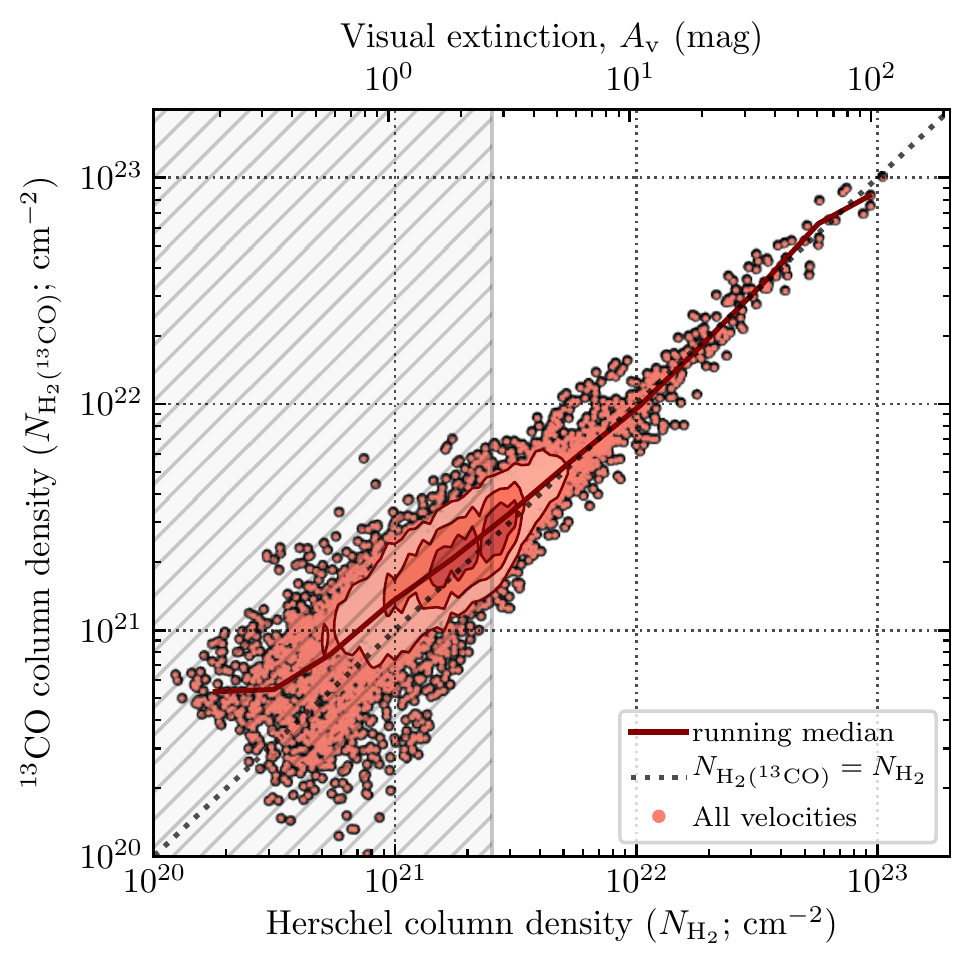}
    	\vspace{-7mm}
    	\caption{A comparison of the molecular hydrogen column density determined from the $^{13}$CO\,(1-0), $N_\mathrm{H_2 (^{13}CO)}$, and {\it Herschel}, $N_\mathrm{H_2}$, observations. The overlaid contours show the pixel density of all the points at levels of 10, 25, 50, and 75 percent. The solid red line shows the running median within equally logarithmically spaced {\it Herschel} column density (extinction) bins. We highlight here values of the column density below $N_\mathrm{H_2}^\mathrm{thresh}$\,=\,$2.5\times10^{21}$\,cm$^{-2}$ ($A_\mathrm{v}^\mathrm{thresh}$\,=\,2.7\,mag) that suffer from fore- and back-ground contamination (see Section\,\ref{subsec:NhTdust}). The diagonal dashed line shows $N_\mathrm{H_2 (^{13}CO)}\,=\,N_\mathrm{H_2}$. This Figure highlights that 1) the {\it Herschel} column density reliably recovers only the molecular gas (i.e. not atomic or Galactic cirrus) down to a column density below $N_\mathrm{H_2}^\mathrm{thresh}$ (Section\,\ref{subsec:ancillary_observations}), and 2) integrating all emission within the molecular line cubes is well correlated with the molecular hydrogen column density (Section\,\ref{sec:IImap}).} 
\label{fig:NH2_13CO}
\end{figure}

Figure\,\ref{fig:NH2_13CO} shows the molecular hydrogen column density determined from the $^{13}$CO($1-0$) observations, $N_\mathrm{H_2 (^{13}CO)}$, as a function of the molecular hydrogen column density determined from the {\it Herschel} observations, $N_\mathrm{H_2}$. The overlaid solid red line shows the running median within logarithmically spaced $N_\mathrm{H_2}$ bins, and the diagonal black dashed line shows $N_\mathrm{H_2 (^{13}CO)}\,=\,N_\mathrm{H_2}$. Comparing these two lines, we find a remarkably good agreement between $N_\mathrm{H_2 (^{13}CO)}$ and $N_\mathrm{H_2}$ across around two orders of magnitude in molecular hydrogen column density. This indicates that the method for subtracting fore- and back-ground emission within in the {\it Herschel} maps has correctly removed the Galactic cirrus emission across the map, leaving only the emission associated with the molecular hydrogen. Moreover, this agreement between $N_\mathrm{H_2 (^{13}CO)}$ and $N_\mathrm{H_2}$ highlights that the same sources within each line-of-sight are recovered within both the integrated intensity and the molecular hydrogen column density maps. This then confirms our choice to integrate over all velocities within the molecular line cubes to produce the integrated intensity maps (Section\,\ref{sec:IImap}).

We find that the $N_\mathrm{H_2 (^{13}CO)}$ and $N_\mathrm{H_2}$ in Figure\,\ref{fig:NH2_13CO} begin to deviate by more than the uncertainty on the {\it Herschel} column density at around $N_\mathrm{H_2} < 10^{21}$\,\cmsq (see Section\,\ref{appendix_colerr}). This value is significantly below $N_\mathrm{H_2}^\mathrm{thresh}$\,=\,$2.5\times10^{21}$\,cm$^{-2}$, which highlights this as a reasonable threshold for the increased uncertainty $>$\,30\,per cent within the {\it Herschel} molecular hydrogen map.

In summary, in this Section, we have been able to differentiate the column density peaks within the mapped region from the systematically lower column density material associated with the background Galactic plane. We defined a column density threshold above which we expect that the measured column density and temperature values can be predominately attributed to a single source along any given line of sight, and, therefore, a more direct comparison can be made to the molecular line integrated intensity maps. This limit is highlighted on all Figures within this work that make use of the column density within the variable plotted on the y-axis. Any values below the threshold should be assumed to have an associated uncertainty of $>$\,30\,per cent and, therefore, be taken with caution.

\subsection{Distribution of the integrated intensities}

\begin{figure}
    	\includegraphics[trim = 1mm 1mm 1mm 0mm, clip, width=0.95\columnwidth]{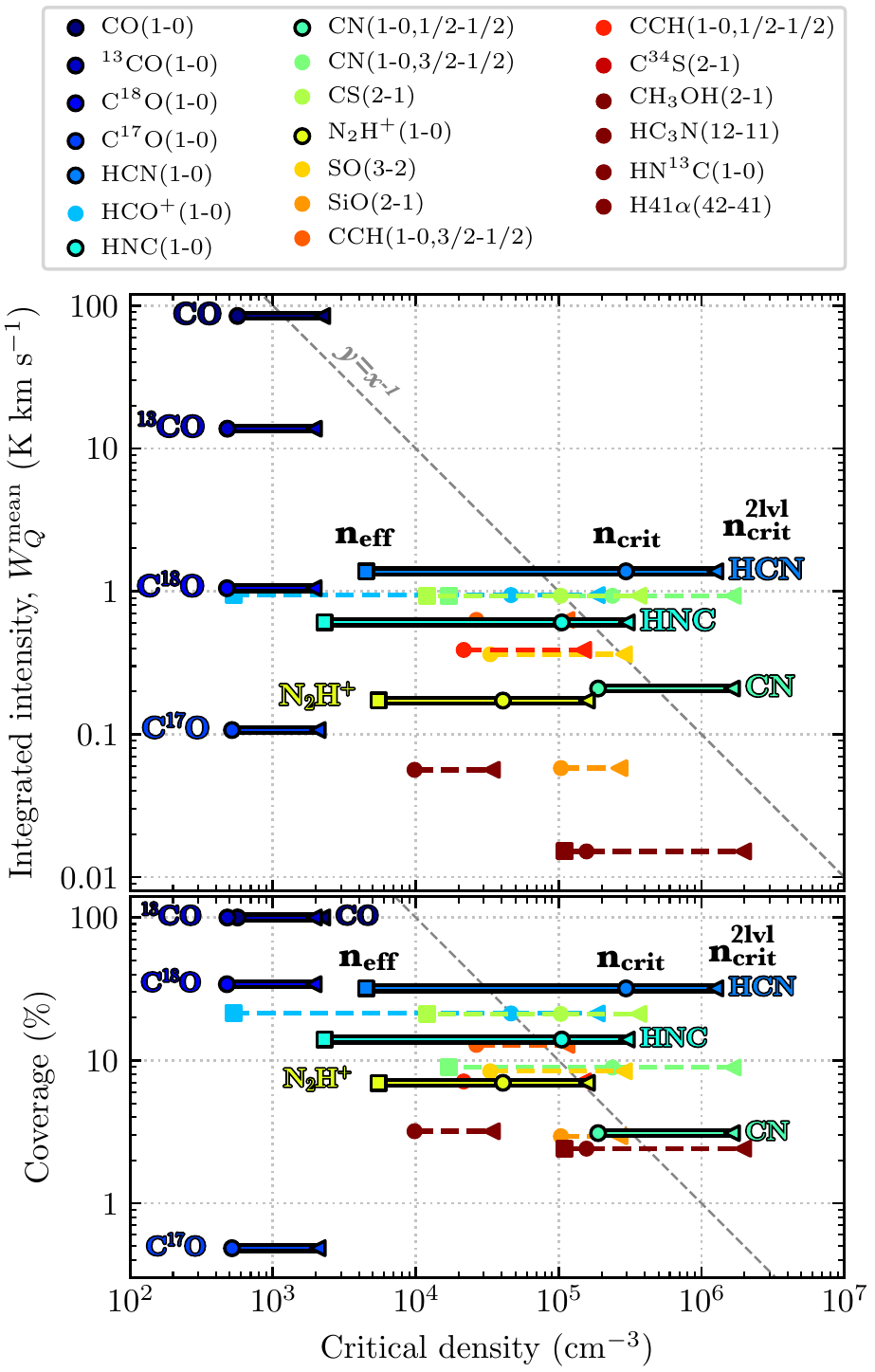}
	\centering
    	\caption{The mean integrated intensity (upper panel) and coverage (lower panel) of each molecular line observed as a function of the critical density. The triangles, circles, and squares show the critical density when using the two-level approximation ($n_{\rm crit}^\mathrm{2lvl}$) and accounting for all the possible collisional rates ($n_{\rm crit}$), and the effective excitation density ($n_{\rm eff}$), respectively (see Table\,\ref{table:obs_info}). The coverage is defined as the percentage of pixels within the mapped region that have an integrated intensity higher than at least five times its associated uncertainty ($W_Q$\,$>$\,$5\sigma_{W_Q}$). The colours of each marker correspond to the molecular lines shown in the legend, and highlighted with labels in the Figure are several molecular lines of interest. A linear $y\,\propto\,x^{-1}$ relationship has been plotted for comparison. This plot highlights that using the critical density as a predictor for line brightness or coverage across a mapped region is not trivial. Rather, the abundance of a given molecule should be taken into account when predicting its emission characteristics.} 
    	\label{fig:ncrit_comp}
\end{figure}

The integrated intensity maps for each molecular transition are shown in Figure\,\ref{fig:momentmaps_smoothmask} (as labelled). The logarithmic colour-scale for each panel has been adjusted to best display the features within each map. Signal-to-noise contours are overlaid at levels of 5\,$\sigma_{W_Q}$, increasing up to the maximum $\sigma_{W_Q}$ value within the mapped region (see Table\,\ref{table:obs_info}).

We find that all the lines have peak integrated intensities towards the centre of W49A, at a position of 19$^\mathrm{h}$10$^\mathrm{m}$13$^\mathrm{s}$, 9\degree06\arcmin16\arcsec. This corresponds to the position of a well studied ring-like cluster of ultra-compact \ion{H}{ii} regions, which is typically referred to as the ``Welch ring'' (e.g. \citealp{welch_1987, rugel_2018b}). This is in agreement with many extensive molecular line studies towards this central region, who find that both higher-$J$ transitions and less abundant isotopologues of the molecule also peak towards this position (e.g. \citealp{nagy_2012, nagy_2015, galvan-madrid_2013}). 

To assess the spatial extent of each line, we define its coverage as the percentage of pixels across the mapped region that have integrated intensities at least five times higher than their associated uncertainty values; $A_\mathrm{cov} = A(W_Q>5\,\sigma_{W_Q})$, where $A$ is the percentage of the total mapped area. We find that the CO lines are the most spatially extended across the mapped region, with both $^{12}$CO and $^{13}$CO covering 100\% and C$^{18}$O covering around $A_\mathrm{cov}\sim$\,35\% of the main mapped region (i.e. that covered with both vertical and horizontal on-the-fly scans). The next most extended lines are HCN, HCO$^{+}$ and CS, which each have coverages of between $A_\mathrm{cov}\sim\,20-30$\%. These are followed by the HNC, CN, SO, CCH, and N$_2$H$^+$ lines that cover around $A_\mathrm{cov}\,\sim\,10$\% of the mapped region. Finally, the remaining 14 lines studied in this work cover only a few per cent of the mapped region. These coverage values, along with the minimum, mean and maximum integrated intensity values measured within each mapped region are summarised in Table\,\ref{table:obs_info}, which is ordered by decreasing coverage values.
 
Figure\,\ref{fig:ncrit_comp} presents the mean integrated intensity and coverage as a function of critical density. Shown as triangles and circles are the critical densities determined using the two-level approximation ($n_{\rm crit}^\mathrm{2lvl}$) and when accounting for all the possible downward collisional rates ($n_{\rm crit}$), respectively. For the lines given by \citet{shirley_2015}, we also plot the effective excitation density ($n_{\rm eff}$) (see Table\,\ref{table:obs_info}). This Figure shows that beyond the $^{12}$CO, $^{13}$CO and C$^{18}$O lines there does not appear to be a clear correlation between the strength or coverage of an observed molecular transition and its critical or effective excitation density. This highlights that the simple approximations we have used to estimate the broad range of characteristic densities are not sufficient to explain the distribution of all the observed molecular lines. Such a result is not surprising, and could be easily explained by the lower abundance and, therefore, optical depth, of rarer molecules. This can be simply seen by studying the CO isotopologues in Figure\,\ref{fig:ncrit_comp}, which show a stark decrease in mean brightness and coverage with increasing rareness despite having very similar critical densities (i.e. $^{12}$CO, $^{13}$CO, C$^{18}$O and C$^{17}$O). This behaviour is discussed further in Sections\,\ref{sec:discussion_densegas} and \ref{sec:model}.

\subsection{Molecular hydrogen column density vs integrated intensities}
\label{subsec:intensity_comp}

\begin{figure*}
	\includegraphics[width=1\textwidth]{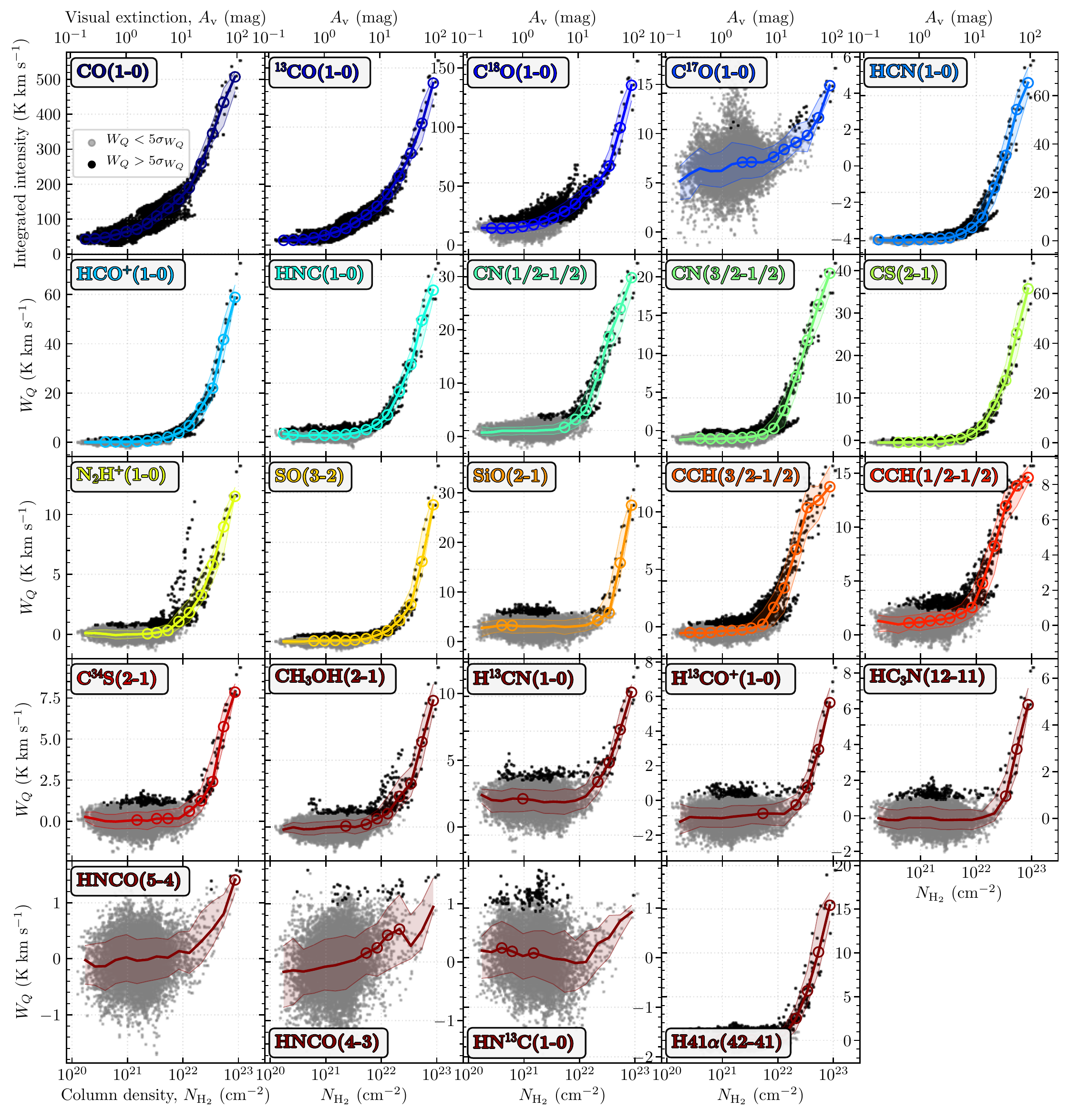}
	\vspace{-6mm}
    \caption{The integrated intensity of each molecular transition as a function of the molecular hydrogen column density derived from the mid-infrared {\it Herschel} observations. The upper x-axis of the top row of the panels shows the visual extinction, $A_\mathrm{v}$ (see Section\,\ref{subsec:ancillary_observations}). The black and grey circular points show the data with an integrated intensity above and below 5$\sigma_{W_Q}$, respectively (as shown in the box in the top left panel). The solid coloured lines show the median values of the integrated intensity within equally spaced logarithmic column density (extinction) bins. The circles show the bins that have a median value above 3\,\sigmabin, where \sigmabin\ represents the propagated uncertainty from all the points contained within the bin (see equation\,\ref{sigmabin}). The shaded region shows the standard deviation of all the points within each bin.}
    \label{fig:W_col_scatter_smoothmask}
\end{figure*}

\begin{figure*}
	\includegraphics[trim = 1mm 1mm 1mm 1mm, clip, width=1\textwidth]{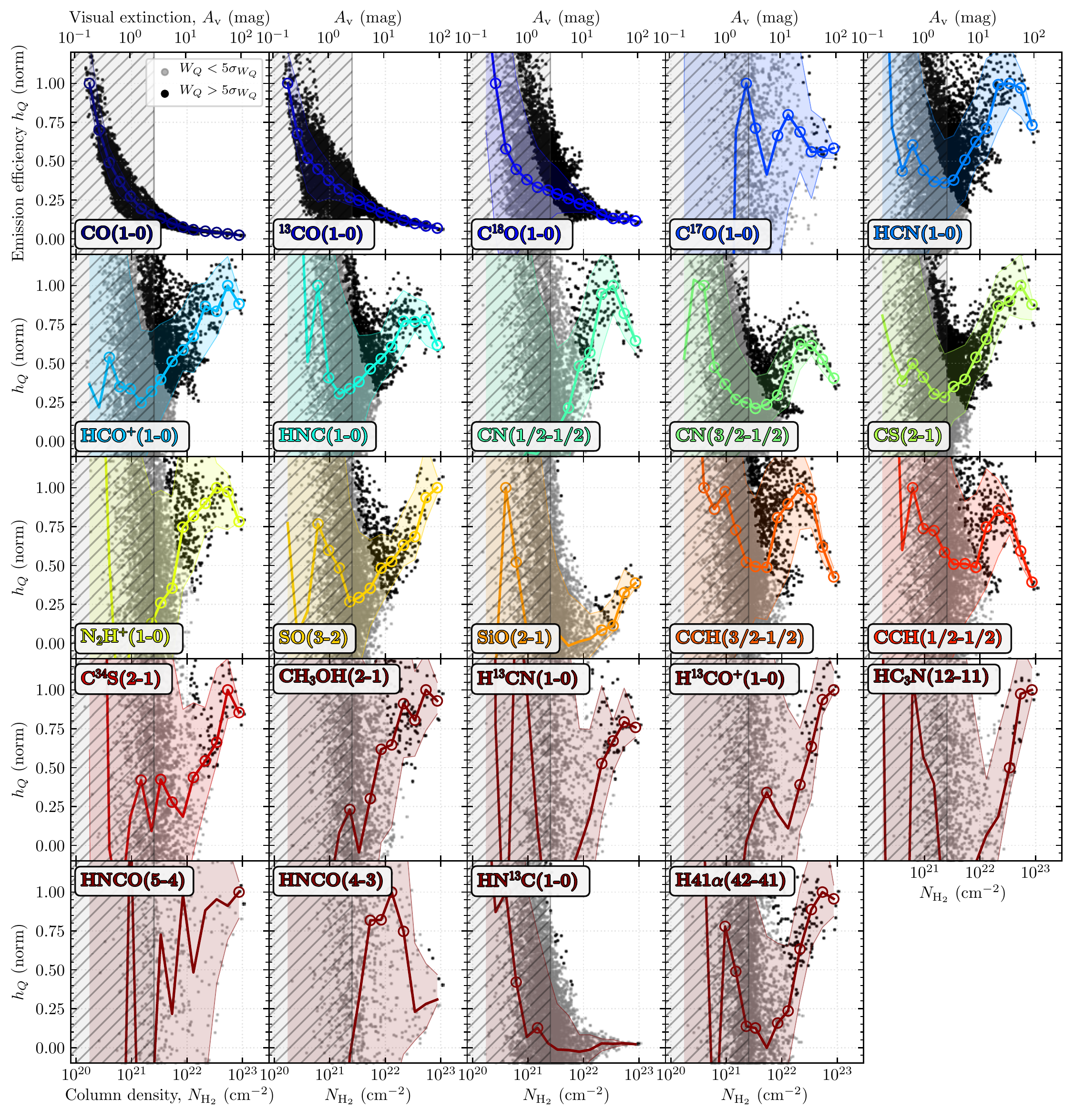}
    \caption{Same as Figure\,\ref{fig:W_col_scatter_smoothmask} except that the y-axis shows the ratio of the integrated intensity to molecular hydrogen column density ($h_{Q} = W(Q)$/\NHtwo). We use this ratio as a proxy for the line emissivity, or the efficiency of an emitting transition per H$_{2}$ molecule. The y-axis has been normalised such that the maximum value of the logarithmically spaced bins is equal to unity. The shaded region highlights values of the column density below $N_\mathrm{H_2}^\mathrm{thresh}$\,=\,$2.5\times10^{21}$\,cm$^{-2}$ ($A_\mathrm{v}^\mathrm{thresh}$\,=\,2.7\,mag) that suffer from increased ($>$\,30\,per cent) uncertainties due to fore- and background contamination (see Section\,\ref{subsec:NhTdust}). Values of the $h_{Q}$ below this threshold value should be taken with caution.}
    \label{fig:W_col_norm_scatter_smoothmask}
\end{figure*}

In Figure\,\ref{fig:W_col_scatter_smoothmask}, we present scatter plots of the integrated intensities as a function of molecular hydrogen column density (or extinction as shown on the upper x-axis; Section\,\ref{subsec:ancillary_observations}). The black and grey circular points show the data with an integrated intensity value above and below 5$\sigma_{W_Q}$, respectively. We plot solid coloured lines that show the median values of the integrated intensity within equally spaced logarithmic column density (extinction) bins (shown at the top of each panel), where the shaded region shows the standard deviation range with each bin. To allow reliable detections of the integrated intensity down to low values of the column density, we stack the individual integrated intensity values within the logarithmic column density (extinction) bins. When doing so, the uncertainty on the median value, $\sigma_\mathrm{bin}(\mathrm{med.})$, is defined as the propagated uncertainty from all the points contained within the bin,
\begin{equation}
\sigma_\mathrm{bin} (\mathrm{med.}) = 1.25 \sigma_\mathrm{bin} (\mathrm{mean}) = 1.25 \frac{f_\mathrm{os}}{n} \sqrt{ \sum^{n}_i \left(\sigma_{W_Q}(i)\right)^2},
\label{sigmabin}
\end{equation}
where $n$ is the number of points within the bin, and $\sigma_{W_Q}(i)$ is the uncertainty associated with the $i^{th}$ value within the bin (see equation\,\ref{sigma}). The oversampling factor is defined as the ratio of the angular beam area to the pixel area, $f_\mathrm{os} = 1.13 \Omega_\mathrm{beam} / \Omega_\mathrm{pixel}$, and corrects for the number of independent measurements within each bin. Using the stacking method we find that even within bins that contain a significant fraction of points with $<5\sigma_{W_Q}$ (grey points), we can still extract a median integrated intensity value with a high level of significance. The result of this can be seen as the open coloured circles, which show the column density bins that have a median value above three times the associated uncertainty. 

The first obvious trend from the binned lines in Figure\,\ref{fig:W_col_scatter_smoothmask} is that on average there is a positive correlation between the integrated intensities and the H$_2$ column density. Indeed, we find that for each line the highest value of the integrated intensity is found at the position of the highest column density. This shows that, in general, the integrated intensities are relatively well correlated with the amount of material along the line-of-sight. However, there are significant deviations from a simple linear relationship. 

To investigate these deviations from a simple linear relation, we normalise the integrated intensity by the H$_2$ column density. \citet{kauffmann_2017} define this fraction $h_{Q}=W(Q)$/\NHtwo, which can be thought of as a proxy for the line emissivity, or the efficiency of an emitting transition per H$_{2}$ molecule. Figure\,\ref{fig:W_col_norm_scatter_smoothmask} shows the emission efficiency ratio as a function of column density (extinction; c.f. figure\,2 of \citealp{kauffmann_2017}). As in Figure\,\ref{fig:W_col_scatter_smoothmask}, we stack within logarithmically spaced column density bins to increase the significant detections at lower column densities. This is shown as the solid coloured lines and shaded regions, and the significant bins are shown with an open circle symbol. In Figure\,\ref{fig:W_col_norm_scatter_smoothmask}, the y-axis of each panel has been normalised such that the maximum value of the significant logarithmically spaced bins is equal to unity.
 
 We find that the $^{12}$CO, $^{13}$CO and \CeO\ show a near-constant decrease in $h_{Q}$ with increasing column density, which is to be expected based on the overall sub-linear relations shown in Figure\,\ref{fig:W_col_scatter_smoothmask}. Focusing on the significant median bins, we find that CS, \CHtOH, HCO$^{+}$, and \NtHp show $h_{Q}$ profiles that quickly rises from low column densities to maximum emission efficiency at a column density of several $10^{22}$\,\cmsq, and slowly decrease towards higher column density values. 

The profiles of HCN, HNC, CN, and CCH show more complex double-peaked profiles. These lines show an initially high value of $h_{Q}$ at the lowest values of the column density ($\sim10^{20-21}$\,\cmsq), that decreases with increasing column density to reach a minimum value at $\sim10^{22}$\,\cmsq. The $h_{Q}$ then rises again to a maximum value of at column density around few $10^{22}$\,\cmsq. We highlight again here, however, that values of the column density below the threshold value of $N_\mathrm{H_2}^\mathrm{thresh}\sim$\,10$^{21}$\cmsq\ suffer from significant fore- and background contamination (see Section\,\ref{subsec:NhTdust}). Therefore, these multiple peaked profiles for the emission efficiency should be taken with caution. 

\section{Analysis}\label{sec:analysis}

\subsection{Fraction of the emission over various column densities}\label{sec:fractional_emission}

\begin{figure}
	\includegraphics[width=0.95\columnwidth]{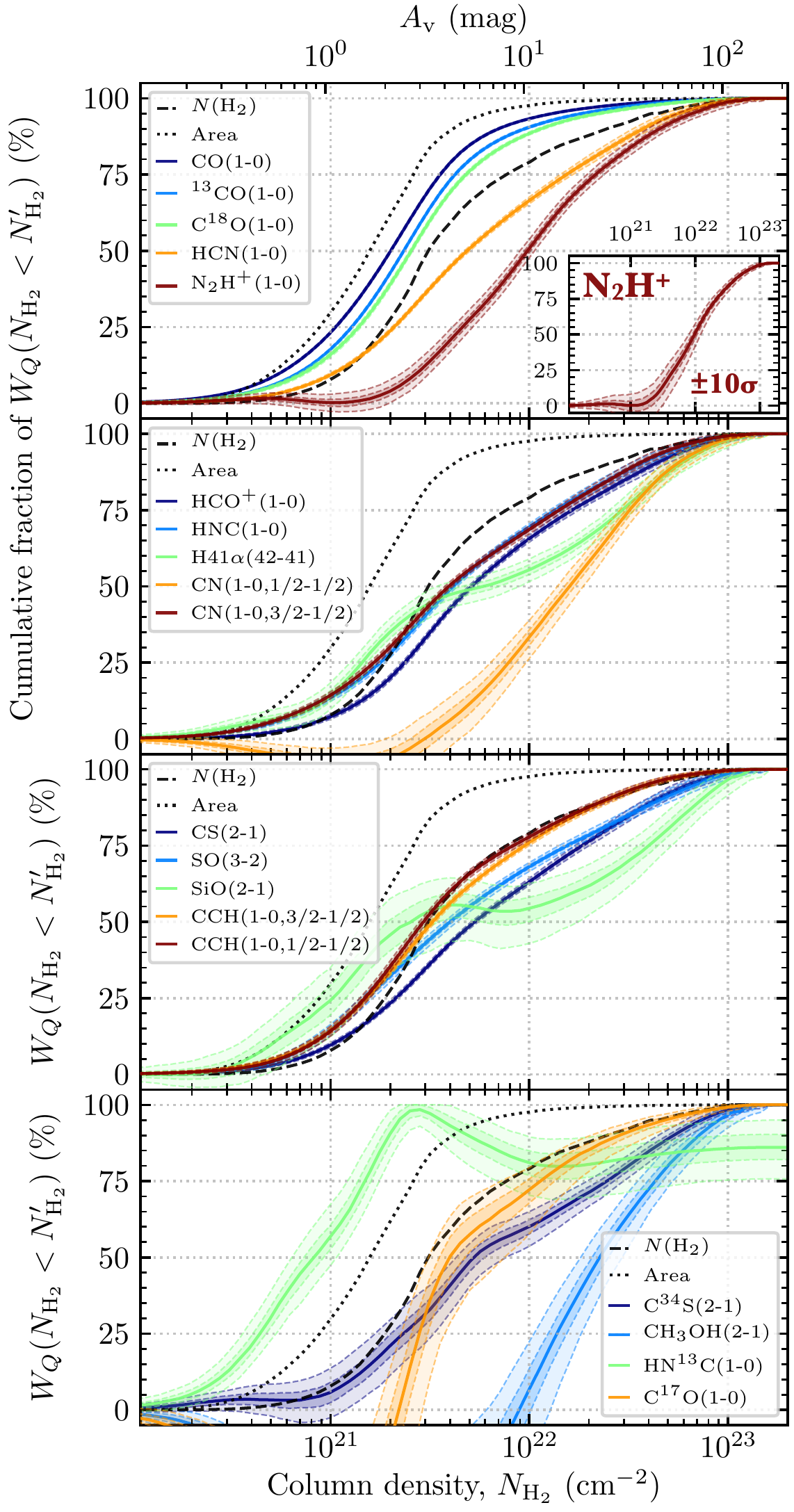} 
    \caption{Cumulative fraction of the integrated intensity below a given column density for each molecular line. The solid line shows the observed value, and the shaded regions show the associated 5, 25, 75, and 95\,percentile uncertainty limits (see Section\,\ref{sec:fractional_emission}). The black dotted line shows the cumulative distribution of the number of pixels, which can be thought of as a completely smooth distribution (i.e. a map containing only a single value). The dashed black line shows the cumulative distribution of the column density values. The lines have been split over four panels for plotting clarity. The inset panel shows the addition of a 10\,$\sigma_{W_Q}$ noise to the \NtHp\ map.}
    \label{fig:W_col_cum_line_all}
\end{figure}

\begin{figure}
	\includegraphics[clip, width=0.975\columnwidth]{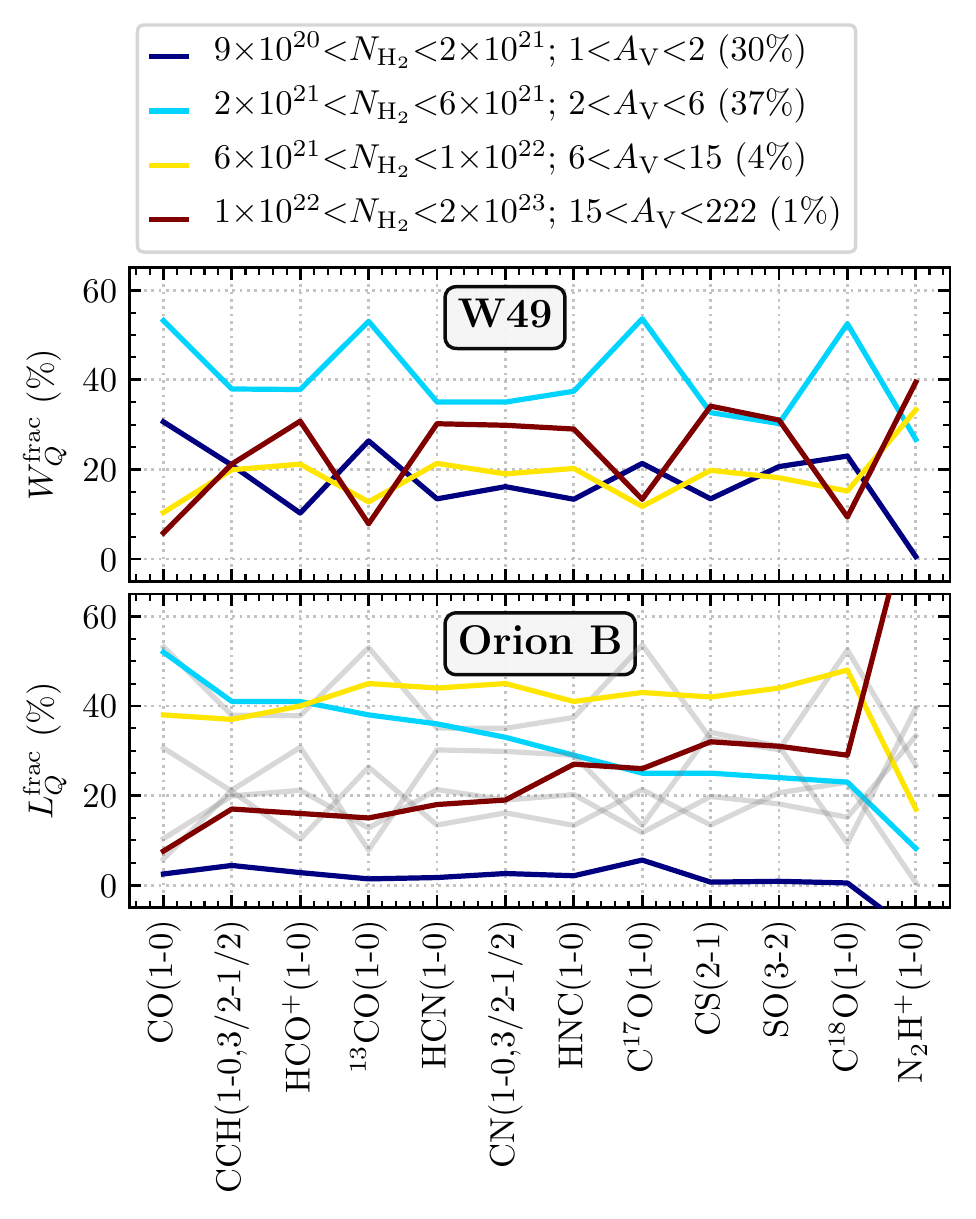} 
    \caption{The fraction of the total intensity, $W_Q^\mathrm{frac}$, from each molecular line within four column density masks, as shown in the legend above the plot (see Section\,\ref{sec:fractional_emission}). Also shown in the legend above the plot is the fraction of the total area covered by each of the column density (extinction) masks. The upper panel shows the intensity fraction determined across the W49 massive star-forming region studied in this work, and the lower panel shows the luminosity fraction, $L_Q^\mathrm{frac}$, from the local star-forming region Orion B taken from \citet[][as shown in their Fig.7]{pety_2017}. For comparison, the W49 results have also been plotted on the lower panel as grey lines.}
    \label{fig:fraction_both}
\end{figure}

In an attempt to determine the typical extinction (column density) that is traced by a given molecular line transition, \citet{kauffmann_2017} assess the cumulative fraction of the line intensity as a function of extinction (column density); $W_Q (N_\mathrm{H_{2}} < N_\mathrm{H_{2}}^\prime)$. Here we follow the same method and determine the cumulative distribution of the integrated intensity for each line. We note that, as we integrate all the intensity along the line of sight without isolating sources (e.g. W49A) in velocity, we do not have a single distance that could be reasonably applied to each pixel. Therefore, instead of producing the cumulative function of the luminosity as done in \citet{kauffmann_2017}, we produce the cumulative function of the integrated intensity values across the mapped region. These results are displayed in Figure\,\ref{fig:W_col_cum_line_all}, where we split the observed molecular lines between four panels for plotting clarity.

Shown as shaded regions in Figure\,\ref{fig:W_col_cum_line_all} is the uncertainty on each cumulative distribution, which is determined as the deviation in the distributions when adding synthetic noise to both the molecular line and column density maps. To produce these noise curves, we first include synthetic noise in the form of a Gaussian distributions centred on zero with widths of $\sigma_{W_Q}$ (Section\,\ref{sec:IImap}) and $\Delta N_\mathrm{H_2}$ to each of the integrated intensity maps and the column density map, respectively. This was repeated 1000 times to produce a sample of noisy maps. The shaded curves in Figure\,\ref{fig:W_col_cum_line_all} show the 5, 25, 75, and 95\,percentiles of the resultant ensemble of cumulative distributions. Here the brightest lines (e.g. the CO lines) show a small uncertainty range (shaded region), as they are not significantly affected by the addition of $\sigma_{W_Q}$. To highlight the rigorous detection of these brighter lines (e.g. \NtHp), the inset panel shows the marginal increase in uncertainty with the addition of 10\,$\sigma_{W_Q}$ noise. The weaker lines (e.g. HC$_3$N), on the other hand, show a larger variation due to the additional uncertainty, as the peak signal within these maps is closer to the $\sigma_{W_Q}$ level (compare integrated intensity uncertainty values given in Table\,\ref{table:obs_info} to integrated intensity statistics given in Table\,\ref{table:obs_stats}). 

We also plot a dotted black line in Figure\,\ref{fig:W_col_cum_line_all} that shows the cumulative distribution of the pixels, and hence would represent the distribution of a line with a completely smooth integrated intensity map (i.e. a map containing pixels with only a single value). We plot the cumulative distribution of the column density values as a dashed black line in Figure\,\ref{fig:W_col_cum_line_all}. Using these, we can assess how much of the molecular line intensity is contained below a given column density; or in the case of W49, how concentrated the emission is towards the column density peak in the W49A star-forming region. At any given column density, if the cumulative distribution profile resides between the smooth (dotted) and column density (dashed) profiles its remaining luminosity should be thought of as still being relatively evenly distributed across the mapped region (i.e. has a smoother integrated intensity map), whilst those to the right of the column density distribution should be thought of as having luminosities that are more strongly concentrated (i.e. has a more peaked integrated intensity map). 

Figure\,\ref{fig:W_col_cum_line_all} shows that the cumulative distributions vary significantly among the different lines. We find that the CO lines are typically less concentrated than the column density (mass) distribution, whereas the remaining lines are typically more concentrated. 

In Figure\,\ref{fig:fraction_both}, we show the fraction of the total intensity, $W_Q^\mathrm{frac}$, from each of the molecular lines within four column density (extinction) masks, which have been chosen for direct comparison with \citet[][their Fig.7]{pety_2017}. These authors suggest that within the Orion molecular cloud these ranges of column density (or extinction) can be used to represent diffuse (1$<$$A_\mathrm{V}$$<$2\,mag), and translucent (2$<$$A_\mathrm{V}$$<$6\,mag) gas, the environment of filaments (6$<$$A_\mathrm{V}$$<$15\,mag), and dense gas (15$<$$A_\mathrm{V}$$<$222\,mag).\footnote{The 1$<$$A_\mathrm{V}$$<$2\,mag and 2$<$$A_\mathrm{V}$$<$6\,mag extinction bins contain emission from below our extinction threshold of $A_\mathrm{v}^\mathrm{thresh}$\,=\,2.7\,mag (see Section\,\ref{subsec:NhTdust}). Figure\,\ref{fig:NH2_13CO} shows that across 1$<$$A_\mathrm{V}$$<$2.7\,mag $N_\mathrm{H_2 (^{13}CO)}$ deviates from $N_\mathrm{H_2}$ by $30-40$\,per cent, which we suggest sets an upper limit to the uncertainty on $W_Q^\mathrm{frac}$ due to fore- and back- ground contamination within this extinction range.} We find that the CO lines have the majority of their intensity ($W_Q^\mathrm{frac}\sim80$\%) within the lowest two column density bins, which combined includes the majority of the mapped area ($\sim$70\%). The HCN, HCO$^+$, HNC, CS, and CN lines show a stark decrease in the intensity within the lowest column density bin ($W_\mathrm{^{12}CO}^\mathrm{frac}\sim30$\%, $W_\mathrm{HCN}^\mathrm{frac}\sim10$\%), and an increase within the highest column density bin ($W_\mathrm{^{12}CO}^\mathrm{frac}\sim5$\%, $W_\mathrm{HCN}^\mathrm{frac}\sim30$\%). The \NtHp\ shows an even higher fraction of emission towards the two highest bins, with $W_\mathrm{N_2H^+}^\mathrm{frac}\sim$40\% of the total emission within the highest column density bin that accounts for only around 1\% of the total mapped area.

These results are compared to the Orion B star-forming region \citep{pety_2017}, which is shown in the lower panel of Figure\,\ref{fig:fraction_both} (also with the W49 results in grey). We see that there are several notable differences between the two regions. First, there appears to be a lower fraction of the intensity within the dense gas for the CO lines in W49 when compared to Orion. This is expected given that we observe a much larger spatial coverage within W49 ($\sim$\,10$^4$\,pc$^{2}$) than the Orion observations ($\sim$\,10\,pc$^{2}$), and hence recover a larger fraction of diffuse gas (relative to the dense gas). Secondly, we see that lines such as CN, HNC, HCO$^{+}$, and HCN have a marginally higher fraction of the intensity within dense gas. Lastly, we observe a significant difference in the fraction of dense gas for the \NtHp\ line, which has more than 80\% of the luminosity within the high-density regime in Orion compared to around 40\% across the mapped region. \citet{watanabe_2017}, and \citet{nishimura_2017} conducted similar analyses of the 3mm lines within the W51 and W3(OH) regions within the Galaxy. These authors also find results broadly in agreement with those found here. For example, \citet{nishimura_2017} classify their observed distributions into the three types: (i) species concentrated just around the centre of the star-forming core in W3(OH), such as \CHtOH, SO, and HC3N, (ii) species loosely concentrated around the centre, such as HCN, HCO$^+$, HNC, and CS, and (iii) species extended widely, such as CCH, \CeO, and \tCO.

\subsection{Line emissivities as a function of both molecular hydrogen column density and dust temperature}\label{sec:col_temp}

\begin{figure*}
\centering
        \includegraphics[width=1.0\textwidth]{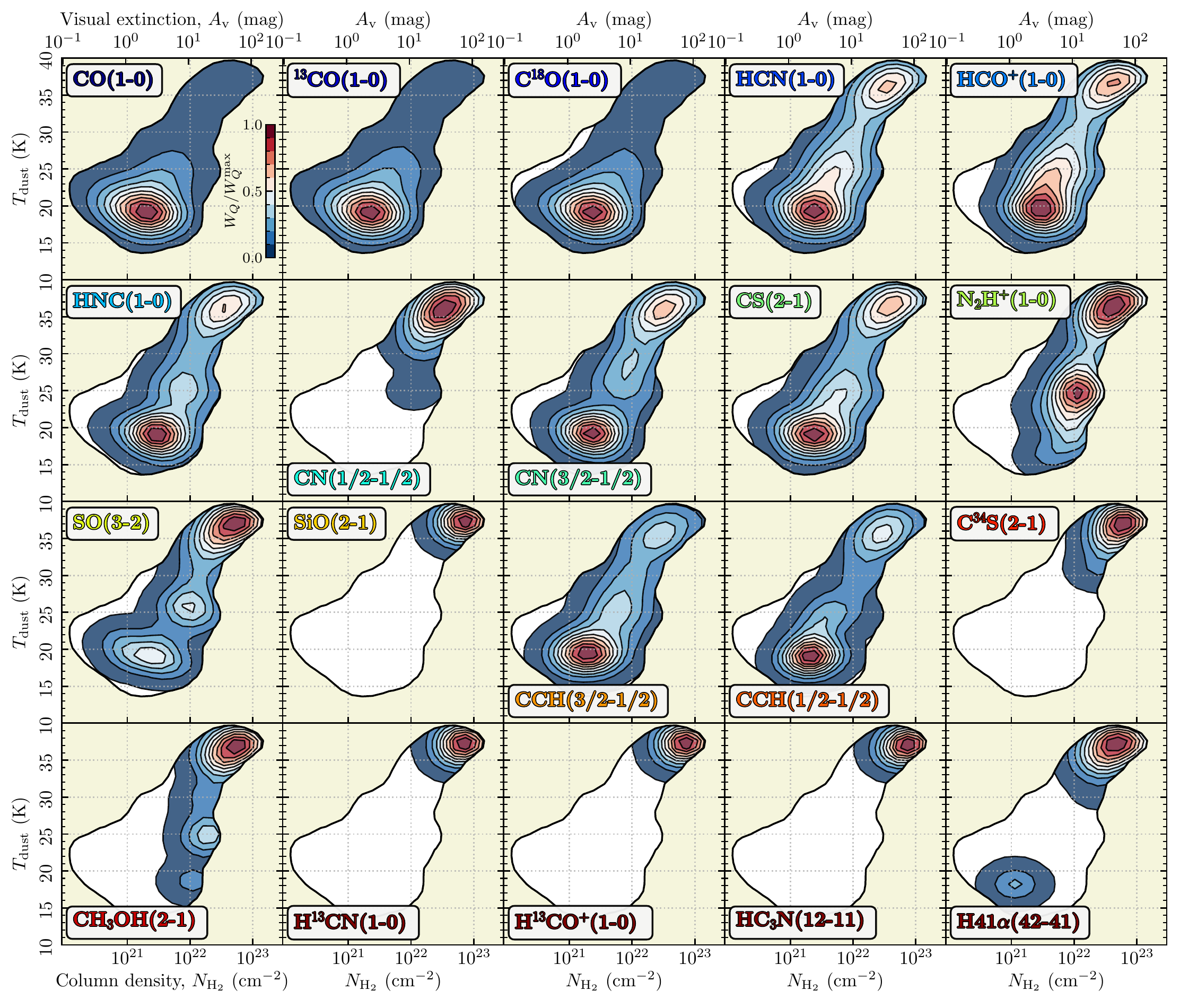}{\vspace{-5mm}}
    \caption{The distribution of integrated intensities as a function of the dust temperature and molecular hydrogen column density (extinction), where the integrated intensities have been binned along each axis to form a contour plot (c.f. scatter plot shown in Figure\,\ref{fig:w49_column-temp}). The contours show the 1, 10, 20, 30 \dots 80, and 90 per cent levels of the total integrated intensity for a given line across the significant bins within the parameter space ($5\sigma_\mathrm{bin}$; see equation\,\ref{sigma}). This plot highlights the properties of the region from which the majority of molecular line emission across the mapped region are originating.} 
    \label{contourplot_Wq}
\end{figure*}

\begin{figure*}
\centering
        \includegraphics[width=1.0\textwidth]{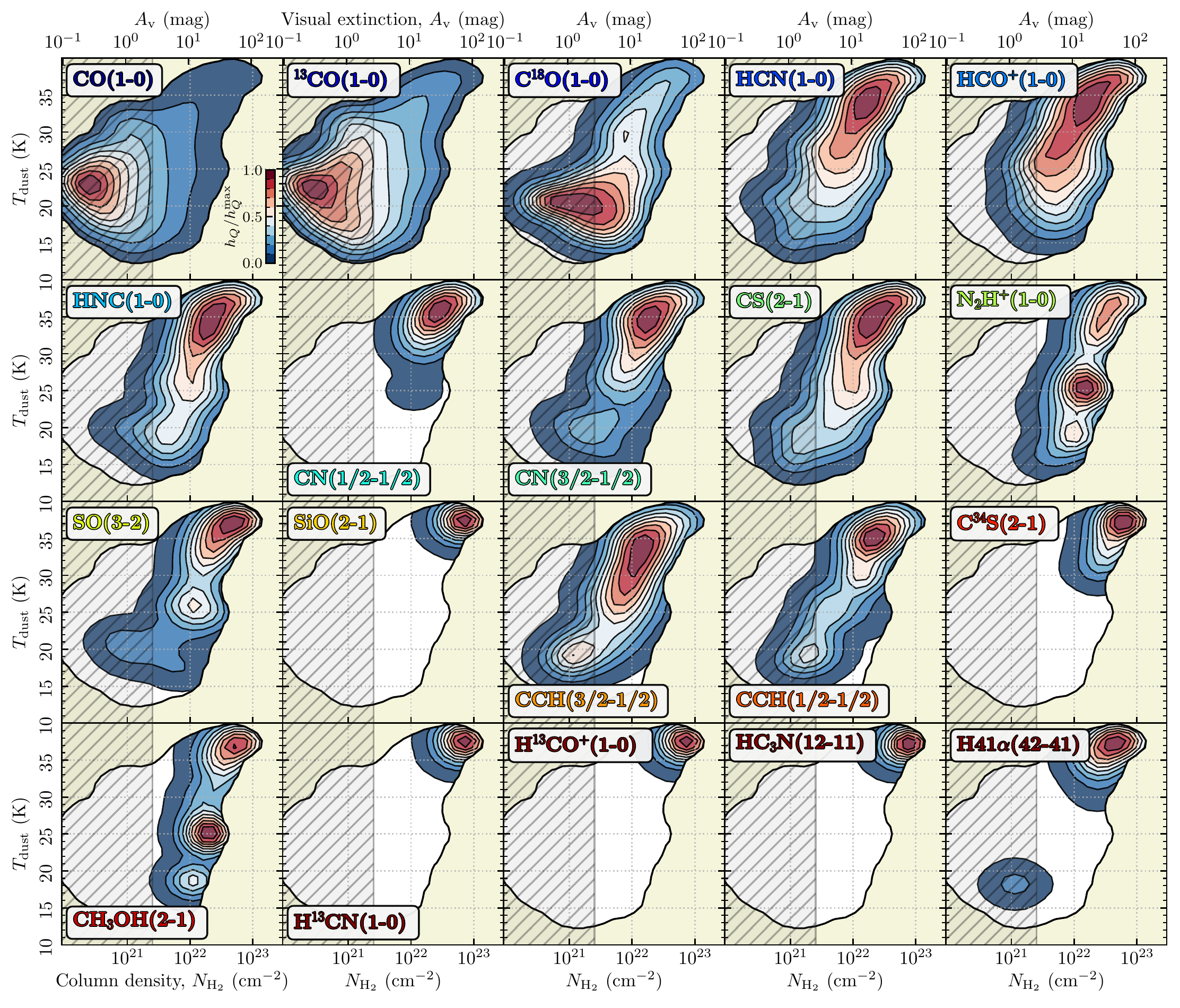}{\vspace{-5mm}}
    \caption{The distribution of emission efficiencies as a function of the dust temperature and molecular hydrogen column density (extinction), where the emission efficiencies have been binned along each axis to form a contour plot (c.f. scatter plot shown in Figure\,\ref{fig:w49_column-temp}) The contours show the 1, 10, 20, 30 \dots 80, and 90 per cent levels of the emission efficiency ($h_{Q}=W(Q)$/\NHtwo) for a given line across the significant bins within the parameter space ($5\sigma_\mathrm{bin}$; see equation\,\ref{sigma}). We highlight here values of the column density below $N_\mathrm{H_2}^\mathrm{thresh}$\,=\,$2.5\times10^{21}$\,cm$^{-2}$ ($A_\mathrm{v}^\mathrm{thresh}$\,=\,2.7\,mag) that suffer from increased ($>$\,30\,per cent) uncertainties due to fore- and background contamination (see Section\,\ref{subsec:NhTdust}). Values of the $h_{Q}$ below this threshold value should be taken with caution. This plot highlights the properties from which the molecular lines are most efficiently emitting across the mapped region are originating.} 
    \label{contourplot_hq}
\end{figure*}

So far in this work, we have investigated the variation of the emission from each molecular line as a function of the column density (extinction) and the critical or effective excitation density. However, as \citet{pety_2017} showed within the Orion B region, the temperature is also a fundamental parameter, as it also is closely linked to both the excitation of molecular lines and the formation chemistry (i.e. abundances). In this Section, we investigate the simultaneous effects of the molecular hydrogen column density and dust temperature on the total line emission and emission efficiency. We remind the reader that spatial distributions of both the molecular hydrogen column density and dust temperature are presented in Figure\,\ref{fig:w49_rgb}, and a comparison of the W49 and a representative portion of the Galactic plane is given in Figure\,\ref{fig:w49_column-temp} (see Section\,\ref{subsec:NhTdust}). It is worth noting here, however, that the dust temperature may not be equal to the kinetic temperature of the gas. Without a probe of the latter within our observations (e.g. NH$_3$ or H$_2$CO), we work under the assumption the dust and gas temperatures are coupled across the mapped region.

To conduct this analysis, we first take the integrated intensity ($W_Q$) and emission efficiency ($h_Q$) maps and identify the value of the column density and dust temperature for each position. We then create evenly spaced bins for the observed range of column density and dust temperature, into which we assign the $W_Q$ and $h_Q$ values based on their column densities and dust temperatures (i.e. similar to the binning process outlined in Section\,\ref{subsec:intensity_comp}). Finally, we calculate the summed $W_Q$ and median $h_Q$ values within each column-density temperature bin, and calculate the uncertainty on this value using equation\,\ref{sigma} (see Section\,\ref{subsec:intensity_comp}). We note that both HNCO lines and the C$^{17}$O and HN$^{13}$C line have been omitted from this analysis due to their lack of significant column density and temperature bins.

Figures\,\ref{contourplot_Wq} and \ref{contourplot_hq} show the parameter space spanned by the dust temperature and column density (extinction). The overlaid contours show the 1, 10, 20, 30 \dots 80, and 90 per cent levels of the maximum summed $W_Q$ and median $h_Q$ for a given line across the significant bins within the parameter space ($5\sigma_\mathrm{bin}$; see equation\,\ref{sigmabin}). These plots highlight the environmental properties which are producing the majority of the molecular line emission, and the conditions which are most efficiently producing this emission, respectively. 

Figure\,\ref{contourplot_Wq} shows that many of the lines emit the majority of their emission from moderate temperatures ($\sim$\,20\,K), and moderate column densities ($\sim$10$^{21-22}$\cmsq) following the peak in the number of pixels within the parameter space (compare to the Figure\,\ref{fig:w49_column-temp}). In addition, many of the lines also peak towards the highest column density and temperature region of the parameter space ($\sim$\,20\,K; $\sim$10$^{23}$\cmsq). This region of the parameter space contains significantly fewer pixels, emphasising that this peak is due to the concentration of bright pixels towards the centre of the W49A massive star-forming region.

Comparing how the total emission (Fig.\,\ref{contourplot_Wq}) and emission efficiency (Fig.\,\ref{contourplot_hq}) are distributed across the observed parameter space, we find the column density and temperature ranges producing the most emission may not do so most efficiently. For example, although CO and $^{13}$CO are producing most of the emission within the moderate density and temperature regime, they have a peak in emission efficiency at almost an order of magnitude lower column density. In addition, we find that many of the lines that showed two peaks in the emission distribution, have a single peak in the emission efficiency distribution (e.g. HCO$^{+}$, HCN, HNC, CS). Interestingly, several of the lines show comparable distributions for the emission and emission efficiency (\NtHp, SO, CCH), highlighting that for these lines the gas properties producing the most emission are also doing so most efficiently.

\section{Discussion}\label{sec:discussion}

In this Section, we discuss the results of this work in the context of our current understanding of the origins of molecular line emission from the interstellar medium. We note that an outline of the uncertainties present within our observations and the caveats that these imply to our interpretations is presented at the end of this Section. 

\subsection{Are ``dense gas'' molecular line tracers preferentially tracing the ``dense'' gas?}\label{sec:discussion_densegas}

\begin{figure*}
	\includegraphics[width=1\textwidth]{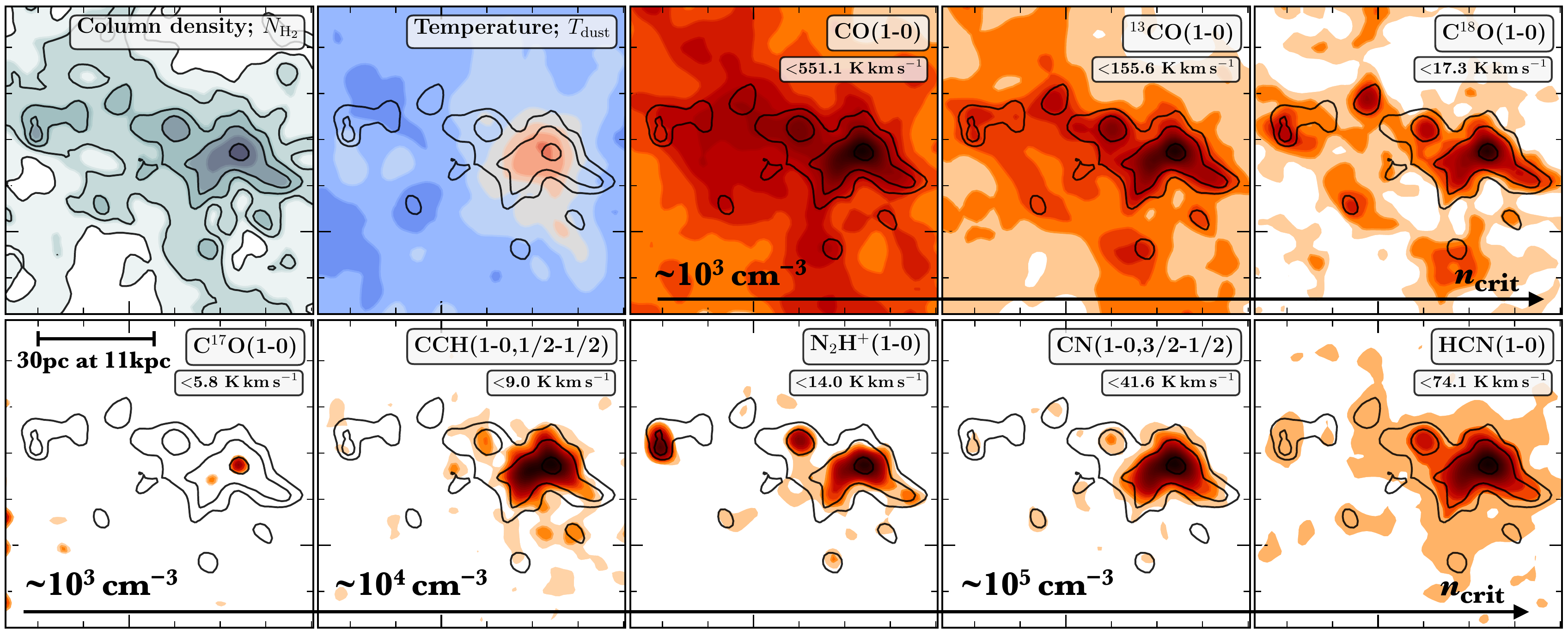}
    \caption{Maps of the integrated intensity for a selection of lines (see comparable figure for Orion A in \citealp{kauffmann_2017}). The upper left two panels show the column density and dust temperature determined from the {\it Herschel} observations. Overlaid on the upper left panel are contours \NHtwo\ in levels of 1 and 2.5\,$\times 10^{21}$~\cmsq, and in all panels levels of 5, 10, 50\,$\times 10^{21}$~\cmsq. The filled contours for the molecular line emission are drawn at signal–to–noise ratios of 3, 5, 10, 30, 50, 100, 150, 200, and 300.  All maps shown here have been smoothed to 60\arcsec ($\sim$\,3\,pc at the source distance of 11\kpc), and the panels are ordered by increasing critical density. Labelled within each of the integrated intensity maps the peak value (see Table\,\ref{table:obs_stats}).}
    \label{toyplot}
\end{figure*}

\begin{figure}
	\includegraphics[width=1\columnwidth]{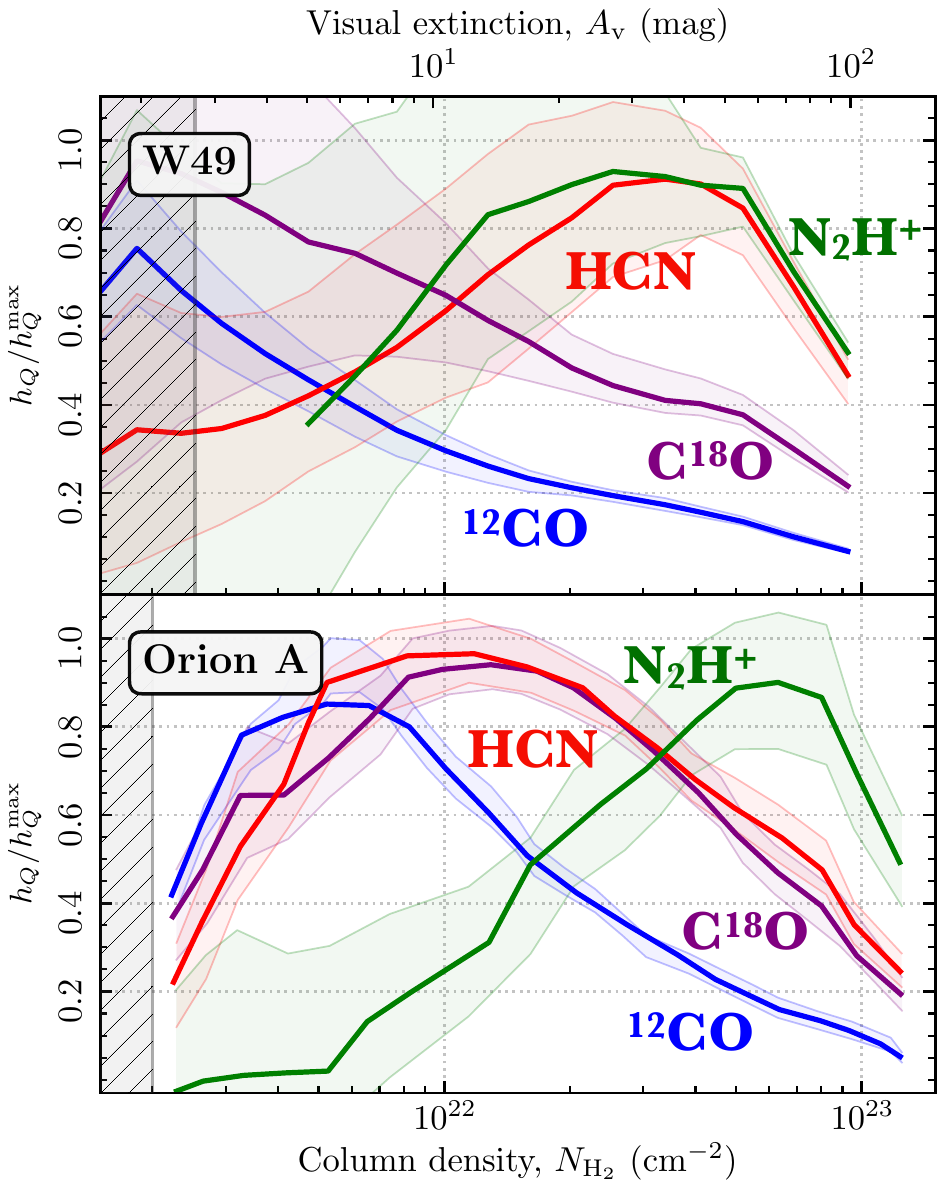}
    \caption{The ratio of the integrated intensity to the molecular hydrogen column density [or emission efficiency ratio; $h_{Q}=W(Q)$/\NHtwo] as a function of column density for four the molecular lines (i.e. similar to Figure\,\ref{fig:W_col_norm_scatter_smoothmask}). Shown in the upper panel are the data presented in this work for the W49 region. To allow a simple comparison, only bins that have a median value above 5\,$\sigma_\mathrm{bin}$ are displayed. The lower panel shows the emission efficiency ratios determined within the Orion A local star-forming region taken from \citet[][their fig. 2]{kauffmann_2017}. We highlight by shading here column densities below the threshold limits for both sources, which are believed to suffer from fore- and background contamination (see Section\,\ref{subsec:NhTdust}). Values of the $h_{Q}$ below this threshold value should be taken with caution.}
    \label{fig:W_col_norm_line_single}
\end{figure}

\begin{figure}
	\includegraphics[width=1\columnwidth]{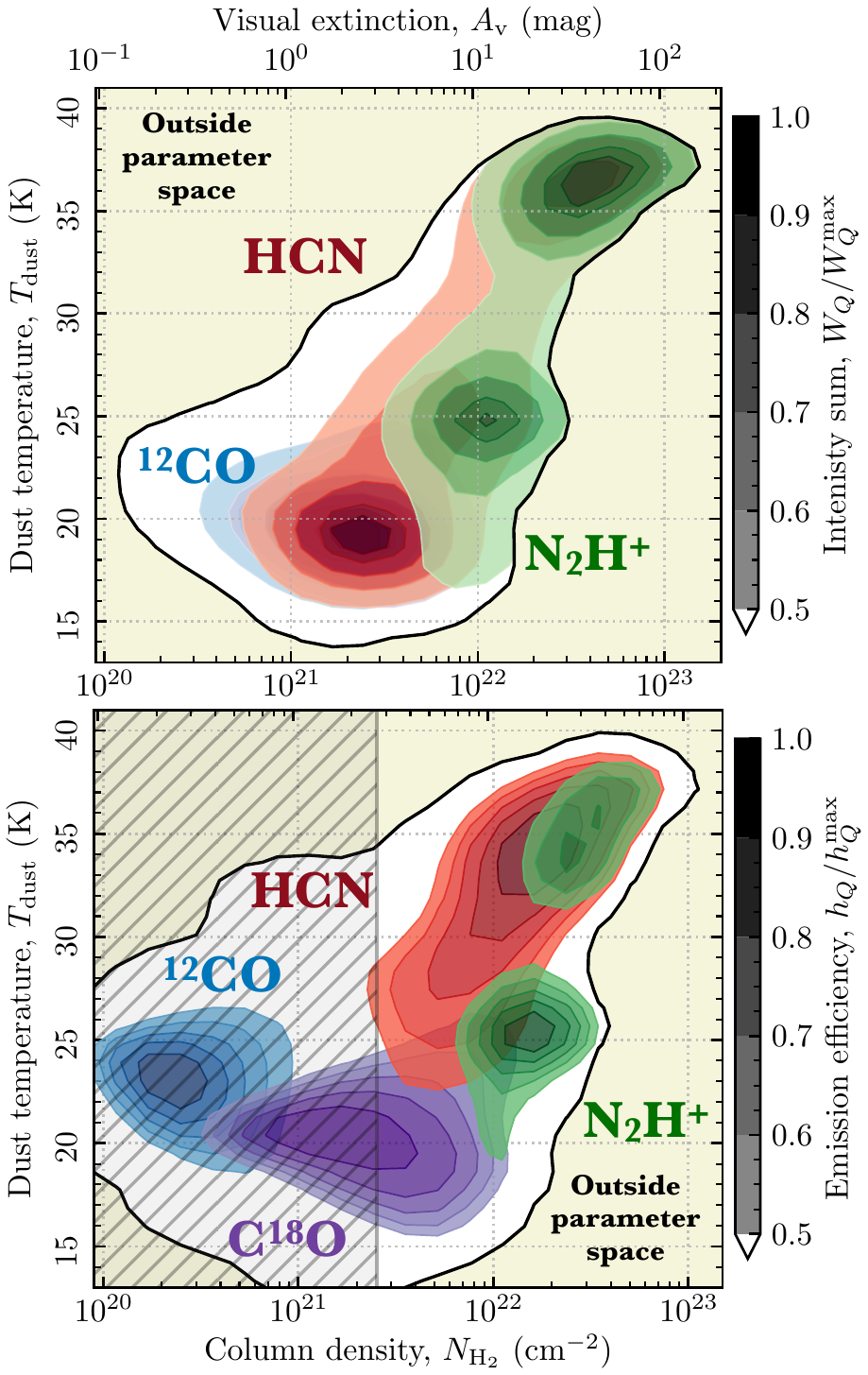}
    \caption{The total integrated intensity and emission efficiency ($h_{Q}=W(Q)$/\NHtwo) as a function of column density and dust temperature for a selection of the molecular lines (see Figures\,\ref{contourplot_Wq} and \ref{contourplot_hq}). The contours show the 50, 60, 70 80, and 90 per cent levels of total intensity in the upper panel, and emission efficiency in the lower panel. The black contour shows the region of the parameter space probed by the observations presented in this work. In the lower panel, we highlight values of the column density below $N_\mathrm{H_2}^\mathrm{thresh}$\,=\,$2.5\times10^{21}$\,cm$^{-2}$ ($A_\mathrm{v}^\mathrm{thresh}$\,=\,2.7\,mag) that suffer from increased ($>$\,30\,per cent) uncertainties due to fore- and background contamination (see Section\,\ref{subsec:NhTdust}). Values of the $h_{Q}$ below this threshold value should be taken with caution. This plot highlights the properties producing the most molecular line emission, and where the emission is being produced most efficiently.}
    \label{contourplot_simple}
\end{figure}

When studying the array of molecular line transitions observable within molecular clouds, the critical density of a given transition is typically used as a proxy of the density from which it is emitting. As such the different molecular line transitions can be used to probe the physical conditions of the various density regimes present throughout the cloud. Of particular interest are the properties of the ``dense gas'' most closely associated with star formation. Much of our understanding of star formation is based on the study of these dense gas properties. For example, the star-forming potential of the molecular clouds is directly determined by the physical conditions (e.g. level of turbulence) of the dense gas, upon which the majority of the current analytic models for star formation are calibrated. Moreover, the mass determined from these dense gas tracers is used to derive empirical scaling relations to the level of star formation, which are widely used to calibrate simulations and estimate the star formation rate within extragalactic observations. The fundamental assumption typically adopted by molecular line studies is that the emission from any dense gas tracer, as inferred from its critical density, is a) emitting from gas at that density (i.e. $n\,\sim\,n_\mathrm{crit}$), and b) this gas is directly linked to the star formation process (e.g. \citealp{gao_2004b}). In this work, we address the first of these assumptions, and ask the question: are ``dense gas" molecular line tracers preferentially tracing the {\bf ``dense"} gas?

In Figure\,\ref{toyplot}, we present integrated intensity maps for a selection of the observed molecular lines, which have been ordered by increasing critical density (all integrated intensity maps are shown in Figure\,\ref{fig:momentmaps_smoothmask}). This plot was made to be comparable to figure 1 of \citet{kauffmann_2017}, which presents the molecular line emission across the local Orion A star-forming region. Here, we can see broadly similar trends to what was found for Orion A. We find CO, $^{13}$CO, have extended emission across all the mapped region, whilst $^{17}$CO emits only towards the centre of W49A. Given that these molecular lines have comparable critical densities, this effect is likely due to the reduced abundance of the rarer isotopologues and insufficient integration time. Or, it could be that the rarer species are destroyed in the lower density material, reducing further their abundance (fractionation effects, e.g. \citealp{langer_1984, rollig_2013}).

Moving beyond these CO lines in Figure\,\ref{toyplot}, we find little correspondence between the extent of the molecular line emission and the increase in critical density. For example, we see comparable spatial morphologies of the CCH, CN and HCN lines, despite them having critical densities differing by almost two orders of magnitude. Moreover, we find that only \NtHp\ appears to also trace the column density peak at the east of the mapped region (towards W49B) at comparable brightness to the peak towards the centre of the W49A; potentially highlighting this as a region of differing excitation or chemistry than W49. These results for all the observed lines are summarised in Figure\,\ref{fig:ncrit_comp}, from which we find that there is a significant scatter around the simple linear relation of the average integrated intensity across the mapped region and increasing critical density; several extreme outliers being \CeO\ and \CHtOH. This scatter is increased when plotting the coverage across the mapped region (i.e. positions with an integrated intensity above five times the uncertainty) as a function of the critical density. Together, Figures\,\ref{fig:ncrit_comp} and \ref{toyplot} then clearly show that the brightness and coverage of a molecular line transition are not completely governed by the critical density; e.g. excitation effects or chemical abundance variations may also be important factors. This is in agreement with the results from the Orion molecular cloud \citep{kauffmann_2017, pety_2017}. 

In Section\,\ref{subsec:intensity_comp}, we investigated the emissivity of the molecular lines. When doing so, we identified large deviations from a simple linear relation for the integrated intensity of a molecular line and the column density. To investigate this, we normalise the integrated intensity by the column density, which we defined as the emission efficiency ratio ($h_{Q}$=$W(Q)$/\NHtwo; \citealp{kauffmann_2017}). In Figure\,\ref{fig:W_col_norm_line_single}, we present the emission efficiency ratio as a function of the column density for a selection of the observed molecular lines, which have plotted on a single axis for comparison (comparable to Figure\,\ref{fig:W_col_norm_scatter_smoothmask}). This plot shows a variety of profiles for the different molecular lines with increasing column density. Also shown in the lower panel of Figure\,\ref{fig:W_col_norm_line_single} are the emission efficiency ratios determined within the Orion A star-forming region taken from \citet[][their fig. 2]{kauffmann_2017}. We find a reasonable first-order agreement between W49 and Orion A. We see that $^{12}$CO has a maximum emission efficiency ratio at the lowest column density. \CeO, HCN and \NtHp show peak emission efficiency ratio values at increasingly higher column densities. There are, however, several second-order differences that should be mentioned. The first is that we do not observe a peaked emission efficiency ratio profile for $^{12}$CO within W49, as is seen within Orion (comparing blue curves in Figure\,\ref{fig:W_col_norm_line_single}). Rather, we find a gradual decline of the $^{12}$CO emission efficiency ratio with increasing column density. The second is that for Orion the HCN emission efficiency ratio peaks at a lower column density than \NtHp, whereas in W49 we find these are comparable, with \NtHp\ even potentially peaking at lower column density values. It is worth noting here that the spatial resolution of the W49 observations is $\sim$\,3\,pc, whereas the beam of the Orion A observations is $\sim$\,0.05\,pc. Therefore, it is encouraging that we are even broadly recovering the trends seen within Orion.

In Section\,\ref{sec:col_temp}, we built upon this analysis by assessing how the total intensity and emission efficiency ratio at a given column density varies as a function of dust temperature. In Figure\,\ref{contourplot_simple} we present a simplified version of Figures\,\ref{contourplot_Wq} and \ref{contourplot_hq}. Here, we show the total intensity in the upper panel, and the emission efficiency ratio in the lower panel, as a function of column density and temperature for $^{12}$CO, \CeO, HCN, and \NtHp\ on a single axis to allow comparison. To highlight the highest intensity peaks and most efficiently emitting gas we only show contours above a 50 per cent level.  

We first see from Figure\,\ref{contourplot_simple} that CO, C$^{18}$O and HCN produce the majority of their total emission at moderate temperatures ($\sim$\,20\,K), and moderate column densities ($\sim$10$^{21-22}$\cmsq). This region of the parameter space contains the most number of pixels (compare to the Figure\,\ref{fig:w49_column-temp}), and hence shows that the majority of the emission from these lines is not strongly affected by smaller scale local intensity enhancements. The majority of \NtHp emission across the mapped region, on the other hand, comes from higher column densities and is split between moderate and high temperatures. These results may then have implications for understanding molecular line emission from unresolved measurements at or above the size scale of the mapped region ($\sim$\,100pc at 11\,kpc). 


Comparing the total intensities to efficiency ratios in the lower panel of Figure\,\ref{contourplot_simple}, however, we find very different distributions. We see that \CO\ has an emissivity that peaks towards the lowest column densities, and at moderate dust temperatures. There are several well-studied mechanisms that could cause this effect. Optical depth effects are most likely the major cause of this for the observed range of column densities ($10^{20-23}$\cmsq): \CO\ for example, is thought to be optically thick in all but the lowest density gas within the Galaxy. An additional effect towards the high values of the column density would be CO depletion. In cold ($T_\mathrm{dust}$<20\,K) and dense ($n_\mathrm{H_2}>10^{4}$\,\cmcb) environments, CO molecules can ``freeze-out'' onto the surfaces of dust grains forming icy mantles (e.g. \citealp{fontani_2012a, giannetti_2014}), such that their observed gas-phase abundances are a factor of several smaller than expected in the case of no depletion (e.g. \citealp{kramer_1999, hernandez_2011}). We note, however, that we do not observe a strong dependence of the CO emission on the temperature at any given column density, and we would expect the freeze-out of CO to be strongly anti-correlated with increasing temperature, which we do not observe. This could be due to the large spatial scale of the beam used to study W49 ($\sim$\,3\,pc). 

Secondly, Figure\,\ref{contourplot_simple} shows that \CeO\ has an emissivity that peaks at $N_\mathrm{H_2}\sim10^{21-22}$\,\cmsq, which spans a temperature range of 15 -- 25\,K. As with the other CO isotopologues, above these densities, the \CeO\ emission saturates, which is likely affected by the optical depth. Interestingly, here we find that for a given column density the emission efficiency ratio appears to peak towards higher dust temperatures (e.g. see between $10^{21} - 2\,\times\,10^{21}$\,\cmsq), which could tentatively point to the previously mentioned freeze-out of CO within the cold environments.

Thirdly, when analysing the emission efficiency ratios as a function of only the column density from Figure\,\ref{fig:W_col_norm_line_single}, we concluded that both HCN and \NtHp\ peak at $N_\mathrm{H_2}\sim3\times10^{22}$\,\cmsq; an order of magnitude above the \CeO\ peak. However, when we now compare with Figure\,\ref{contourplot_simple}, we find that around this peak in column density, the distribution of HCN and \NtHp\ lines split into two discrete temperature peaks. We find that the \NtHp\ peaks towards the lowest observed temperatures at the column density within the observed parameter space ($\sim\,$20\,K), whereas the HCN peaks towards moderate-to-high temperatures of around $\sim$\,30-25\,K. This then highlights the usefulness of the dust temperature as an additional variable to further investigate the emissivity profiles within the regions even at a low spatial resolution ($\sim$\,3\,pc for these W49 observations). It is not surprising that both HCN and \NtHp\ peak towards the densest gas as the abundance of N-bearing species is boosted within dense gas (e.g. \citealp{caselli_1999}). However, it is somewhat surprising that they peak so clearly at different temperatures. Nonetheless, we propose that this could be explained by the following. The abundance of \NtHp\ is known to be particularly enhanced towards low temperatures, within regions where the CO molecules are frozen-out of the gas phase. This is due to CO being an effective destructor of the precursors in the formation pathway of \NtHp; removing molecules such as NH$^+$ and H$^{+}_{3}$. On the other hand, \citet{pety_2017} showed that the molecules such as HCN (also HCO$^+$, CN, and HNC) are sensitive to the far-UV radiation produced from star formation, where star-forming regions can be identified as having elevated dust temperatures. For example, the HCN\,($1-0$) line is thought to be pumped by mid-infrared photons and/or the effects of X-ray Dominated Regions (e.g. \citealp{alto_2007}), which can both be present within actively star-forming regions (see Figure\,\ref{fig:w49_rgb}). 

\subsection{Focusing on the ``dense'' gas within W49A}\label{sec:model}

\begin{table}
\centering
	\caption{Properties determined from the cumulative distributions. Tabulated is the molecule, and its characteristic column density, \Nchar, which is defined as the column density that contains half the total line intensity; $W_Q (N_\mathrm{H_{2}} < N_\mathrm{H_{2}}^\prime) = 50\%$. The uncertainties show the range of \Nchar\ after adding a synthetic noise of 1\,$\sigma_{W_Q}$ (see shaded region in Figure\,\ref{fig:W_col_cum_line_all}). We provide the results of the analysis across the whole mapped region (i.e. using the cumulative distributions shown in Section\,\ref{sec:fractional_emission}), and when limited to W49A (i.e. using the cumulative distributions shown in Section\,\ref{sec:model}). Also shown for W49A is the characteristic density, which we define as \nchar\,=\,n(\Nchar) (see Section\,\ref{sec:model}). This Table has been ordered by increasing critical density (see Table\,\ref{table:line_info}). The full, machine-readable version of this Table can be obtained from the supplementary online material.}
	\begin{tabular}{cccc}
\hline
Molecule & \Nchar & \Nchar &  \nchar \\
  & (whole map) & (W49A) &  (W49A)  \\ 
 & ($\times 10^{21}$cm$^{-2}$) & ($\times 10^{21}$cm$^{-2}$)  & ($\times 10^{3}$cm$^{-3}$)  \\
\hline

C$^{18}$O(1-0) & 2.7$\pm$0.1 & 6.7$\pm$0.2 & 2.0$\pm$1.7 \\
$^{13}$CO(1-0) & 2.4$\pm$0.0 & 5.3$\pm$0.2 & 1.6$\pm$1.3 \\
C$^{17}$O(1-0) & 4.2$\pm$2.2 & 9.0$\pm$4.0 & 2.7$\pm$2.2 \\
CO(1-0) & 2.1$\pm$0.1 & 4.5$\pm$0.1 & 1.3$\pm$1.1 \\
N$_{2}$H$^{+}$(1-0) & 10.1$\pm$0.6 & 14.2$\pm$0.8 & 4.2$\pm$3.5 \\
HCN(1-0) & 5.0$\pm$0.1 & 11.3$\pm$1.0 & 3.4$\pm$2.8 \\
\dots & \dots & \dots & \dots \\


\hline
	\end{tabular}
	\label{table_Nchar}
\end{table}

\begin{figure}
	\includegraphics[width=1\columnwidth]{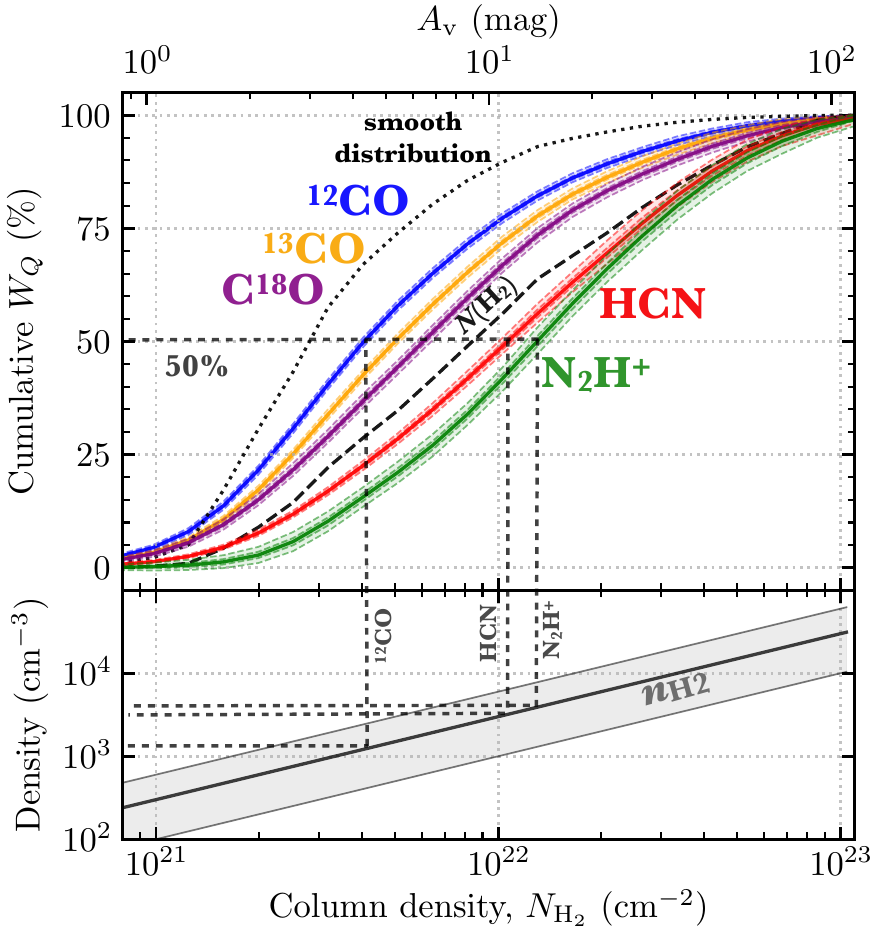}
    \caption{Cumulative fraction of the emission from each molecular line as a function of column density across the W49A region only. The lines and shaded regions within the upper panel have the same meaning as plotted for Figure\,\ref{fig:W_col_cum_line_all}. The lower panel shows the model density as a function of the column density. The grey shaded region shows one standard deviation of the distribution of model densities for a given column density, whilst the solid line shows the median of this distribution. The $W_Q (N_\mathrm{H_{2}} < N_\mathrm{H_{2}}^\prime) = 50\%$, \Nchar\, and \nchar\ values for \CO, HCN, and \NtHp are overlaid as black dashed lines.}
    \label{fig:model_map_cumulative}
\end{figure}

To better understand the line emission properties across the region we define a density model. To do so we focus on the central $\sim40$\,pc (or 0.2\degree) of W49A,\footnote{We choose to define the central position of the W49A region as RA, Dec (J2000) = 19$^\mathrm{h}$10$^\mathrm{m}$14$^\mathrm{s}$, 9\degree06\degree17\arcsec, and take only the pixels within a box of 0.2\degree around this position for the analysis presented in this Section.} where we are the most confident that the majority of the molecular line emission and column density is associated to the material at or close to the distance of the W49A star-forming region ($11.11^{+0.79}_{-0.69}$\,kpc; \citealp{gwinn_1992, zhang_2013}). As the true three-dimensional source structure of this region is not well understood, and there is almost certainly a significant level of sub-structure present on scales much smaller than the $\sim\,$3\,pc beam size used throughout this work, we choose to define the most basic possible density model. We propose that the number density using the spatial resolution of the {\it Herschel} observations can also be obtained by limiting the three-dimensional structure to a single spatial element along the line-of-sight: or in other words, by assuming that the measured column density at each position originates from a $20\arcsec\times20\arcsec\times20\arcsec$ cube (or $\sim$\,1\,pc$^{3}$ at $\sim$\,11\,kpc). We determine an uncertainty on the model density by varying the cube size by 1) the linear size based on the uncertainty from the distance estimate, and 2) varying the depth of the box between $0.5 - 3$\,pc, which represent the average size of the sub-structures identified by \citet{eden_2018}, and the size of the angular resolution of the molecular line observations ($\sim$60\arcsec), respectively.

The lower panel of Figure\,\ref{fig:model_map_cumulative} shows the distribution of the number density as a function of the observed column density. We find minimum, median and maximum values for the model density across the W49A region of 1.9$\times$10$^{2}$\cmcb, 1.1$\times$10$^{3}$\cmcb\ and 6.4$\times$10$^{4}$\cmcb, respectively. The upper panel of Figure\,\ref{fig:model_map_cumulative} shows the cumulative distributions for a selection of the molecular lines as a function of column density when limited to the W49A region. We find that this plot recovers many of the trends seen in the cumulative distribution function of the entire mapped region (c.f. Figure\,\ref{fig:W_col_cum_line_all}). 

To then condense the cumulative distributions shown in Figure\,\ref{fig:model_map_cumulative} into a single value for each line, \citet{kauffmann_2017} define a characteristic column density, \Nchar, as the column density that contains half the total line intensity; $W_Q (N_\mathrm{H_{2}} < N_\mathrm{H_{2}}^\prime) = 50\%$. Highlighted in Figure\,\ref{fig:model_map_cumulative} is the \Nchar\ for \CO, HCN and \NtHp, and the corresponding characteristic density: \nchar\,=\,n(\Nchar). The characteristic column densities and densities for all lines are summarised within Table\,\ref{table_Nchar}, which we can compare to the critical densities listed in Table\,\ref{table:line_info}. We find that the characteristic densities and critical densities for the low density tracing CO lines are comparable (\nchar$\sim$\,10$^{3}$\cmcb). However, we find that the higher density tracing molecular lines, such as HCN and \NtHp, have characteristic densities only marginally higher than the CO lines (\nchar$\sim$\,$4\times10^{3}$\cmcb), and hence significantly lower than their critical densities ($n_{\rm crit}\,\sim$\,10$^{5-6}$\cmcb). A similar analysis was carried out for the Orion A molecular cloud by \citet[][see their figure\,3]{kauffmann_2017}. Assuming a cylindrical density distribution, these authors determine a characteristic density for HCN of \nchar$\approx$\,$900^{+1240}_{-550}$\cmcb, which is comparable to within the uncertainty of the value calculated here (3400\,$\pm$2800\,\cmcb).
 
A potential explanation for the excess emission at characteristic densities below the critical density could be sub-thermal excitation (as previously mentioned for the extended map). Indeed, \citet{evans_1989} highlighted that the subthermal emission from gas below the critical density can be significant for molecular lines within the sub-mm regime. Additionally, this could be partly a beam dilution effect. As the W49 region is known to contain significantly higher densities on smaller spatial scales that are not resolved by the observations presented in this work (e.g. \citealp{ vastel_2001, eden_2018}). The line emission could then emit from these compact regions that are heavily beam diluted in the observations presented in this work. For lines such as HCN or \NtHp, which have critical densities around three orders of magnitude higher than the observed characteristic density, we estimate that observations with a spatial resolution of around a factor of $\sqrt{10^{3}}=30$ (i.e. 1\arcsec\ or 0.05pc) would be required to investigate these beam dilution effects.

\subsection{Column density and mass conversions factors}\label{sec:conversions}

There are two commonly adopted conversion factors within the literature that are used to estimate the mass of the material within a source from molecular line emission. Such conversions are useful for distant objects -- such as extragalactic sources -- where other methods of determining the gas mass of a system are not possible (e.g. dust extinction/continuum measurements). In this Section, we investigate how the typically assumed canonical conversion factors adopted within the literature hold within the W49 region.  

\subsubsection{Column density conversion; X-factor ($X_{Q}$)}

The first of these is typically referred to as the ``X-factor'', and is used to convert the integrated intensity of a line to the molecular hydrogen column density. This can be defined as, 
\begin{equation}
N_\mathrm{H_2} = X_{Q} W_{Q},
\end{equation}
where $X_{Q}$ is the X-factor of a given molecular line transition in units of \cmsq\,(\Kkms)$^{-1}$. This conversion is typically used to estimate the total molecular gas mass from CO emission, which has an assumed canonical conversion value of $X_{\mathrm{CO} (1-0)}=2\times10^{20}$\cmsq\,(\Kkms)$^{-1}$ \citep{bolatto_2013}. Variations of the $X_{\mathrm{CO} (1-0)}$ are, however, widely seen in both observations (see table\,1 of \citealp{bolatto_2013}) and simulations (e.g. \citealp{shetty_2011a, shetty_2011b, clark_2015, gong_2018}). 

We calculate the $X_{\mathrm{CO} (1-0)}$ factor using the mean molecular hydrogen column density across the mapped region, and the mean integrated intensity. Treating the whole mapped region as a single pixel, for this analysis we do not apply any uncertainty threshold clipping to the molecular line dataset (i.e. $X_\mathrm{CO} = N_\mathrm{H_2}^\mathrm{mean} / W_\mathrm{CO}^\mathrm{mean}$; see Table\,\ref{table:Xalpha_info2} for the corresponding calculation for other lines). When doing so we find a mean $X_{\mathrm{CO} (1-0)}=0.3\,\times10^{20}$\cmsq\,(\Kkms)$^{-1}$ across the mapped region (error on the mean shown as uncertainty), which is significantly below the value found within the literature (\citealp{bolatto_2013}). \citet{pety_2017} found a similarly low value of $X_{\mathrm{CO} (1-0)}$ towards the Orion B star-forming region. These authors attributed this difference to the UV illumination of the gas by massive stars, where only diffuse gas or the UV shielded dense gas have $X_{\mathrm{CO} (1-0)}$ close to the typically assumed value. Manual inspection of $X_{\mathrm{CO} (1-0)}$ on a pixel-by-pixel basis show elevated value towards the highest column density and temperature positions; towards the W49A region where embedded high-mass stars are known to reside. It is then difficult to attribute the $X_{\mathrm{CO} (1-0)}$ discrepancy observed across W49 to UV illumination.  

\subsubsection{Mass conversions; alpha-factor ($\alpha_{Q}$)}

\begin{figure}
\centering
    \includegraphics[width=1\columnwidth]{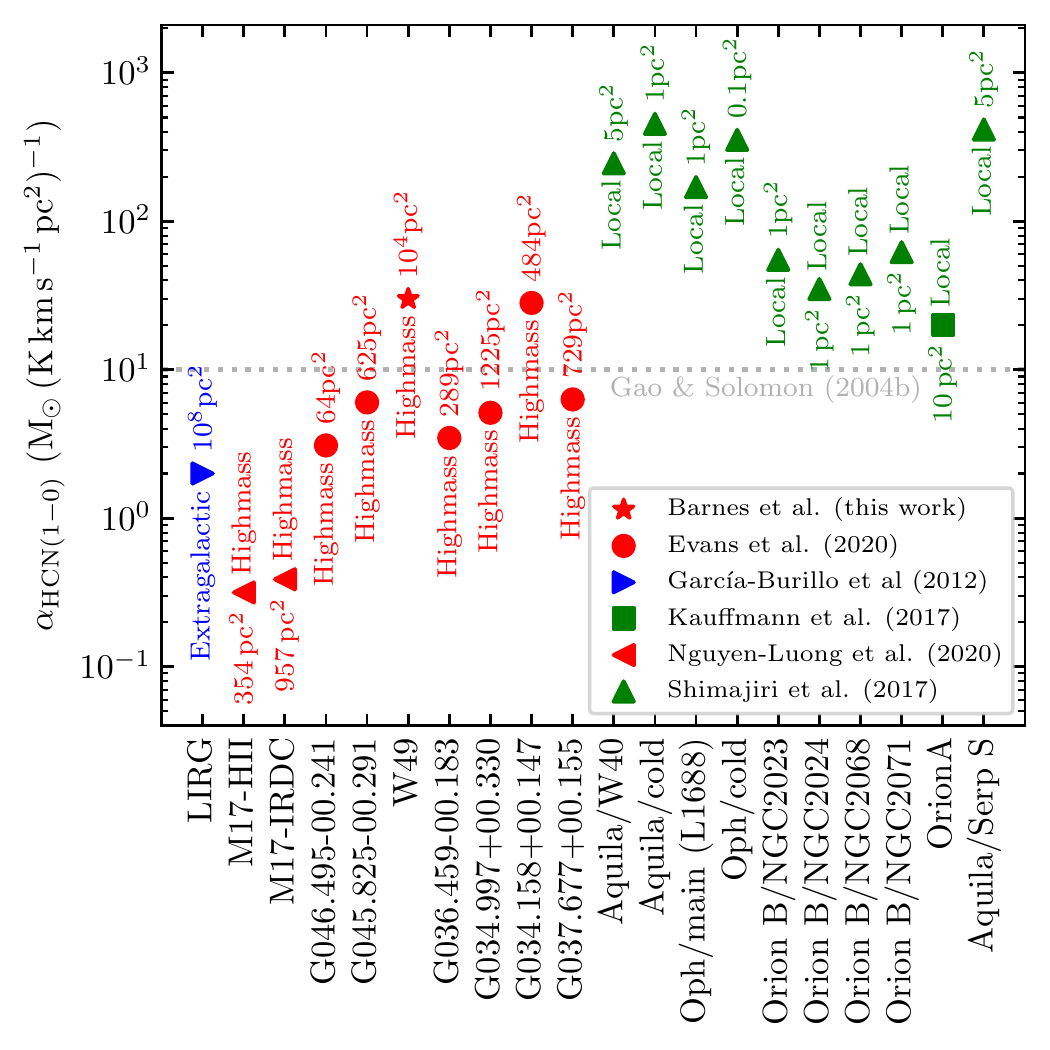} 
    \vspace{-5mm}
    \caption{A comparison of $\upalpha_{\mathrm{HCN} (1-0)}$ relevant for unresolved star forming regions measured across various sources taken from the literature \citep{garcaburillo_2012, kauffmann_2017, shimajiri_2017, evans_2020, nguyenluong_2020}. The points have been coloured by the type of source studied by each work, which have been classified as extragalactic (blue), high-mass (red), and local star-forming regions (green). The horizontal dashed line shows the typically adopted value determined of $\upalpha_{\mathrm{HCN} (1-0)} = 10$\Msol\,(\Kkms\,pc$^2$)$^{-1}$ \citep{gao_2004b}. Note that $\upalpha_{\mathrm{HCN} (1-0)}$ has been calculated for all sources using the mass of gas above an extinction threshold of $A_\mathrm{v}\gtrapprox\mathrm{8\,mag}$, divided by the total HCN luminosity across the region ($\upalpha_{Q} (\mathrm{all}) = M^\mathrm{sum}_{A_\mathrm{v}>\mathrm{8\,mag}}/L_Q^\mathrm{sum}$). For each point, we label the size of the map used to determine both $M^\mathrm{sum}_{A_\mathrm{v}>\mathrm{8\,mag}}$ and $L_Q^\mathrm{sum}$. This Figure highlights that the contribution of HCN emission from lower density gas present within large maps (i.e. towards more distance higher mass star forming regions) could be a potential cause for lower than typically assumed $\upalpha_{\mathrm{HCN} (1-0)}$ values. Additionally, the higher levels of star formation and, hence, radiation field have also been suggested to lower $\upalpha_{\mathrm{HCN} (1-0)}$ (see \citealp{shimajiri_2017}).}
    \label{fig:alphaHCN_litcomp}
\end{figure}

The second conversion factor we investigate here is between the gas mass and the luminosity of the molecular line, $\alpha_{Q}$. This can be defined as, 
\begin{equation}
M = \upalpha_{Q} L_{Q},
\end{equation}
where $\upalpha_{Q}$ is in units of \Msol\,(\Kkms\,pc$^2$)$^{-1}$. This conversion is typically used for molecular lines that are thought to trace only the dense gas, and hence to calculate the mass of dense gas within the region (e.g. using HCN; \citealp{gao_2004a}). \citet{gao_2004b} outlined that the conversion for a virialised parcel of gas with a volume-averaged molecular hydrogen number density of $n_\mathrm{H_2} \sim 3\times10^{4}$, and optically thick HCN\,$(1-0)$ emission with a brightness temperature of $T_\mathrm{MB}$\,=\,35\,K, yields the value $\upalpha_{\mathrm{HCN} (1-0)} = 10$\Msol\,(\Kkms\,pc$^2$)$^{-1}$. This calculation contains several fundamental assumptions for the HCN emission, which are not confirmed by the results presented in this work. For example, towards the centre of W49A we find peak main beam brightness temperatures for HCN of $T_\mathrm{MB}\,\sim\,$4\,K, and characteristic number densities of $\sim\,10^{3-4}$\,\cmcb\ (Section\,\ref{sec:model}), which would then give a significantly higher than standard value of $\upalpha_{\mathrm{HCN} (1-0)} \sim 15 - 50$\Msol\,(\Kkms\,pc$^2$)$^{-1}$ [see Appendix\,\ref{appendix_alphacal} for the full derivation of $\upalpha_{\mathrm{HCN} (1-0)}$]. Indeed, higher values of the $\upalpha_{\mathrm{HCN} (1-0)}$ have been calculated for larger regions containing less efficiently emitting HCN\,(1-0). \citet{kauffmann_2017}, for example, found $\upalpha_{\mathrm{HCN} (1-0)} = 20$\Msol\,(\Kkms\,pc$^2$)$^{-1}$ for the Orion A molecular cloud (also see e.g. \citealp{wu_2010}). Most recently, \citet{evans_2020} studied a sample of $\sim$\,20\,pc size molecular clumps within the Milky Way, and found $\upalpha_{\mathrm{HCN} (1-0)} = 20$\Msol\,(\Kkms\,pc$^2$)$^{-1}$ within the gas above a visual extinction of $A_\mathrm{v}>\mathrm{8\,mag}$, and $\upalpha_{\mathrm{HCN} (1-0)} = 6$\Msol\,(\Kkms\,pc$^2$)$^{-1}$ when integrating the HCN emission across the whole clumps.

We calculate $\upalpha_{Q}$ in both the total luminosity of the bulk molecular gas, and limiting the analysis to only the luminosity from the ``dense'' molecular gas following the $\upalpha_{Q}$ definitions of \citet{evans_2020}. To do so, we firstly convert the integrated intensity maps to luminosity assuming a single distance of 11.11\,kpc \citep{zhang_2013}. We then sum the luminosity across the whole mapped region without imposing an extinction threshold ($L_Q^\mathrm{sum}$). In addition, we determine the luminosity within a $A_\mathrm{v}>\mathrm{8\,mag}$ contour ($L_{Q, {A_\mathrm{v}>\mathrm{8\,mag}}}^\mathrm{sum}$), which is typically referred to as a limit for dense gas (e.g. see \citealp{kauffmann_2017}). We impose the same extinction threshold for the molecular gas ($M^\mathrm{sum}_{A_\mathrm{v}>\mathrm{8\,mag}}$). The mass conversion is then calculated as the mass of $A_\mathrm{v}>\mathrm{8\,mag}$ gas divided by the total luminosity of the given line: $\upalpha_{Q} = M^\mathrm{sum}_{A_\mathrm{v}>\mathrm{8\,mag}}/L_Q^\mathrm{sum}$. This definition is most appropriate for determining the dense gas mass from unresolved (extra-galactic) observations that measure the total HCN luminosity including emission below and $A_\mathrm{v}$ of $\mathrm{8\,mag}$. Additionally, we calculate $\upalpha_{Q}$ using both the mass and luminosity above the $A_\mathrm{v}>\mathrm{8\,mag}$ threshold: $\upalpha_{Q} ({A_\mathrm{v}>\mathrm{8\,mag}}) = M^\mathrm{sum}_{A_\mathrm{v}>\mathrm{8\,mag}} / L_{Q, {A_\mathrm{v}>\mathrm{8\,mag}}}^\mathrm{sum}$. This second method is appropriate for resolved (galactic) observations, where the HCN emission can be measured exclusively within the dense gas. 

We find $\upalpha_{\mathrm{HCN} (1-0)}\,=\,30\,\pm\,1$\,\Msol\,(\Kkms\,pc$^2$)$^{-1}$, and $\upalpha_{\mathrm{HCN} (1-0)}({A_\mathrm{v}>\mathrm{8\,mag}})\,=\,73\,\pm\,4$\,\Msol\,(\Kkms\,pc$^2$)$^{-1}$ (see Table\,\ref{table:Xalpha_info} for additional conversion factors such as $\upalpha_{\mathrm{CO} (1-0)}$).\footnote{Table\,\ref{table:Xalpha_info2} provides $\upalpha_{Q}$ values that have been determined without imposing extinction thresholds on either the mass or luminosity measurements.} These $\upalpha_{\mathrm{HCN} (1-0)}$ values are both significantly higher than the typically adopted value ($\sim\,10-20$\,\Msol\,(\Kkms\,pc$^2$)$^{-1}$), and more similar to the value determined analytically based on the observed source properties ($15-50$\,\Msol\,(\Kkms\,pc$^2$)$^{-1}$; Appendix\,\ref{appendix_alphacal}). Figure\,\ref{fig:alphaHCN_litcomp} shows a comparison of the measured $\upalpha_{\mathrm{HCN} (1-0)}$ to values available within the literature. The points have been coloured by extragalactic (blue; \citealp{garcaburillo_2012}), high-mass (red; \citealp{evans_2020, nguyenluong_2020}) and local star-forming regions (green; \citealp{kauffmann_2017, shimajiri_2017}). We see that there is a large scatter in $\upalpha_{\mathrm{HCN} (1-0)}$ ranging $\sim0.3-300$\,\Msol\,(\Kkms\,pc$^2$)$^{-1}$ across all sources. Within this scatter, we find the high mass and extragalactic sources appear to be systematically lower compared to local star-forming regions. We propose that the lower $\upalpha_{\mathrm{HCN} (1-0)}$ within higher mass regions could be due to the contribution of HCN emission from lower density gas present within subsequently large maps (i.e. towards more distance higher mass star-forming regions). Also highlighted in Figure\,\ref{fig:alphaHCN_litcomp} is the map size used to determine each $\upalpha_{\mathrm{HCN} (1-0)}$ value, and indeed we see that higher mass star-forming regions are studies across two to four orders of magnitude large areas compared to local star-forming regions. An alternative explanation, as suggested by \citet{shimajiri_2017}, is that the higher levels of star formation and, hence, the radiation field could lower $\upalpha_{\mathrm{HCN} (1-0)}$.

The analysis presented here highlights that significant caution should be exercised when adopting mass conversion factors from the literature. For example, if we take the mapped region as a typical bright 100\,pc\,$\times$\,100\,pc pixel within an extragalactic observation, using the typically adopted dense gas conversion factor would cause an underestimation of the dense gas mass by a factor of up to $\sim$\,3 with respect to the extragalactic conversion factor $\upalpha_{\mathrm{HCN} (1-0)}$ determined in this work. This would cause an underestimation of the dense gas fraction ($f_\mathrm{dense} = M_\mathrm{dense} / M_\mathrm{tot}$), and in turn, any property that contains $f_\mathrm{dense}$ (e.g. the dense gas star formation efficiency: $\mathrm{SFE}_\mathrm{dense} = f_\mathrm{dense} M_\mathrm{tot} / \mathrm{SFR}$). This would then cause significant scatter in any comparison to the dense gas star formation efficiency relations (e.g. \citealp{lada_2010}). 

\subsection{Detection limits for extragalactic studies}

In this Section, we assume that the 100\,pc$\times$100\,pc region of W49 studied here is representative of most massive star-forming regions within the Galaxy. We then pose a question: if this region were a single pixel in an extragalactic observation, would we be able to detect the molecular lines based on the brightnesses observed in this work? \citet{pety_2017} performed a similar experiment for the Orion B star-forming region, yet this had a very limited spatial coverage of 5\,pc$\times$5\,pc, which would correspond to an angular resolution at extragalactic distances that is currently difficult to obtain with even the largest mm-telescopes (e.g. ALMA; 1\,pc at 10\,Mpc is $\sim$0.02\arcsec). The observations presented here cover a much larger spatial area that corresponds to a projected angular size of $\sim$2\arcsec\ at 10\,Mpc, which is becoming much more routinely observed over large areas of extragalactic systems (e.g. \citealp{schinnerer_2013, leroy_2016}). 

We estimate that with 10\,hours of Atacama Large Millimeter Array observing time, we could reach a $3\sigma_{W_Q}\,\sim$\,0.3\,\Kkms\ line sensitivity within a $\sim$2\arcsec\ beam (100\,pc at 10\,Mpc) across a 200$\times$200\arcsec\ mosaic (or 10\,kpc at 10\,Mpc).\footnote{This time estimate has been performed using the ALMA Cycle 8 observing tool, and accounts for only the (C43-3 configuration) 12\,m array observations. For these estimates, we choose the rest frequency of the HCN($1-0$) transition, a representative bandwidth of 282\,kHz (or 0.96\kms), and Nyquist sampling of 39 12\,m array pointings that create a 200$\times$200\arcsec\ mosaic. The quoted time includes the on-source time to reach $3\sigma_{W_Q}\,\sim$\,0.3\,\Kkms\ of around 7\,hours, and calibration and overhead time of around 4.12\,hours.} We compare this sensitivity to the mean integrated intensities across the mapped region listed in Table\,\ref{table:obs_info}. We find that the CO ($W^\mathrm{mean}_Q \sim 80$\,\Kkms) and \tCO\ ($\sim$\,10\,\Kkms) molecular lines are significantly ($>100\sigma_{W_Q}$) above this value, and therefore would be easily detectable within the proposed ALMA observations. Additionally, lines such as HCN (1.4\,\Kkms), \CeO\ (1.0\,\Kkms) and HCO$^+$ (0.9\,\Kkms), are above this sensitivity estimate and should, therefore, also be moderately ($10\sigma_{W_Q}$) detectable. To significantly detect e.g. \NtHp\ (0.17\,\Kkms), however, would require an increase in sensitivity of a factor of around two, which translates to a factor of around four more integration time ($\sim$\,46\,hours).\footnote{The on-source only 12\,m array observing time required to reach $3\sigma_{W_Q}\,\sim$\,0.15\,\Kkms\ would be 27.6\,hours.} It is worth noting that these estimates have been determined using the W49 region, which is one of the brightest regions of molecular line emission within the Milky Way. Therefore, these estimate may not be representative of the detection expected across the whole disc for the proposed set of extragalactic observations.

\note{In this section, we have discussed the potential to detect weak molecular lines (e.g. \NtHp) across the complete discs of nearby spiral galaxies thought to be comparable in their star formation properties to the Milky Way. However, it is worth highlighting that these lines have already been detected towards extragalactic sources using deep single-pointing or small mosaic observations (e.g. \citealp{sage_1995}). However, these detections are typically towards galactic centres, or starburst systems that are known to have extended and elevated e.g. HCN or \NtHp\ emission relative to typical star-forming regions in disc galaxies (e.g. \citealp{meier_2005, sakamoto_2010}). For example, \citet{ginard_2015} studied the central $\sim$kpc of the starburst galaxy M82 at scales of $\sim$50-100\,pc, and found typical \NtHp\,(1-0) integrated intensities in the range of $\sim$\,5-10\,\Kkms\ (see their figure\,2). This is significantly higher than the mean \NtHp\,(1-0) integrated intensity measured across the whole mapped region of W49 (see Table\,\ref{table:obs_info}). The integrated intensity range of $\sim$\,5-10\,\Kkms\ is limited to the central $\sim10$\,pc of W49A (see Figure\,\ref{fig:momentmaps_smoothmask} and Table\,\ref{table:obs_stats}), which may be indicative of the ``mini-starburst'' nature of W49A (e.g. \citealp{peng_2010, stock_2014}).}

\subsection{Caveats}\label{sec_caveats}

At several points throughout the discussion, we mention the possible systematic uncertainties in the observations that may have affected our results. In this Section, we summarise the most potentially significant of these caveats.

{\it Limited spatial resolution}: The observations presented throughout this work have been smoothed to a spatial resolution of $\sim$\,3\,pc at the assumed source distance (11\,kpc; \citealp{zhang_2013}). Higher resolution observations have shown that much sub-structure exists within W49A on scales much smaller than this beam size, which would then not be resolved with these observations \citep{galvan-madrid_2013}. For example, using James Clerk Maxwell Telescope and higher-resolution {\it Herschel} observations, \citet{eden_2018} identified $\sim$\,50 compact sources within the central $\sim$\,60\,pc of W49A that have effective radii ranging from $10-20$\arcsec ($\sim$\,0.2 -- 0.3\,pc). These cores have masses of $>10^{3}$\,\Msol, which when assuming a spherical geometry approximately have column densities of $>10^{23}$\,\cmsq. Column density values of $10^{23}$\,\cmsq\ are only observed towards the centre of W49A within the $\sim$\,60\arcsec smoothed {\it Herschel} maps (see Figure\,\ref{fig:w49_rgb}). The dense region identified by \citet{eden_2018} would then be beam diluted within our smoothed maps, along with any molecular line emission originating from these regions. It is plausible that this would cause molecular line emission predominately originating from isolated compact, high (column) dense regions to appear to originate from low column densities when diluted. This is still, however, speculative and it is difficult to determine exactly how this would affect the results presented in this work. 

We should note, however, that it is promising that the studies of the Orion star-forming regions find similar results to this work; e.g. emission efficiency ratio as a function of column density (Section\,\ref{fig:W_col_norm_scatter_smoothmask}; \citealp{kauffmann_2017, pety_2017}). We previously mentioned that it is surprising that we even recover these trends, given that the spatial resolution of the Orion studies is $\sim$\,0.1\,pc; an order of magnitude higher than this work ($\sim$\,3\,pc). This might then highlight that limited spatial resolution is not such an issue, and it will be interesting to investigate if this holds for other distant sources within the LEGO survey. 

{\it Large spatial coverage}: One of the fundamental improvements of this work, and the LEGO survey as a whole, over previous attempts to understand line emissivities of weak molecular line transitions is the large spatial coverage at high sensitivity. In the case of W49A, we have a total mapped area of $\sim$\,100$\times$100\,pc at the source distance (11\,kpc; \citealp{zhang_2013}). This has allowed us to analyse the molecular line emission across a very broad range of column densities. In this first work, we have considered the mapped region as a whole, and have not considered the individual environments contained within the map. These environments could, for example, differ considerably in their chemistry, and therefore have very different abundances of the various molecular species. Given the large distance to W49A, it is almost certain that some emission within the mapped area will not be physically associated and have very different chemical properties. For example, the W49B supernova remnant is within the mapped region (see Figure\,\ref{fig:w49_rgb}). However, it is not currently clear if W49B is in any way linked to W49A; with some surveys placing them up to several kilo parsecs away from each other \citep{radhakrishnan_1972, chen_2014, zhu_2014}. One may assume that the physical and chemical properties of any molecular clouds interacting with this supernova could be different from those within actively star-forming region W49A. 

{\it Multiple sources along line-of-sight}: This final caveat is somewhat linked to both the previous two, and concerns the fact that we do not separate the distinct sources along each line-of-sight within our map. We proposed that this may cause scatter in our results via two plausible scenarios: a) if two (or more) dense clouds are present along with a single line-of-sight, b) if one dense cloud is present with a large amount of foreground material. The first of these scenarios is most likely the case towards the centre of the mapped region for the majority of molecular lines. W49 is known to be a very kinematically complex region \citep{galvan_madrid_2010}, and indeed manual inspection of the spectra show that multiple velocity components are present within this region. Away from the centre of W49A, the CO lines suffer the widespread overlap of velocity components at $\sim$\,10\,\kms, $\sim$\,40\,\kms\ and $\sim$\,60\,\kms, which are all extended across the mapped region. The remaining lines do not suffer from such large-scale spatial overlap; only peaking towards W49A at $\sim$\,10\,\kms (albeit with several distinct components around this velocity), and then to the east of the mapped region at $\sim$\,60\,\kms. An in-depth investigation of this velocity structure will be presented in future work. The second of the above-mentioned scenarios would cause an issue if there was a single dense cloud and a significant amount of diffuse foreground emission. This foreground material would contribute to the total column density along the line-of-sight, and to the more easily excited lines (e.g. CO). This material may not, however, emit strongly from e.g. HCN, which would cause an over (under) estimation in $\upalpha_\mathrm{HCN}$ ($h_\mathrm{HCN}$) value.

\section{Conclusions}\label{sec:conclusions}

Molecular line emission has long been used to probe the properties of molecular clouds within the Milky Way, and now with the advent of the current generation of (sub)mm-telescopes, they are also quickly becoming major tools for the study of extragalactic systems. It is typically assumed that emission from carbon monoxide (CO) and its isotopologues allows us to probe the bulk gas properties in galaxies, whilst molecular species such as HCN and HCO$^{+}$ (and \NtHp\ in future) potentially allow us to selectively characterise the densest gas. This is based on fundamental assumptions of the abundance and critical densities of these molecules across the density regimes within molecular clouds. It has, however, become clear such assumptions may not hold for all environments \note{within individual, resolved star-forming regions} (e.g. \citealp{kauffmann_2017, pety_2017, evans_2020}). For example, the so-called ``dense gas tracers'' are assumed to emit at or above their critical densities (i.e. density when the collisional de-excitation rate is equal to the rate of spontaneous emission of photons for a molecular transition), however, additional mechanisms could play a major role in the total luminosity of the molecular line measured over a large region (e.g. \citealp{evans_1999, shirley_2015, goldsmith_2018}). Before we can exploit extragalactic line emission data to its full potential, we first need to develop a detailed understanding of these --- and many other --- emission lines in Milky Way molecular clouds. With this in mind, the “Line Emission to assess Galaxy Observations” project (LEGO) aims at developing a first comprehensive picture of how the 3mm-band emission lines in Milky Way molecular clouds depend on factors like cloud density, star formation feedback, galactic environment, and metallicity.

In this paper, we present a study of emissivity of a selection of commonly observed molecular lines within the 3mm-band across a large mapped region containing the W49A massive star-forming source ($\sim$\,100\,$\times$\,100\,pc at 11\,kpc). The main results are summarised below.

\begin{enumerate}
    \item[i)] We detect bright, extended emission from the \CO\ and \tCO\ ($1-0$) lines, which have significant emission -- i.e. pixels with integrated intensities higher than three times their associated uncertainty --  across the entire mapped region (Section\,\ref{sec:fractional_emission}). The \CeO\,($1-0$) is the next most extended line, which has significant emission over half the mapped region (spatial coverage of $\sim$\,35\%). We find that the higher density tracing lines such as HCN, HCO$^{+}$\,($1-0$) cover just under a third of the mapped region (spatial coverage of $\sim$\,30\%), and, therefore, are fairly extended \note{yet much less than \CO\ and \tCO\,($1-0$)}. Lastly, the lines such as \NtHp\,($1-0$) are the most spatially compact (spatial coverage of $<$\,10\%). These results show the critical density of the observed molecular lines cannot be used as trivial predictors for how extended the emission is across a star-forming region (Figure\,\ref{fig:ncrit_comp}). Such a result is not surprising and could be easily explained by the lower abundance and, therefore, optical depth, of rarer molecules. In other words, it would not be fair to suggest that molecular line X should be more extended or brighter than molecular line Y across a given parcel of molecular gas solely because of the lower critical density of molecular line Y. Rather, the relative abundances of the molecules X and Y should be taken into account. As if the molecular Y is significantly more abundant than molecule X, molecular line Y may be brighter and more extended than molecular line X regardless of its lower critical density.
\smallskip

    \item[ii)] It is desirable to distil the emission characteristics of each molecular line into a single number. To do so, following the analysis of \citet{kauffmann_2017}, we define the characteristic column density at which half the line intensity is observed (\Nchar = $N_\mathrm{H_{2}} (W_Q = 50\%)$; Section\,\ref{sec:fractional_emission}). \note{We find that the dense gas tracers (e.g. HCN and \NtHp) have higher \Nchar\ relative to the CO lines.} We also outline a simple number density model for the central region of W49A (Section\,\ref{sec:model}), which we then use to define the characteristic density for each line (\nchar\,=\,n(\Nchar)). We find that the CO lines have comparable characteristic densities and critical densities (\nchar$\sim$\,10$^{3}$\cmcb). However, the higher density tracing molecular lines, such as HCN and \NtHp, have characteristic densities only marginally higher than the CO lines (\nchar$\sim$\,$4\times10^{3}$\cmcb), and hence significantly lower than their critical densities ($n_{\rm crit}\,\sim$\,10$^{5-6}$\cmcb). A potential explanation for the excess emission at characteristic densities below the critical density could be sub-thermal excitation, which can be significant for molecular lines within the sub-mm regime (e.g. \citealp{evans_1989}). \note{This confirms that the critical density of a line can not be used as a probe of the physical volume density in an absolute sense. That said, the use of emission from dense gas tracers, such as HCN and \NtHp, relative to CO emission may still be a useful tool to probe the proportion of relatively denser gas. Future works from the LEGO survey will aim at expanding the dynamic range in \nchar\ between CO and the dense gas tracers.}
 \smallskip
  
    \item[iii)] The ratio of the integrated intensity over the column density can be thought as a proxy for the line emissivity, or the efficiency of an emitting transition per H$_{2}$ molecule (also previously called the emission efficiency). One may assume that the emissivity is a function of number density, and, therefore, a function of column density, which would be the case if the molecular abundance or excitation is directly linked to the density \citep{kauffmann_2017}. We find, however, significant variations in the emissivity of the various molecular line transitions (Section\,\ref{subsec:intensity_comp}). The \CO\ and \tCO\ transitions show an overall decrease with increasing column density (see Figure\,\ref{fig:W_col_norm_scatter_smoothmask}). The \CeO\ transition shows a peak at $\sim10^{21}$\,\cmsq, and decreases towards higher column density values. We find that the remaining lines of HCN, HCO$^{+}$, HNC, CN, CS, \CHtOH, and \NtHp have similar peaked profiles, albeit towards higher values of the column density (few $10^{22}$\,\cmsq). These trends are broadly in agreement with the results found towards the Orion star-forming regions (e.g. \citealp{kauffmann_2017, pety_2017}), which is somewhat surprising given the observations presented here have over an order of magnitude larger spatial resolution ($\sim$\,2-3\,pc at the source distance of 11\,kpc). 
 \smallskip
    
    \item[iv)] Direct comparison to the Orion A results (\citealp{kauffmann_2017}), however, highlight several relative differences between the emission efficiency profiles between the two regions (see Figure\,\ref{fig:W_col_norm_line_single}). For example, within Orion the HCN emission efficiency peaks at a lower column density than \NtHp, whereas in W49 we find these are comparable, with \NtHp\ even potentially peaking at lower column density values. To further investigate this, we assess how the emission efficiency at a given column density varies as a function of dust temperature (Section\,\ref{sec:col_temp}). We find that around the single column density emission efficiency peak, the distribution for HCN and \NtHp\ lines split into two discrete emission efficiency temperature peaks. We find that the \NtHp\ peaks towards the lowest observed temperatures at the highest column density within the observed parameter space ($\sim\,$20\,K), whereas the HCN peaks towards moderate-to-high temperatures of around $\sim$\,30-25\,K. We believe this could be caused by the chemical formation pathways of these molecules that favour different temperature regimes, or the enhancement in the emission due to transition specific excitation mechanisms (e.g. the mid-IR pumping of HCN\,($1-0$)). 
\smallskip

    \item[v)] We also investigate several commonly used conversion factors between molecular line emission and molecular gas mass and column density across the W49 region (Section\,\ref{sec:conversions}). First, we focus on the conversion factor between the integrated intensity of CO and column density ($N_\mathrm{H_2} = X_{Q} W_{Q}$; Section\,\ref{sec:conversions}). We calculate an average value across the mapped region of
    \begin{equation*} 
    X_{\mathrm{CO} (1-0)} = 0.297\,\pm\,0.006 \times\,10^{20}\,\mathrm{cm^{-2}\,(K\,km\,s^{-1})^{-1}},
    \end{equation*}
    which is significantly lower than the canonical value within the literature [$\sim$\,2\,$\times\,10^{20}$\,cm$^{-2}$\,(K\,km\,s$^{-1}$)$^{-1}$; \citealp{bolatto_2013}]. Secondly, we investigated the conversion between the total luminosity of HCN and gas mass above a visual extinction of \Av\,>\,8\,mag within the region (i.e. $\upalpha_{Q} = M^\mathrm{sum}_{A_\mathrm{v}>\mathrm{8\,mag}}/L_Q^\mathrm{sum}$). We find 
    \begin{equation*}  
    \upalpha_{\mathrm{HCN} (1-0)}  = 30\,\pm\,1\,\mathrm{M_\odot\,(K\,km\,s^{-1}\,pc^2)^{-1}},
    \end{equation*}
    which is significantly higher than the typically adopted value within the literature ($10$\,\Msol\,(\Kkms\,pc$^2$)$^{-1}$; \citealp{gao_2004a,gao_2004b}). If we assume that the 100\,pc$\times$100\,pc region of W49 studied here is a typical massive star-forming complex within the Milky Way, or within an extragalactic system, these results have serious implications for our understanding of star-formation. For example, if we take our mapped region as a single 100\,pc$\times$100\,pc pixel in an extragalactic survey, and use the typically assumed conversion factors as opposed to the conversion factors presented in this work. We would overestimate the total molecular column density and molecular gas mass by a factor of seven, and underestimate the dense gas mass by a factor of three. This would then cause over an order of magnitude underestimation of a dense gas mass fraction ($f_\mathrm{dense} = M_\mathrm{dense} / M_\mathrm{tot}$), and in turn, cause a large systematic offset in any property that contains $f_\mathrm{dense}$ (e.g. the dense gas mass star formation efficiency: $\mathrm{SFE}_\mathrm{dense} = f_\mathrm{dense} M_\mathrm{tot} / \mathrm{SFR}$). \note{It is worth noting that here we take the $\upalpha_{\mathrm{HCN} (1-0)}$ measured over the mapped region as an example single pixel in an extragalactic observation, and, in reality, the statistics achieved with large maps of extragalactic systems {\it may} cause deviations in $\upalpha_{\mathrm{HCN} (1-0)}$ to be averaged-out over various mass and evolutionary stage star-forming regions.} Nonetheless, this example should highlight the need for further resolved measurements of mass conversion factors to precisely measure such properties as the dense gas star formation efficiency, which are needed to accurately constrain current star formation theories.
\end{enumerate}

\section*{Acknowledgements}

We would like to thank the referee for their constructive feedback that helped improve the paper. ATB and FB would like to acknowledge funding from the European Research Council (ERC) under the European Union’s Horizon 2020 research and innovation programme (grant agreement No. 726384/Empire). The LEGO survey is made possible by, and the contributions of JK are enabled via, support from the National Science Foundation under Grant Number AST--1909097. This work is based on observations carried out under project number 183--17 with the IRAM 30m telescope. IRAM is supported by INSU/CNRS (France), MPG (Germany), and IGN (Spain). This work was carried out in part at the Jet Propulsion Laboratory, which is operated for NASA by the California Institute of Technology. VW acknowledges the CNRS program "Physique et Chimie du Milieu Interstellaire" (PCMI) co-funded by the Centre National d’Etudes Spatiales (CNES). SCOG acknowledges financial support from the German Research Foundation (DFG) via the collaborative research center (SFB 881, Project-ID 138713538), ``The Milky Way System'' (subprojects B1, B2, and B8), and through Germany’s Excellence Strategy project EXC-2181/1 - 390900948 (the Heidelberg STRUCTURES Cluster of Excellence). \note{YN is supported by NAOJ ALMA Scientific Research grant No. 2017-06B and JSPS KAKENHI grant No. JP18K13577. DC acknowledges support by the \emph{Deut\-sche For\-schungs\-ge\-mein\-schaft, DFG\/} project number SFB956A. MW acknowledges funding from the European Union's Horizon 2020 research and innovation programme under the Marie Skłodowska-Curie grant agreement No 796461.}

\section*{Data availability}

The full machine-readable format of several Tables within this article are available in the online supplementary material. The data underlying this article will be shared on reasonable request to the corresponding author.


\bibliographystyle{mnras}
\bibliography{./references}
\appendix

\section{LEGO molecular line database information}\label{appendix_LEGOline}

\begin{table*}
\centering
	\caption{Details on the emission lines covered by the LEGO survey, and processed by our pipeline. Section~\ref{sec:line-selection} explains how lines examined in this study were selected. Each of the selected lines was given a LEGO reference code in the reduction pipeline, which are presented in the first column of the Table. The rest frequency recorded in this table, $\nu_\mathrm{rest}$, refers to the specific full transition \citep{lovas_2004}. These transitions are sometimes part of larger groups, as explained in Sec.~\ref{sec:line-selection} (see the final column of this Table). In those cases, $\nu_\mathrm{min}$ and $\nu_\mathrm{max}$ state the minimum and maximum frequencies of the lines considered to form a group. The full, machine-readable version of this Table can be obtained from the supplementary online material.}
	\begin{tabular}{ccccccc}
\hline
LEGO ID & Species & Full transition & \multicolumn{3}{c}{ Frequencies ($\mathrm{GHz}$) } & Notes \\
& & & $\nu_\mathrm{rest}$ & $\nu_\mathrm{min}$ & $\nu_\mathrm{max}$ & \\
\hline

H13CN\_86 & H$^{13}$CN & $J=1-0, F=2-1$ & 86.3401764 & 86.3387367 & 86.3422551 & Group of 3 transitions \\
H13CO+\_87 & H$^{13}$CO$^{+}$ & $J=1-0$ & 86.7542880 & \dots & \dots & \dots \\
SiO\_87 & SiO & $J=2-1$ & 86.8469950 & \dots & \dots & \dots \\
N13C\_87 & HN$^{13}$C & $J=1-0, F=2-1$ & 87.0908590 & 87.090735 & 87.090942 & Group of 3 transitions \\
CCH\_87a & CCH & $N=1-0, J=3/2-1/2, F=2-1$ & 87.3169250 & 87.284156 & 87.328624 & Group of 3 transitions \\
 CCH\_87b & CCH & $N=1-0, J=1/2-1/2, F=1-1$ & 87.4020040 & 87.402004 & 87.446512 & Group of 3 transitions \\
 HNCO\_88 & HNCO & $J_{K_{a},K_{c}} = 4_{0,4} - 3_{0,3}$ & 87.9252380 & 87.898416 & 87.925238 & Group of 3 transitions \\
 HCN\_89 & HCN & $J=1-0, F=2-1$ & 88.6318473 & 88.6304157 & 88.633936 & Group of 3 transitions \\
 HCO+\_89 & HCO$^{+}$ & $J=1-0$ & 89.1885260 & \dots & \dots & \dots \\
 HNC\_91 & HNC & $J=1-0, F=2-1$ & 90.6635640 & 90.66345 & 90.663656 & Group of 3 transitions \\
 H\_41a & H & 41$\alpha$\,$(m-n=42-41)$ & 92.0344340 & \dots & \dots & Recombination line \\
 N2H+\_93 & N$_{2}$H$^{+}$ & $J=1-0, F_1=2-1, F=2-1$ & 93.1737770 & 93.171621 & 93.176265 & Group of 7 transitions \\
 C34S\_96 & C$^{34}$S & $J=2-1$ & 96.4129500 & \dots & \dots & \dots \\
 CH3OH\_97-E & CH$_3$OH-E & $J_K = 2_{-1} - 1_{-1}$ & 96.7393630 & 96.739363 & 96.755507 & Group of 3 transitions \\
 CH3OH\_97-A & CH$_3$OH-A & $J_K = 2_{0} - 1_{0}$ & 96.7413770 & \dots & \dots & \dots \\
 CS\_98 & CS & $J=2-1$ & 97.9809530 & \dots & \dots & \dots \\
 SO\_99 & SO & $J_{K} = 3_{2} - 2_{1}$ & 99.2999050 & \dots & \dots & \dots \\
 HC3N\_109 & HC$_{3}$N & $J=12-11$ & 109.1736380 & \dots & \dots & \dots \\
 C18O\_110 & C$^{18}$O & $J=1-0$ & 109.7821760 & \dots & \dots & \dots \\
 HNCO\_110 & HNCO & $J_{K_{a},K_{c}} = 5_{0,5} - 4_{0,4}$ & 109.9057530 & 109.833489 & 109.905753 & Group of 5 transitions \\
 13CO\_110 & $^{13}$CO & $J=1-0$ & 110.2013540 & \dots & \dots & \dots \\
 C17O\_112 & C$^{17}$O & $J=1-0, F=7/2-5/2$ & 112.3589880 & 112.35878 & 112.360005 & Group of 3 transitions \\
 CN\_113a & CN & $N=1-0, J=1/2-1/2, F=3/2-3/2$ & 113.1913170 & 113.123337 & 113.191317 & Group of 4 transitions \\
 CN\_113b & CN & $N=1-0, J=3/2-1/2, F=5/2-3/2$ & 113.4909820 & 113.48814 & 113.520414 & Group of 5 transitions \\
 CO\_115 & CO & $J=1-0$ & 115.2712020 & \dots & \dots & \dots \\

\hline
	\end{tabular}
	\label{table:line_database}
\end{table*}

Tables~\ref{table:line_info} and \ref{table:line_database} record the information on molecular emission lines the LEGO survey uses to observe and characterise emission lines. Line frequencies are from \citet{lovas_2004}, while energy levels, Einstein coefficients and downward collisional rate coefficients are taken from the Leiden Atomic and Molecular Database (\citealp{Schoier_2005}; LAMDA, accessed in November 2019). Section~\ref{sec:line-selection} provides further information on how Tables~\ref{table:line_info} and \ref{table:line_database} were constructed.

\section{Integrated intensity map statistics}\label{appendix_IIstats}

\begin{table*}
\centering
	\caption{Integrated intensity statistics across the mapped region (i.e. that covered with both vertical and horizontal on-the-fly scans). The columns show the molecule name, the minimum values of all pixels within the map, and the minimum, 5, 16, 50, 84, and 95 percentile ranges for the intensities at positions with $W_Q$\,$>$\,$3\sigma_{W_Q}$. The full, machine-readable version of this Table can be obtained from the supplementary online material.}
	\begin{tabular}{lcccccccc}
\hline
Line & $W^\mathrm{min}_Q$ (all) & $W^\mathrm{min}_Q$ & $W^\mathrm{5\%}_Q$ &  $W^\mathrm{16\%}_Q$ & $W^\mathrm{median}_Q$ & $W^\mathrm{84\%}_Q$ & $W^\mathrm{95\%}_Q$ & $W^\mathrm{max}_Q$ \smallskip \\
 & \multicolumn{8}{c}{ (\Kkms) } \\
\hline


 CO\,(1-0) & 25.19 & 25.19 & 41.49 & 54.19 & 75.69 & 113.52 & 147.71 & 551.06 \\
 $^{13}$CO\,(1-0) & 0.49 & 0.86 & 3.57 & 6.19 & 11.89 & 19.98 & 28.71 & 155.55 \\
 C$^{18}$O\,(1-0) & -1.36 & 0.60 & 0.77 & 0.91 & 1.34 & 2.29 & 3.51 & 17.31 \\
 HCN\,(1-0) & -1.84 & 0.54 & 0.71 & 0.82 & 1.26 & 2.69 & 6.91 & 74.13 \\
 HCO$^{+}$\,(1-0) & -2.07 & 0.58 & 0.72 & 0.82 & 1.25 & 2.74 & 6.39 & 72.50 \\
 CS\,(2-1) & -1.86 & 0.51 & 0.64 & 0.73 & 1.05 & 2.09 & 5.54 & 72.03 \\
 HNC\,(1-0) & -2.04 & 0.59 & 0.70 & 0.78 & 1.07 & 1.83 & 4.10 & 32.40 \\
 CCH\,(1-0,3/2-1/2) & -1.38 & 0.47 & 0.64 & 0.72 & 0.94 & 1.72 & 4.39 & 15.69 \\
 CN\,(1-0,3/2-1/2) & -1.69 & 0.76 & 1.01 & 1.15 & 1.50 & 2.60 & 8.44 & 41.63 \\
 SO\,(3-2) & -1.61 & 0.51 & 0.63 & 0.71 & 0.94 & 1.43 & 2.88 & 35.42 \\
 CCH\,(1-0,1/2-1/2) & -1.36 & 0.53 & 0.65 & 0.72 & 0.92 & 1.42 & 3.29 & 9.04 \\
 N$_{2}$H$^{+}$\,(1-0) & -1.26 & 0.39 & 0.45 & 0.49 & 0.69 & 1.45 & 4.40 & 14.05 \\
 H41$\alpha$\,(42-41) & -1.55 & 0.54 & 0.64 & 0.70 & 0.89 & 1.27 & 3.04 & 19.45 \\
 CH$_3$OH\,(2-1) & -1.86 & 0.51 & 0.61 & 0.68 & 0.86 & 1.42 & 3.26 & 11.90 \\
 CN\,(1-0,1/2-1/2) & -2.20 & 0.87 & 1.00 & 1.10 & 1.52 & 3.93 & 10.97 & 21.69 \\
 SiO\,(2-1) & -1.99 & 0.62 & 0.70 & 0.75 & 0.93 & 1.25 & 1.54 & 14.00 \\
 C$^{34}$S\,(2-1) & -1.86 & 0.52 & 0.63 & 0.70 & 0.87 & 1.19 & 1.78 & 9.31 \\
 HC$_{3}$N\,(12-11) & -1.41 & 0.48 & 0.54 & 0.59 & 0.73 & 1.00 & 1.27 & 6.43 \\
 H$^{13}$CO$^{+}$\,(1-0) & -2.01 & 0.60 & 0.67 & 0.74 & 0.92 & 1.27 & 1.64 & 8.27 \\
 H$^{13}$CN\,(1-0) & -2.97 & 0.65 & 0.76 & 0.83 & 1.02 & 1.41 & 1.99 & 7.63 \\
 HN$^{13}$C\,(1-0) & -1.94 & 0.58 & 0.64 & 0.70 & 0.84 & 1.08 & 1.30 & 1.75 \\
 HNCO\,(4-3) & -1.57 & 0.59 & 0.65 & 0.69 & 0.81 & 1.03 & 1.18 & 1.56 \\
 C$^{17}$O\,(1-0) & -4.37 & 1.08 & 1.26 & 1.40 & 1.84 & 2.76 & 4.63 & 5.75 \\
 HNCO\,(5-4) & -1.70 & 0.63 & 0.69 & 0.72 & 0.83 & 1.01 & 1.20 & 1.57 \\

\hline
	\end{tabular}
	\label{table:obs_stats}
\end{table*}

In this Section, we present statistics determined from the molecular line integrated intensity maps (see Section\,\ref{sec:IImap}). Shown in Table\,\ref{table:obs_stats} is the minimum value of all pixels across each map (i.e. without taking a significance threshold). We see that CO and $^{13}$CO have positive minimum values, which shows that they are detected across all positions within the map (also see Figure\,\ref{table:obs_info}). The other molecular lines, however, are negative, showing that they are detected above a threshold intensity. This is given as the minimum value above $W_Q$\,$>$\,$3\sigma_{W_Q}$ also given in Table\,\ref{table:obs_stats}. In addition, we also give the minimum, 5, 16, 50, 84, and 95 percentile ranges when taking only the significant positions. Note that Table\,\ref{table:obs_stats} has been ordered by decreasing $A_\mathrm{cov}$ values (see Table\,\ref{table:obs_info}). We can see that there is little correlation to the line brightness and its coverage across the region (e.g. see $W^\mathrm{max}_Q$). Some line are, therefore, see as extended as relative faint (e.g. CCH\,(1-0,1/2-1/2)), whilst other are seen as compact and bright (e.g. SiO\,(2-1); see Figure\,\ref{fig:W_col_scatter_smoothmask}).

\section{Random uncertainties on the H$_2$ column density and dust temperature maps}\label{appendix_colerr}

We follow the weighted least squared fitting procedure outlined in \citet{guzman_2015} to calculate the molecular hydrogen column densities and dust temperatures from the {\it Herschel} Hi-Gal observations (see Section\,\ref{sec:observations}). This involves fitting the {\it Herschel} filter fluxes to the modified blackbody model, and adjusting the free parameters (surface density and dust temperature) in order to minimise the weighted least squared difference:
\begin{equation}
 \chi^2 = \sum_i \frac{(\mathrm{data}_i-\mathrm{model}_i)}{\sigma_i^2},
\end{equation}
where $i$ runs through each filter, $\sigma_i^2=(0.1*\mathrm{data}_i)^2+\sigma_{i,\mathrm{N}}^2$ and $\sigma_{i,\mathrm{N}}$ is given by the fifth column of table\,1 in \citet{guzman_2015}. 

\begin{figure}
\centering
\includegraphics[width=0.9\columnwidth]{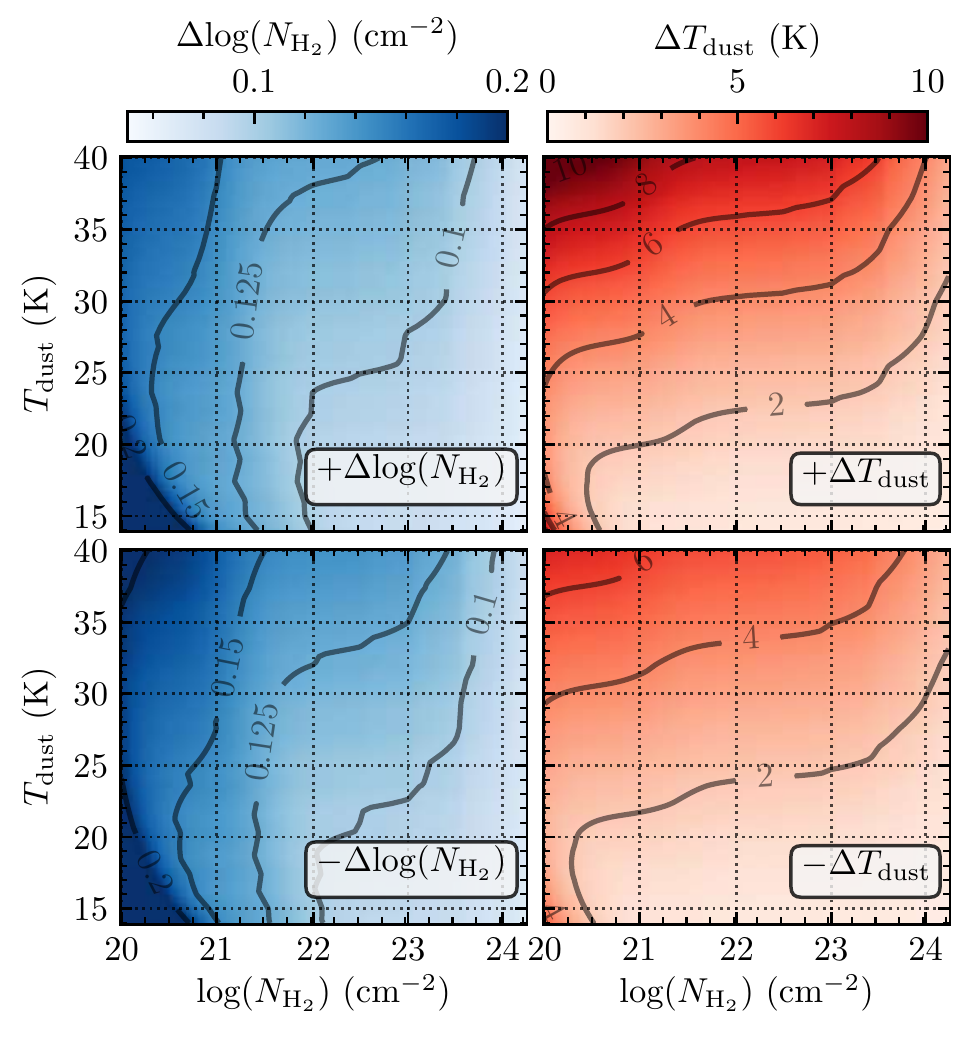}
    \caption{The random uncertainty associated with the H$_2$ column densities and dust temperatures. Shown from left to right in colour scale is the positive and negative uncertainties on the (log) H$_2$ column density, and dust temperature as a function of the H$_2$ column density and dust temperature. Overlaid and labelled on each panel are contours of the background colour scale for reference.}
    \label{realnoise_pspace1}
\end{figure}

To get the random uncertainty associated with the fit at each position, we follow the method outlined by \citet{lampon_1976}. We test how ``sensitive'' is the minimisation function to variations in the input parameters. In practice, this means finding the $\chi^2 = \chi^2_\mathrm{min}+2.3$ contour (2.3 is adequate to get 1-$\sigma$ confidence intervals with two free parameters). This contour (usually close to an ellipse) defines an uncertainty region, whose projections onto the parameter axes define the 1-$\sigma$ confidence intervals (see figure\,3 in \citealp{guzman_2015}). Figure\,\ref{realnoise_pspace1} shows the size of these confidence intervals towards the positive and negative side of the best fitting parameter. The parameters shown here are the dust temperature, and the logarithm of the H$_2$ column density.

\begin{figure}
\centering
\includegraphics[width=1\columnwidth]{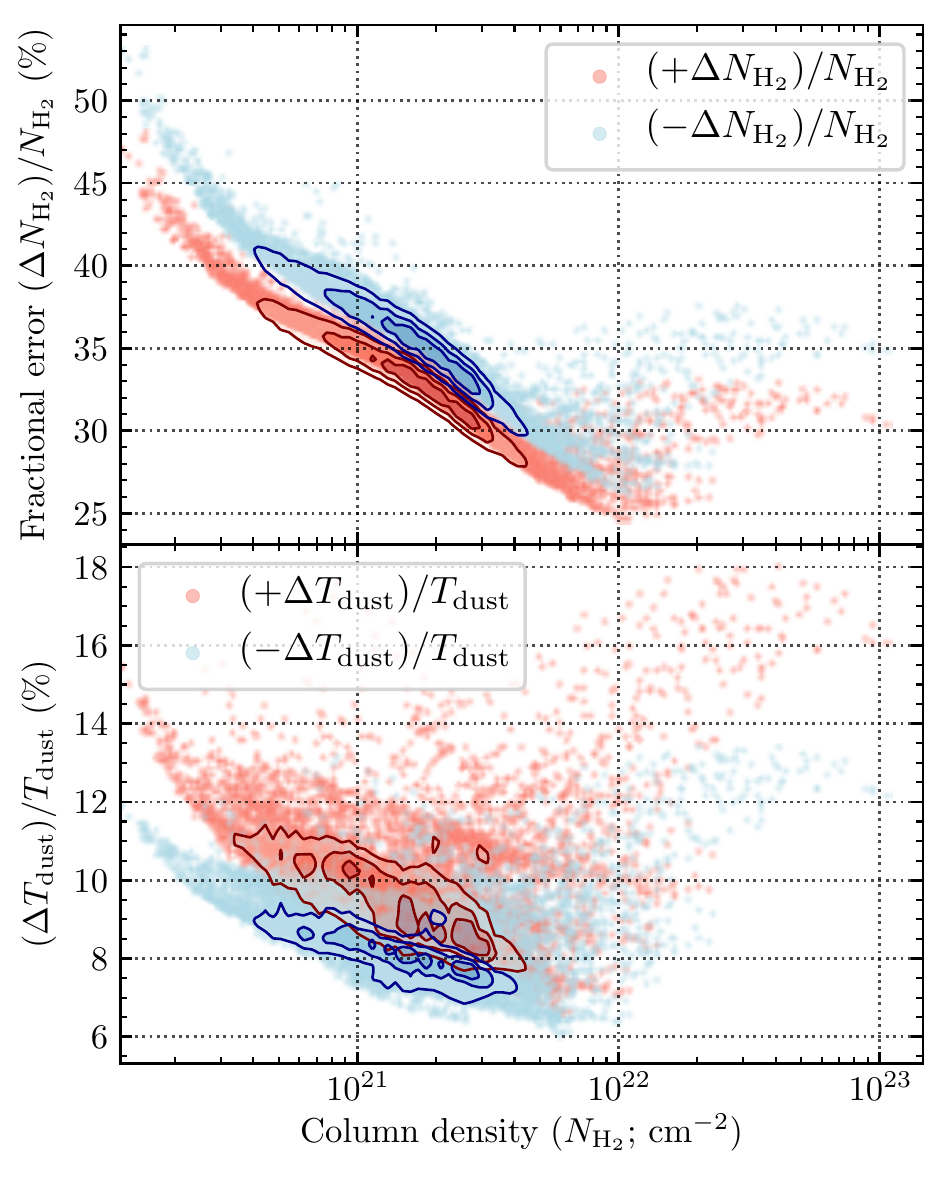}
    \caption{The fractional random uncertainty associated with the H$_2$ column density (upper panel) and dust temperature (lower panel) measurements across the mapped region. The fractional uncertainty is defined at each position as the ratio of the uncertainty over the measured value, and plotted as a function of the H$_2$ column density. The blue and red points show positive and negative uncertainties, respectively. The overlaid contours show the density of plotted points in increasing levels of 25, 50 and 75\,per cent.}
    \label{realnoise_pspace2}
\end{figure}

We investigate the fractional uncertainty across the mapped region by taking the ratios of the positive and negative uncertainty at each position over the corresponding measured value at the same position. The distribution of the H$_2$ column density and dust temperature fractional uncertainties are plotted in Figure\,\ref{realnoise_pspace2}. We find that the mean fractional uncertainties produced by random noise on the column density and dust temperature maps are $\frac{\Delta N_\mathrm{H_2}}{N_\mathrm{H_2}} = _{-35\%}^{+33\%}$ and $\frac{\Delta T_\mathrm{dust}}{T_\mathrm{dust}} = _{-8\%}^{+9\%}$. 

\section{Line of sight components in spectral cubes}\label{appendix_los}

\begin{figure*}
\centering
\includegraphics[width=1\textwidth]{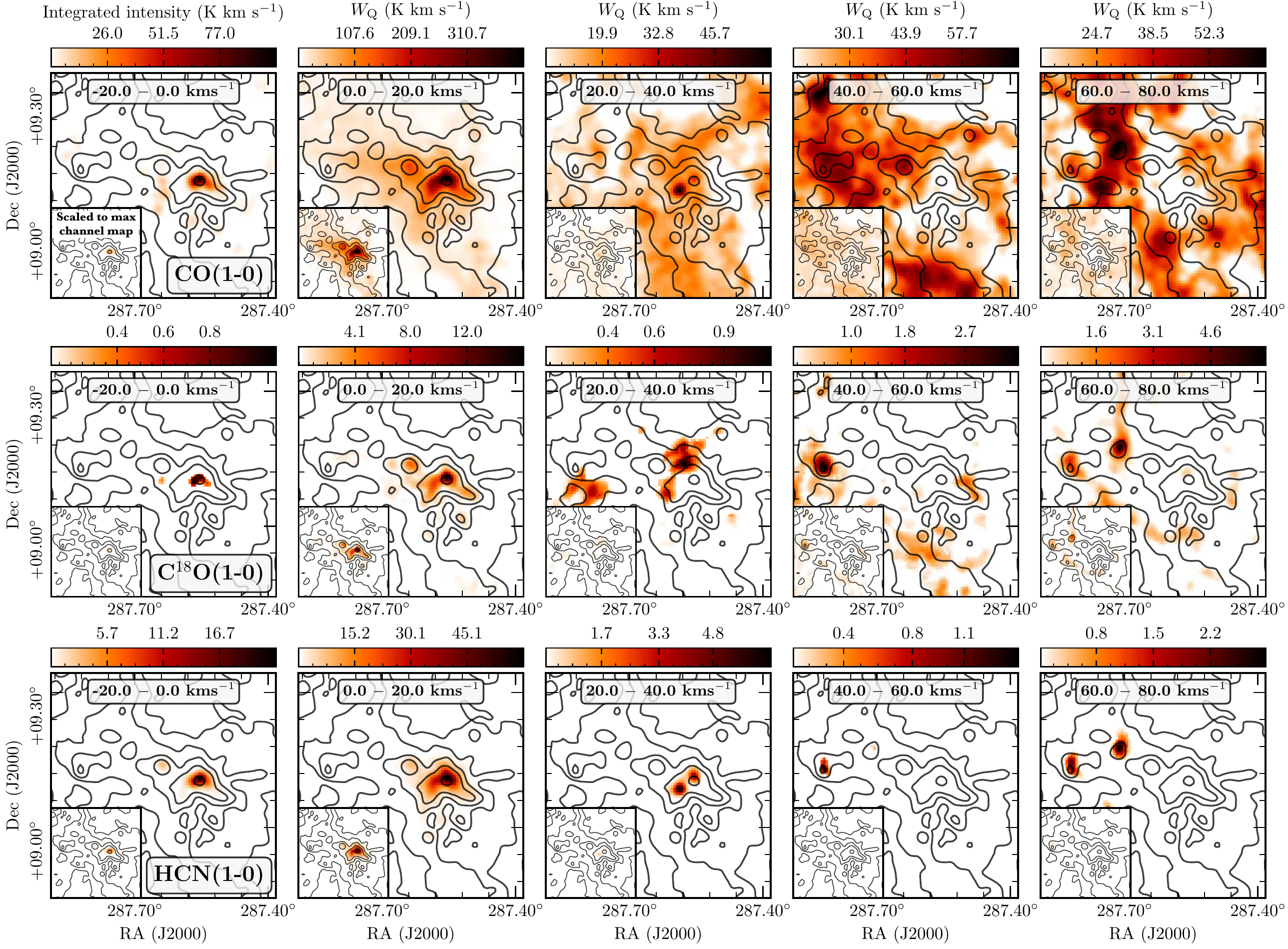}
    \caption{Channel maps for CO\,($1-0$), \CeO\,($1-0$), and HCN\,($1-0$) across the mapped region of W49. The panels from left to right show the intensity integrated in velocity windows of -20\,\kms\ to 80\,\kms\ in steps of 20\,\kms\ (see Section\,\ref{sec:IImap}). The maps shown in the main plot of each panel have a colour scale scaled to the values within that velocity window, whilst the maps shown in the lower left of each panel have the identical colour scale which has been scaled to the window with the highest integrated intensity. The black contours are of the {\it Herschel} derived molecular hydrogen column density, in levels of 1, 2.5, 5, 10, 50, 100\,$\times 10^{22}$~\cmsq.}
    \label{channel_maps}
\end{figure*}

In this Section, we present a brief inspection of the velocity structure observed within the molecular line data cubes. To do so, we integrate the line intensity within discrete velocity windows across the velocity range where the significant emission is seen in the average spectra (Figure\,\ref{fig:w49_spec}). Figure\,\ref{channel_maps} displays the result when using velocity windows of -20\,\kms\ to 80\,\kms\ in steps of 20\,\kms\ for the CO\,($1-0$), \CeO\,($1-0$), and HCN\,($1-0$) maps. We note that to produce these maps we use cubes that have all spectral channels below 5\,$\sigma_\mathrm{rms}$ masked (see Table\,\ref{table:line_info}), rather than using the masked cubes produced by the CO masking routine (see Section\,\ref{sec:IImap}). These plots highlight the three broad velocity components that are present across the mapped region, which peak at velocities of $\sim$\,10\kms, $\sim$\,40\kms, $\sim$\,60\kms. The brightest of these is the $\sim$\,10\kms component, which is seen in all the displayed lines and appears to be spatially correspondent with the W49A star-forming region (see Figure\,\ref{fig:w49_rgb}).  

\section{Understanding the mass conversion factors}\label{appendix_alphacal}

In this Section, we review the calculation of the mass conversion factor discussed in Section\,\ref{sec:conversions}; namely $\upalpha_{Q} = L_{Q} / M$. The determination of a constant value for this conversion is fundamentally based on two assumptions: 1) the region is in virial equilibrium, i.e. not undergoing collapse or expansion, 2) the observed emission is optically thick.

A parcel of gas is defined as being in virial equilibrium when twice the internal kinetic energy equals the potential energy, which following \citet{solomon_1987} can be expressed as, 
\begin{equation}
M_\mathrm{vir} \propto R \sigma^2,
\label{virial}
\end{equation}
where $M_\mathrm{vir}$ is the virial mass, $R$ the radius, and $\sigma$ the observed velocity dispersion. The virial mass can also be given as a product of its density, $\rho$, and volume,
\begin{equation}
M_\mathrm{vir} \propto \rho R^3. 
\label{virial_mod1}
\end{equation}
Empirically, molecular clouds are observed to follow a size–line width relation \citep{larson_1981},
\begin{equation}
\sigma \propto R^{0.5}
\label{linewidthsize}
\end{equation}
Equating equation\,\ref{virial} and equation\,\ref{virial_mod1}, and substituting equation\,\ref{linewidthsize} gives ($R \propto \sigma^{2}$),
\begin{equation}
\sigma \propto \rho ^{-0.5}. 
\label{sigma_rho}
\end{equation}
Equating equations\,\ref{linewidthsize} and \ref{sigma_rho} gives,
\begin{equation}
R \propto \rho ^{-1}, 
\label{Rrho}
\end{equation}
and substituting this into equation\,\ref{virial_mod1}, 
\begin{equation}
M_\mathrm{vir} \propto \rho^{-2}.
\label{Mvir_rho}
\end{equation}
The luminosity of a molecular line transition, $L$, can be approximated as the product of the integrated intensity ($W=T_\mathrm{MB} \sqrt{2} \sigma$), and the area over which the intensity has been measured ($A = \pi R^{2}$),  
\begin{equation}
L = A W \propto T_\mathrm{MB} \sigma R^{2}
\end{equation}
Substituting equations\,\ref{sigma_rho} and \ref{Rrho} in the above gives,  
\begin{equation}
L \propto T_\mathrm{MB} \rho^{-2.5}.
\label{luminosity_mod1}
\end{equation}
Finally, taking the ratio of the virial mass in equation\,\ref{Mvir_rho}, and the luminosity in equation\,\ref{luminosity_mod1} yields, 
\begin{equation}
\alpha = M L^{-1} \propto \rho^{-0.5}\,(T_\mathrm{MB}\,\rho^{-2.5})^{-1} \propto \frac{\rho^{0.5}} {T_\mathrm{MB}},
\label{luminosity_mod2}
\end{equation}
which in terms of molecular hydrogen number density, $n_\mathrm{H_2}$, can be approximated as,
\begin{equation}
\upalpha \approx 2.1\,\frac{n_\mathrm{H_2}^{0.5}}{T_\mathrm{MB}}.
\label{luminosity_mod3}
\end{equation}
\citet{gao_2004a, gao_2004b} approximate that a virialised core with a volume averaged molecular hydrogen number density of $n_\mathrm{H_2} \sim 3\times10^{4}$\cmcb, and optically think HCN\,$(1-0)$ emission with a brightness temperature of $T_\mathrm{MB}$=35\,K, gives $\upalpha_{\mathrm{HCN} (1-0)} = 10$\Msol\,(\Kkms\,pc$^2$)$^{-1}$. Alternatively, in this work we measure a peak main beam brightness temperatures for HCN of $T_\mathrm{MB}\,\sim\,$4\,K, and characteristic number densities of $\sim\,10^{3-4}$\,\cmcb\ (Section\,\ref{sec:model}), which would then give a significantly higher than standard values of $\upalpha_{\mathrm{HCN} (1-0)} \sim 15 -- 50$\Msol\,(\Kkms\,pc$^2$)$^{-1}$ (see Appendix\,\ref{appendix_alphacal} for the full derivation of $\upalpha_{\mathrm{HCN} (1-0)}$).

The column density conversion $X_{Q} = W_{Q} / N_\mathrm{H_2}$ follows a very similar calculation, and can be found in full in e.g. \citet{dickman_1986}. 

\clearpage

\begin{table*}
\centering
	\caption{Calculated conversion factors for the various molecular line transitions (see Section\,\ref{sec:conversions}). We highlight the two most commonly adopted conversion factors: $X_{\mathrm{CO} (1-0)}$ and $\upalpha_{\mathrm{HCN} (1-0)}$. The values shown in the upper half of the Table have been calculated using the column density and mass of $A_\mathrm{v}>\mathrm{8\,mag}$ gas divided by the mean integrated intensity and total luminosity of the given line across the region (e.g. $\upalpha_{Q} = M^\mathrm{sum}_{A_\mathrm{v}>\mathrm{8\,mag}}/L_Q^\mathrm{sum}$), whilst those within the lower half have been calculated using both the column density and mass, and integrated intensity and luminosity above the $A_\mathrm{v}>\mathrm{8\,mag}$ threshold ($\upalpha_{Q} ({A_\mathrm{v}>\mathrm{8\,mag}}) = M^\mathrm{sum}_{A_\mathrm{v}>\mathrm{8\,mag}} / L_{Q, {A_\mathrm{v}>\mathrm{8\,mag}}}^\mathrm{sum}$). The full, machine-readable version of this Table can be obtained from the supplementary online material.}
	\begin{tabular}{ccccc}
\hline
Line & log($W_Q^\mathrm{mean}$) & log($L_Q^\mathrm{sum}$) & $X_Q$ &  $\upalpha_Q$ \\
 & \Kkms & \Kkms\,pc$^2$ & \cmsq\,(\Kkms)$^{-1}$ & \Msol\,(\Kkms\,pc$^2$)$^{-1}$ \\
 &  & & ($\times\,10^{20}$) & \\
\hline
\multicolumn{5}{c}{$W^\mathrm{mean}_{Q}$; $L^\mathrm{sum}_{Q}$; $N_\mathrm{H_2, A_\mathrm{v}>\mathrm{8\,mag}}^\mathrm{mean}=1.7\times10^{21}$\,cm$^{-2}$;  $M^\mathrm{sum}_{A_\mathrm{v}>\mathrm{8\,mag}}=4.0\times10^{5}$\,\Msol} \smallskip \\


 CO(1-0) & 1.928 $\pm$ 0.003 & 5.911 $\pm$ 0.002 & 2.02 $\pm$ 0.094 & 0.491 $\pm$ 0.005 \\
 HCN(1-0) & 0.15 $\pm$ 0.02 & 4.13 $\pm$ 0.014 & 124 $\pm$ 6.9 & 30.1 $\pm$ 0.98 \smallskip \\
 $^{13}$CO(1-0) & 1.139 $\pm$ 0.004 & 5.122 $\pm$ 0.004 & 12.4 $\pm$ 0.58 & 3.02 $\pm$ 0.035 \\
 C$^{18}$O(1-0) & 0.022 $\pm$ 0.006 & 4.006 $\pm$ 0.005 & 162 $\pm$ 7.7 & 39.5 $\pm$ 0.57 \\
 C$^{17}$O(1-0) & -0.97 $\pm$ 0.05 & 3.02 $\pm$ 0.049 & 1600 $\pm$ 200 & 390 $\pm$ 44 \\
 HCO$^{+}$(1-0) & -0.02 $\pm$ 0.02 & 3.96 $\pm$ 0.015 & 190 $\pm$ 11 & 45 $\pm$ 1.6 \\
 HNC(1-0) & -0.21 $\pm$ 0.02 & 3.77 $\pm$ 0.013 & 290 $\pm$ 16 & 69 $\pm$ 2.2 \\
 CN(1-0,1/2-1/2) & -0.67 $\pm$ 0.04 & 3.31 $\pm$ 0.031 & 820 $\pm$ 69 & 200 $\pm$ 15 \\
 CN(1-0,3/2-1/2) & -0.03 $\pm$ 0.02 & 3.96 $\pm$ 0.013 & 190 $\pm$ 10 & 45 $\pm$ 1.4 \\
 CS(2-1) & -0.03 $\pm$ 0.02 & 3.96 $\pm$ 0.016 & 190 $\pm$ 11 & 45 $\pm$ 1.7 \\
 N$_{2}$H$^{+}$(1-0) & -0.76 $\pm$ 0.03 & 3.22 $\pm$ 0.023 & 990 $\pm$ 69 & 250 $\pm$ 13 \\
 SO(3-2) & -0.44 $\pm$ 0.02 & 3.55 $\pm$ 0.016 & 470 $\pm$ 28 & 115 $\pm$ 4.5 \\
 SiO(2-1) & -1.23 $\pm$ 0.06 & 2.75 $\pm$ 0.056 & 3000 $\pm$ 410 & 720 $\pm$ 94 \\
 CCH(1-0,3/2-1/2) & -0.19 $\pm$ 0.01 & 3.784 $\pm$ 0.009 & 270 $\pm$ 14 & 66 $\pm$ 1.6 \\
 CCH(1-0,1/2-1/2) & -0.41 $\pm$ 0.01 & 3.58 $\pm$ 0.01 & 440 $\pm$ 23 & 107 $\pm$ 2.6 \\
 C$^{34}$S(2-1) & -1.05 $\pm$ 0.04 & 2.93 $\pm$ 0.033 & 2000 $\pm$ 180 & 480 $\pm$ 37 \\
 CH$_3$OH(2-1) & -1.24 $\pm$ 0.06 & 2.74 $\pm$ 0.058 & 3100 $\pm$ 430 & 740 $\pm$ 99 \\
 HC$_{3}$N(12-11) & -1.8 $\pm$ 0.2 & 2.2 $\pm$ 0.15 & 12000 $\pm$ 3900 & 2800 $\pm$ 940 \\
 HN$^{13}$C(1-0) & -1.3 $\pm$ 0.05 & 2.68 $\pm$ 0.047 & 3500 $\pm$ 410 & 840 $\pm$ 92 \\

\hline

\multicolumn{5}{c}{$W^\mathrm{mean}_{Q,A_\mathrm{v}>\mathrm{8\,mag}}$; $L^\mathrm{sum}_{Q,A_\mathrm{v}>\mathrm{8\,mag}}$; $N_\mathrm{H_2, A_\mathrm{v}>\mathrm{8\,mag}}^\mathrm{mean}=1.7\times10^{22}$\,cm$^{-2}$;  $M^\mathrm{sum}_\mathrm{A_\mathrm{v}>\mathrm{8\,mag}}=4.0\times10^{5}$\,\Msol} \smallskip \\


 CO(1-0) & 2.32 $\pm$ 0.01 & 4.9 $\pm$ 0.01 & 0.82 $\pm$ 0.042 & 5.2 $\pm$ 0.13 \\
 HCN(1-0) & 1.18 $\pm$ 0.03 & 3.75 $\pm$ 0.023 & 11.5 $\pm$ 0.8 & 73 $\pm$ 3.9 \smallskip \\
 $^{13}$CO(1-0) & 1.68 $\pm$ 0.02 & 4.25 $\pm$ 0.011 & 3.6 $\pm$ 0.19 & 22.6 $\pm$ 0.62 \\
 C$^{18}$O(1-0) & 0.65 $\pm$ 0.02 & 3.22 $\pm$ 0.013 & 39 $\pm$ 2.1 & 242 $\pm$ 7.4 \\
 C$^{17}$O(1-0) & 0 $\pm$ 0.03 & 2.57 $\pm$ 0.024 & 180 $\pm$ 13 & 1090 $\pm$ 61 \\
 HCO$^{+}$(1-0) & 1.02 $\pm$ 0.03 & 3.59 $\pm$ 0.026 & 17 $\pm$ 1.3 & 104 $\pm$ 6.3 \\
 HNC(1-0) & 0.78 $\pm$ 0.03 & 3.35 $\pm$ 0.023 & 29 $\pm$ 2.1 & 181 $\pm$ 9.7 \\
 CN(1-0,1/2-1/2) & 0.64 $\pm$ 0.03 & 3.22 $\pm$ 0.025 & 39 $\pm$ 2.9 & 250 $\pm$ 15 \\
 CN(1-0,3/2-1/2) & 0.97 $\pm$ 0.03 & 3.54 $\pm$ 0.023 & 19 $\pm$ 1.3 & 116 $\pm$ 6.2 \\
 CS(2-1) & 1.04 $\pm$ 0.03 & 3.61 $\pm$ 0.026 & 16 $\pm$ 1.3 & 100 $\pm$ 6.2 \\
 N$_{2}$H$^{+}$(1-0) & 0.46 $\pm$ 0.03 & 3.03 $\pm$ 0.022 & 61 $\pm$ 4.1 & 380 $\pm$ 20 \\
 SO(3-2) & 0.56 $\pm$ 0.04 & 3.14 $\pm$ 0.032 & 47 $\pm$ 4.2 & 300 $\pm$ 23 \\
 SiO(2-1) & -0.12 $\pm$ 0.07 & 2.45 $\pm$ 0.062 & 230 $\pm$ 35 & 1500 $\pm$ 210 \\
 CCH(1-0,3/2-1/2) & 0.71 $\pm$ 0.02 & 3.28 $\pm$ 0.017 & 34 $\pm$ 2.1 & 211 $\pm$ 8.6 \\
 CCH(1-0,1/2-1/2) & 0.45 $\pm$ 0.02 & 3.03 $\pm$ 0.019 & 61 $\pm$ 3.9 & 390 $\pm$ 18 \\
 C$^{34}$S(2-1) & 0.01 $\pm$ 0.04 & 2.58 $\pm$ 0.039 & 170 $\pm$ 17 & 1060 $\pm$ 96 \\
 CH$_3$OH(2-1) & 0.23 $\pm$ 0.03 & 2.81 $\pm$ 0.028 & 101 $\pm$ 7.9 & 640 $\pm$ 41 \\
 H$^{13}$CN(1-0) & -0.16 $\pm$ 0.05 & 2.41 $\pm$ 0.048 & 250 $\pm$ 30 & 1600 $\pm$ 180 \\
 H$^{13}$CO$^{+}$(1-0) & -0.24 $\pm$ 0.06 & 2.34 $\pm$ 0.056 & 300 $\pm$ 41 & 1900 $\pm$ 250 \\
 HC$_{3}$N(12-11) & -0.45 $\pm$ 0.08 & 2.12 $\pm$ 0.073 & 500 $\pm$ 85 & 3100 $\pm$ 520 \\
 HNCO(5-4) & -0.61 $\pm$ 0.05 & 1.97 $\pm$ 0.042 & 700 $\pm$ 75 & 4400 $\pm$ 430 \\
 HNCO(4-3) & -0.42 $\pm$ 0.04 & 2.15 $\pm$ 0.032 & 460 $\pm$ 41 & 2900 $\pm$ 220 \\
 HN$^{13}$C(1-0) & -1.5 $\pm$ 0.5 & 1 $\pm$ 0.48 & 7000 $\pm$ 7200 & 50000 $\pm$ 45000 \\

\hline
	\end{tabular}
	\label{table:Xalpha_info}
\end{table*}

\begin{table*}
\centering
	\caption{Calculated conversion factors for the various molecular line transitions (identical to Table\,\ref{table:Xalpha_info}). The values shown have been calculated without imposing any extinction threshold on the gas column density and gas mass, and the integrated intensity and luminosity of the given line (such that e.g. $\upalpha_{Q} = M^\mathrm{sum}/L_Q^\mathrm{sum}$). The full, machine-readable version of this Table can be obtained from the supplementary online material.}
	\begin{tabular}{ccccc}
\hline
Line & log($W_Q^\mathrm{mean}$) & log($L_Q^\mathrm{sum}$) & $X_Q$ &  $\upalpha_Q$ \\
 & \Kkms & \Kkms\,pc$^2$ & \cmsq\,(\Kkms)$^{-1}$ & \Msol\,(\Kkms\,pc$^2$)$^{-1}$ \\
 &  & & ($\times\,10^{20}$) & \\
\hline
\multicolumn{5}{c}{$W^\mathrm{mean}_Q$; $L^\mathrm{sum}_Q$; $N_\mathrm{H_2}^\mathrm{mean}=2.5\times10^{21}$\,cm$^{-2}$;  $M^\mathrm{sum}=5.4\times10^{5}$\,\Msol} \smallskip \\


 CO(1-0) & 1.928 $\pm$ 0.003 & 5.911 $\pm$ 0.002 & 0.297 $\pm$ 0.0057 & 0.67 $\pm$ 0.013 \\
 HCN(1-0) & 0.15 $\pm$ 0.02 & 4.13 $\pm$ 0.014 & 18.2 $\pm$ 0.67 & 41 $\pm$ 1.5 \smallskip \\
 $^{13}$CO(1-0) & 1.139 $\pm$ 0.004 & 5.122 $\pm$ 0.004 & 1.83 $\pm$ 0.037 & 4.1 $\pm$ 0.082 \\
 C$^{18}$O(1-0) & 0.022 $\pm$ 0.006 & 4.006 $\pm$ 0.005 & 23.9 $\pm$ 0.53 & 54 $\pm$ 1.2 \\
 C$^{17}$O(1-0) & -0.97 $\pm$ 0.05 & 3.02 $\pm$ 0.049 & 240 $\pm$ 27 & 530 $\pm$ 60 \\
 HCO$^{+}$(1-0) & -0.02 $\pm$ 0.02 & 3.96 $\pm$ 0.015 & 27 $\pm$ 1.1 & 60 $\pm$ 2.4 \\
 HNC(1-0) & -0.21 $\pm$ 0.02 & 3.77 $\pm$ 0.013 & 42 $\pm$ 1.5 & 93 $\pm$ 3.3 \\
 CN(1-0,1/2-1/2) & -0.67 $\pm$ 0.04 & 3.31 $\pm$ 0.031 & 121 $\pm$ 8.8 & 270 $\pm$ 20 \\
 CN(1-0,3/2-1/2) & -0.03 $\pm$ 0.02 & 3.96 $\pm$ 0.013 & 27 $\pm$ 0.94 & 61 $\pm$ 2.2 \\
 CS(2-1) & -0.03 $\pm$ 0.02 & 3.96 $\pm$ 0.016 & 27 $\pm$ 1.1 & 61 $\pm$ 2.5 \\
 N$_{2}$H$^{+}$(1-0) & -0.76 $\pm$ 0.03 & 3.22 $\pm$ 0.023 & 146 $\pm$ 8.1 & 330 $\pm$ 19 \\
 SO(3-2) & -0.44 $\pm$ 0.02 & 3.55 $\pm$ 0.016 & 70 $\pm$ 3 & 156 $\pm$ 6.6 \\
 SiO(2-1) & -1.23 $\pm$ 0.06 & 2.75 $\pm$ 0.056 & 440 $\pm$ 57 & 1000 $\pm$ 130 \\
 CCH(1-0,3/2-1/2) & -0.19 $\pm$ 0.01 & 3.784 $\pm$ 0.009 & 40 $\pm$ 1.2 & 90 $\pm$ 2.6 \\
 CCH(1-0,1/2-1/2) & -0.41 $\pm$ 0.01 & 3.58 $\pm$ 0.01 & 65 $\pm$ 1.9 & 145 $\pm$ 4.3 \\
 C$^{34}$S(2-1) & -1.05 $\pm$ 0.04 & 2.93 $\pm$ 0.033 & 290 $\pm$ 23 & 650 $\pm$ 51 \\
 CH$_3$OH(2-1) & -1.24 $\pm$ 0.06 & 2.74 $\pm$ 0.058 & 450 $\pm$ 61 & 1000 $\pm$ 140 \\
 HC$_{3}$N(12-11) & -1.8 $\pm$ 0.2 & 2.2 $\pm$ 0.15 & 1700 $\pm$ 570 & 4000 $\pm$ 1300 \\
 HN$^{13}$C(1-0) & -1.3 $\pm$ 0.05 & 2.68 $\pm$ 0.047 & 510 $\pm$ 56 & 1200 $\pm$ 130 \\

\hline
	\end{tabular}
	\label{table:Xalpha_info2}
\end{table*}

{\footnotesize \it \noindent
$^{1}$Argelander-Institut f\"{u}r Astronomie (AIfA), Universit\"{a}t Bonn, Auf dem H\"{u}gel 71, 53121, Bonn, Germany \\
$^{2}$Haystack Observatory, Massachusetts Institute of Technology, Westford, MA 01886, USA \\
$^{3}$Max-Planck-Institut für Radioastronomie, Auf dem Hügel 69, D-53121 Bonn, Germany \\
$^{4}$National Astronomical Observatory of Japan, National Institutes of Natural Sciences, 2-21-1 Osawa, Mitaka, Tokyo 181-8588, Japan \\
$^{5}$Instituto de Radioastronom\'{i}a Milim\'{e}trica (IRAM), Granada, Spain \\
$^{6}$Max-Planck-Institute for Extraterrestrial Physics (MPE), Giessenbachstr. 1, D-85748 Garching, Germany\\
$^{7}$Laboratoire d’astrophysique de Bordeaux, Univ. Bordeaux, CNRS, B18N, all\'ee Geoffroy Saint-Hilaire, 33615 Pessac, France\\
$^{8}$Department of Space, Earth and Environment, Chalmers University of Technology, Onsala Observatory, 439 92 Onsala, Sweden \\
$^{9}$Department of Astronomy The University of Texas at Austin 2515 Speedway, Stop C1400 Austin, Texas 78712-1205, USA \\
$^{10}$Institut für Theoretische Astrophysik, Zentrum für Astronomie der Universität Heidelberg, Albert-Ueberle-Strasse 2, 69120 Heidelberg, Germany \\
$^{11}$Jet Propulsion Laboratory, California Institute of Technology, 4800 Oak Grove Drive, Pasadena CA 91109, USA \\
$^{12}$Institut de Radioastronomie Millimétrique, 300 rue de la Piscine, Domaine Universitaire 38406 Saint Martin d'Hères, France \\
$^{13}$Institute of Astronomy, The University of Tokyo, 2-21-1, Osawa, Mitaka, Tokyo 181-0015, Japan \\
$^{14}$National Astronomical Observatory of Japan, 2-21-1, Osawa, Mitaka, Tokyo 181-8588, Japan \\
$^{15}$Department of Physics and Astronomy, UCL, Gower St., London WC1E 6BT, UK \\
$^{16}$Materials Science and Engineering, College of Engineering, Shibaura Institute of Technology, 3-7-5 Toyosu, Koto-ku, Tokyo 135-8548, Japan \\
$^{17}$ University of Exeter, Stocker Road, Exeter EX4 4QL, UK}

\label{lastpage}
\end{document}